\documentclass[12pt]{JHEP3}
\newcommand{\intd}{\mathrm{d}}

\usepackage{amsmath,epsfig}
\usepackage{placeins}
\usepackage{amssymb,amsfonts}
\usepackage{latexsym}
\usepackage{slashed}
\usepackage{empheq}
\numberwithin{equation}{section}
\usepackage{dsfont}
\usepackage{cancel}
\usepackage{overpic}
\usepackage{subcaption}
\usepackage{subeqnarray}
\usepackage{diagbox}
\usepackage{xcolor}
\let\normalcolor\relax
\usepackage{graphicx}
\usepackage{amsmath}
\usepackage{longtable}
\usepackage{color}
\usepackage{bbm}
\usepackage{bm}

\newcommand{\bw}{\boldsymbol\omega}
\newcommand{\bk}{\boldsymbol k}

\newcommand{\N}{\mathfrak N}
\newcommand{\Dg}{\mathfrak D}
\newcommand{\NT}{\widetilde{\mathfrak N}}
\newcommand{\DT}{\widetilde{\mathfrak D}}
\newcommand{\Lam}{\Lambda}

\usepackage{longtable}
\usepackage{caption}

\usepackage[force]{feynmp-auto}

\allowdisplaybreaks

\usepackage[multiple]{footmisc}

\usepackage{booktabs}

\newcommand{\exclude}[1]{}
\def\nn{\nonumber}

\def\a#1{\alpha_{#1}}

\def\beq{\begin{equation}}
	\def\eeq{\end{equation}}
\def\be{\begin{equation}}
	\def\ee{\end{equation}}
\def\bea{\begin{eqnarray}}
	\def\eea{\end{eqnarray}}
\def\bal{\begin{align}}
	\def\eal{\end{align}}

\def\2b2[#1,#2][#3,#4]{\left( \begin{array}{cc} #1 & #2 \\ #3 & #4 \end{array}
	\right)}
\def\3b3[#1,#2,#3][#4,#5,#6][#7,#8,#9]{\left( \begin{array}{ccc} #1 & #2 #3 \\
		#4 & #5 & #6\\#7&#8&#9\end{array} \right)}

\newcommand\fverb{\setbox\pippobox=\hbox\bgroup\verb}
\newcommand\fverbdo{\egroup\medskip\noindent%
	\fbox{\unhbox\pippobox}\ }
\newcommand\fverbit{\egroup\item[\fbox{\unhbox\pippobox}]}

\usepackage{float}

\newcommand{\bear}{\begin{eqnarray}}
	\usepackage{comment}
	
	\newcommand{\eear}{\end{eqnarray}}

\newcommand{\bsea}{\begin{subeqnarray}}
	\newcommand{\esea}{\end{subeqnarray}}
\newbox\pippobox

\def\6{\partial}
\def\a{\alpha}
\def\nn{\nonumber}

\def\e{\epsilon}

\def\sq
\def\a{\alpha}

\def\e{\epsilon}

\def\D{\Delta}

\def\AA{{\cal A}}

\def\NN{{\cal N}}
\def\OO{{\cal O}}

\title{NLO corrections to charged-current correlators near extremality:\\ non-Abelian hydrodynamics meets AdS$_2$
}
\author{
	M. J\"arvinen$^*$, E. Kiritsis$^{\natural,\flat}$,  F. Nitti$^\natural$, A. Olzi$^{\natural}$, E. Pr\'eau$^\dagger$
	~\\
	$^*$ \href{http://www.itp.cas.cn}{Institute of Theoretical Physics, Chinese Academy of Sciences}, Beijing 100190, China \\
	$^\natural$ \href{http://www.apc.univ-paris7.fr}{Universit\'e  Paris Cit\'e, CNRS, Astroparticule et Cosmologie,  F-75006 Paris, France}\\
	$^\flat$ \href{http://hep.physics.uoc.gr}{Crete Center for Theoretical Physics}, Institute for Theoretical and Computational Physics,
	Department of Physics,\\
	University of Crete, 70013, Heraklion, Greece\\
	$\dagger$ Institute for Theoretical Physics, Utrecht University, 3584 CC Utrecht, The Netherlands
}

\preprint{	CCTP-2026-13\\
	ITCP-2026/13}
\abstract{ We analytically compute the longitudinal charged-current correlator in a
	near-extremal holographic plasma with quark and isospin chemical potentials, $\mu_q$ and $\mu_3$ respectively. To this end, we study fluctuations of gauge fields around a low-temperature AdS$_5$ Reissner-Nordstr\"om black hole charged both under the quark number and the isospin.
	The result is obtained at next-to-leading order in a hydrodynamic expansion where apart from the temperature and isospin chemical potential, the frequency and momenta are small compared to the quark chemical potential, $\omega,k,T,\mu_3\ll\mu_q$. This approach captures the interplay
	between the hydrodynamic-like pole and the tower of infrared modes produced by
	the AdS$_2$ throat controlling the geometry of the black hole near extremality. At zero isospin chemical potential, the resulting
	quasinormal-mode formula reproduces the collision of the diffusive pole with
	the infrared spectrum, while the finite isospin chemical potential lifts this
	collision. The analytic spectrum agrees precisely with the full
	numerical results. The next-to-leading-order correlator also substantially
	improves the spectral function beyond the standard hydrodynamic regime $\omega,k,\mu_3\ll T$,
	extending the agreement with numerics to larger frequencies and momenta.
}

\begin{document}
\maketitle

\section{Introduction}
\label{sec:introduction}

Retarded correlation functions of conserved currents characterize the transport of charge and the relaxation of fluctuations in interacting matter. At long time and length scales, their structure is constrained by hydrodynamics: conservation laws and constitutive relations determine the collective modes and relate their dispersion relations to transport coefficients~\cite{Kovtun:2012rj}. Holography provides a framework in which these observables can be computed at strong coupling. The retarded correlators follow from bulk fluctuations satisfying ingoing boundary conditions at a black-brane horizon \cite{Son:2002sd}, and their poles are the corresponding quasinormal modes (QNMs) \cite{Horowitz:1999jd,Birmingham:2001pj,Kovtun:2005Quasinormal}. This allows one to study both the hydrodynamic excitations and the additional modes that control the departure from the hydrodynamic regime.

At low temperature and finite density, the separation between the hydrodynamical and the  nonhydrodynamical modes becomes subtle. Charged black branes often develop an $\mathrm{AdS}_2$ near-horizon region at extremality. The associated infrared (IR) dynamics produces non-analytic frequency dependence in the boundary correlators \cite{Faulkner:2009wj,Edalati:2009bi,Edalati:2010hk,Edalati:2010pn,Gursoy:2021vpu}. At small non-zero temperature, this non-analytic structure is generically replaced by a sequence of poles with characteristic spacing of order the temperature $T$, which accumulates into a branch cut as $T\to 0$. These modes therefore become parametrically softer than the density scale as the temperature decreases. Consequently, a hydrodynamic pole can interact with the IR spectrum while its frequency and momentum are still small compared with the microscopic scale set by the chemical potential.

The existence of this hierarchy raises the question of how far a hydrodynamic description can be extended into cold, dense matter. Analyses of near-extremal black branes \cite{Edalati:2009bi,Edalati:2010hk,Edalati:2010pn,Davison:2013bxa,Moitra:2020dal,Arean:2020eus,Jarvinen:2023xrx,Gouteraux:2025kta}, have shown that collective modes with hydrodynamic dispersion can persist beyond the conventional thermal regime $\omega,k\ll T$. Their persistence does not, however, imply that a finite-order local derivative expansion captures the complete low-energy response. In particular, the interaction of diffusion with the $\mathrm{AdS}_2$ modes can limit the hydrodynamic expansion at a momentum scale that vanishes as $T\to0$~\cite{Arean:2020eus}. An analytic charge correlator resolving this interaction was constructed in~\cite{Gouteraux:2025kta}. It describes the crossover from diffusion to the low-temperature regime, including successive pole collisions and the non-analytic dispersion of the modes that survive at zero temperature. More generally, the convergence of hydrodynamic dispersion relations
is controlled by singularities in complex momentum associated with
collisions of modes in~\cite{Grozdanov:2019ComplexLife,Jansen:2020ComplexMomentum}.

Another analytic treatment of low-temperature correlators in the
five-dimensional Reissner--Nordstr\"om geometry was developed
in~\cite{Arnaudo:2024LowTemperature}. By solving the Heun connection
problem for charged scalar fluctuations, that work obtains Green's
functions and quasinormal frequencies in the double-scaling limit
$T/\mu,\omega/\mu\to0$ at fixed $\omega/T$ and $k/\mu$, without
requiring a small-momentum expansion. The thermal IR contribution
also exhibits how the zero-temperature branch cut emerges.

A complementary approach to thermal correlators is provided by the thermal product formula. Under appropriate analyticity conditions, it expresses holographic thermal two-sided correlators as products over their poles~\cite{Dodelson:2023vrw}. Its near-extremal application clarifies how low-energy collective modes and IR conformal dynamics determine spectral functions, and identifies the regimes in which these contributions factorize~\cite{Preau:2025rex}. These developments motivate an analytic description that retains the thermal IR response to account for phenomena that are invisible in the leading hydrodynamic approximation.

An additional ingredient arises when the currents carry a non-Abelian charge. A chemical potential for a non-Abelian generator acts as a temporal background gauge field, so currents charged under that generator obey covariant conservation equations \cite{Torabian:2009qk}. In this work, we consider a plasma with a quark chemical potential $\mu_q$ and an isospin chemical potential $\mu_3$. The latter selects a Cartan generator $T_3$ of the flavour group. The flavor components of the current charged under the residual $U(1)_{3}$, associated with the generator $T_3$, are referred to as \textit{charged} currents $J^\pm_\mu$. The $\pm$ refers to sign of the charge under $U(1)_{3}$. Due to their covariant conservation law, their longitudinal correlators (with respect to the three-momentum) involve the shifted frequencies $\omega\pm\mu_3$~\cite{Erdmenger:2007ja,Erdmenger:2008yj,Apostolidis:2026efq}. With Fourier convention $e^{-i\omega t}$, the leading hydrodynamic poles associated with the charged current longitudinal correlator take the form
\begin{equation}
	\omega^\pm_{\mathrm{hydro,LO}}(k)
	=\mp\mu_3-iDk^2,
	\label{intro:hydroLO}
\end{equation}
where $k=|\vec{k}|$ and $D$ is the leading diffusion coefficient. This treatment is valid as long as $\mu_3$ is small compared with the microscopic scale of the theory. At $\mu_3=0$, these reduce to ordinary diffusive poles. At finite $\mu_3$, they describe diffusion together with flavour precession, whose zero-momentum frequency is fixed by covariant charge conservation. We refer to their continuation beyond the conventional hydrodynamic regime as hydro-like modes.

The charged current  retarded correlators were studied in~\cite{Apostolidis:2026efq} using a five -dimensional Einstein--Yang--Mills model with quark and isospin densities.\footnote{Apparent singularities of the charged current fluctuation equation have been recently studied in \cite{Preau:2026hge}.} The background geometry is an asymptotically $\mathrm{AdS}_5$ Reissner--Nordstr\"om black brane, with total charge
\begin{equation}
	\mu\equiv\sqrt{\mu_q^2+\mu_3^2}.
	\label{intro:mu}
\end{equation}
The charged current fluctuations decouple from the metric fluctuations at linear order. At leading order in $\omega,k,T,\mu_3\ll\mu$, the correlators reproduce non-Abelian hydrodynamics, while the exact numerical spectrum reveals the interaction of the hydro-like modes with the IR tower. Finite isospin moves the principal charged pole away from the imaginary frequency axis and changes the collision pattern. An analytic description of these effects requires going beyond leading order.

In this work, we compute longitudinal component of charged current retarded correlator $\left<J^\pm J^\mp\right>^R$ in the Reissner--Nordstr\"om black brane background of~\cite{Apostolidis:2026efq}, to next-to-leading-order (NLO) in a specific hydrodynamic expansion. That is, we work in the near-extremal low-energy regime $	\omega,\ k,\ T,\ \mu_3\ll\mu$, where we adopt more precisely the following scalings
\begin{equation}
	\omega r_H\sim(k r_H)^2\sim T r_H\sim\mu_3 r_H
	=\mathcal{O}(\epsilon),
	\qquad \epsilon\ll1,
	\label{intro:scaling}
\end{equation}
where $r_H$ is the horizon radius and $r_H^{-1}$ is of order $\mu$ near extremality. The relation above corresponds to the diffusive power counting: the rescaled frequency, temperature, isospin chemical potential and squared momentum are held fixed as \(\epsilon\to0\). In particular, we retain the full dependence on \(\omega/T\), and therefore the resulting expressions extend beyond the regime \(\omega\ll T\) of standard hydrodynamics.

Besides its relevance to the structure of near-extremal dynamics, this problem is motivated by weak probes of dense matter. Flavour-current spectral functions enter neutrino transport in strongly coupled matter, where low temperatures and flavour asymmetries are both relevant~\cite{Jarvinen:2023xrx}. The present model provides an analytically tractable setting in which to isolate these effects. Quantitative applications to neutron-star matter require additional physical ingredients beyond those included here.

\subsection{Summary of results}
\label{sec:summary}

Our main result is the analytic NLO longitudinal charged current retarded correlator obtained using the method of matched asymptotic expansion in holography \cite{Faulkner:2009wj}. In this method, the fluctuation equation is solved in two different regions of finite overlap: the \textit{inner} region, which resolves the finite-temperature $\mathrm{AdS}_2$ throat near the black hole horizon, and the \textit{outer}, which contains the rest of the bulk, up to the  $\mathrm{AdS}_5$ boundary. Matching the two solutions at an intermediate holographic coordinate in the overlap region fixes the source and response coefficients of the fluctuation. To determine the correlator to NLO, we compute the inner solution at NLO and the outer solution at NNLO, retaining all terms in the expansion that contribute at the required order after matching in the overlap region. The matched response-to-source ratio is given in formulas \eqref{BAepsilon}--\eqref{Dpm}.

The NLO source coefficient, when set to zero, determines an equation for the QNM frequencies. When the hydro-like pole and the IR poles are sufficiently separated, we obtain explicit perturbative expressions for their frequencies. For the hydro-like pole, the dispersion relation through NLO is
\begin{align}
	\omega^\pm_{\mathrm{hydro,NLO}}
	&={}\mp\mu_3-\frac{i r_H k^2}{2}+
	\nonumber\\
	&+\frac{r_H^2k^2}{24}
	\bigl(\pm2\mu_3+i r_H k^2\bigr)
	\left[
	\gamma_E+
	\psi\!\left(1+\frac{\pm2i\mu_3-r_H k^2}{4\pi T}\right)
	+\log\!\left(\frac{2\pi T r_H}{9}\right)
	\right],
	\label{intro:hydroNLO}
\end{align}
where $\gamma_E$ is the Euler--Mascheroni constant and $\psi(z)=\Gamma'(z)/\Gamma(z)$ is the digamma function. Thus at leading order the diffusion constant reads $D=r_H/2$, and the NLO correction to the dispersion relation vanishes at $k=0$.\footnote{Note that this agrees with an observation of \cite{Preau:2026hge}.}

The IR poles are organized around the frequencies $p^{(n)} =-2\pi iTn$, with $n=1,2,\ldots$. The frequencies $p^{(n)}$ are singularities of the digamma function in \eqref{Npm} and \eqref{Dpm}, not generically exact poles of the full correlator: in order to obtain the precise locations, the singular contributions to its numerator and denominator must be considered together. Their isolated-pole corrections, i.e., the corrections when the IR poles are far from the hydro-like pole, are given in \eqref{eq:NLOIR}:
\begin{align}\nonumber
\omega_\text{IR,NLO}^{\pm,(n)} = &-2\pi i\,T\,n+\\ \label{intro5}
&+\pi\,T\,r_H\,\dfrac{r_H^2k^4+2\pi\,Tr_Hk^2n+2(-2\pi i\,T\,n\pm\mu_3)(-2\pi i\,T\,n\pm2\mu_3)}{6\left[i\,r_Hk^2+2(-2\pi i\,T\,n\pm\mu_3)\right]}\,.
\end{align}

At $\mu_3=0$, the hydro-like pole approaches the $n^{th}$ IR pole and collides with it when $r_Hk^2\simeq4\pi Tn$. The first interaction therefore occurs at a momentum of order $k\sim\sqrt{T\mu}$. Perturbation theory around isolated poles ceases to be reliable in this \textit{interaction} region. In this region, we derive a \textit{local} quadratic equation for the two interacting frequencies.\footnote{We stress that here and after, local is referred to being local in the interacting region.} Its solutions, given in \eqref{int:roots}, remain finite in the local interaction region and reproduce the isolated-pole expansions to NLO when the levels are sufficiently separated. This approximation assumes that the other IR levels remain well separated from the selected pair.

At non-zero real \(\mu_3\), the real-momentum collision is lifted, and the local two-pole expression captures the resulting separation of the modes. A very small non-zero isospin does not by itself guarantee an accurate isolated-pole expansion: near-degenerate modes can still require the interacting description.

At zero temperature, the IR tower becomes a branch cut along the negative imaginary frequency axis, and at non-zero isospin the position of the hydro-like pole relative to this cut fixes the determination of the logarithm in its dispersion relation. As a result, at small momentum each charge channel has a single hydro-like pole, whereas at $\mu_3=0$ there are two. Above a threshold momentum  $r_ek_c\simeq(24\,r_e|\mu_3|/\pi)^{1/4}$, where $r_e$ is the extremal horizon radius, a second hydro-like pole emerges from the cut with a finite damping rate. At small non-zero temperature, and at leading order, it emerges from the IR tower near the level $n_c\simeq r_ek_c^2/(4\pi T)$.

In section \ref{sec:numerics}, we compare our analytic results for both the quasinormal spectrum and the imaginary part of the retarded correlator with the corresponding exact numerical results. Near extremality, the equation that determines at NLO the QNMs reproduces the principal hydro-like branch and the first IR modes with good accuracy at small momentum. The local interacting-pole formulas also capture the first collision at zero isospin and its lifting in the finite-isospin examples considered. The agreement deteriorates as the momentum or temperature increases, since we are moving out of the perturbative region where we derived the NLO expressions. 
For the imaginary part of the retarded correlator, the NLO expression significantly improves agreement with the numerical results over a broader region of the frequency–momentum plane than the LO expression, both at zero and finite isospin.

The paper is organized as follows. Section~\ref{sec:holomodel} introduces the holographic model, the charged fluctuation equations and the method for computing the retarded correlators. Section~\ref{sec:NLOlong} presents the matched asymptotic calculation and the NLO polarization function, including a form in which its zero-temperature limit is manifest. Section~\ref{sec:QNMs} derives its analytic quasinormal spectrum: the hydro-like pole and its zero-temperature limit, the emergence of a second hydro-like pole at non-zero isospin, the IR poles and the interacting poles. Section~\ref{sec:numerics} compares these results with numerical computations of the quasinormal modes and retarded correlators. Appendix~\ref{app:details} collects the technical details of the inner and outer solutions and their matching, appendix~\ref{sec:NLO-residues} computes the residues of the NLO correlator, while appendix \ref{app:otherdef} gives more details about the comparison of the NLO QNMs with the exact results.

\subsection{Outlook}
\label{sec:outlook}

The results obtained in this work suggest several extensions. A natural continuation is to extend the calculation to the transverse charged currents and to the neutral sector of the currents. Together with the longitudinal result, the transverse calculation would determine both polarization functions of the charged-current correlator through NLO.\footnote{The transverse sector will require a slightly different procedure to obtain the NLO correlator. In particular, we would need to go to NNLO in solving the inner equations of motion in order to derive the NLO correlator.}
In the neutral sector, the gauge field fluctuations couple to metric fluctuations, so the matching problem involves a matrix of sources and responses. This extension would allow one to investigate the sector of coupled charge and energy-momentum transport, and to determine which aspects of the pole interaction survive for this system.

It would also be interesting to include a Chern-Simons term in the holographic model to incorporate chiral anomalies. This would shed light on the interplay between anomaly-induced transport, non-Abelian hydrodynamics and the infrared dynamics in the near-extremal limit.

Finally, the analytic NLO current correlators can be used to improve the computation of weak-interaction observables, including neutrino transport \cite{Jarvinen:2023xrx}, or weak reaction rates \cite{Hoyos:2024pkl}.

\section{The holographic model}
\label{sec:holomodel}
In this section, we briefly introduce the holographic model that is used in this work to compute thermal correlators of flavor currents.

We consider a bottom-up holographic model of a strongly interacting plasma with \(U(N_f)_L\times U(N_f)_R\) chiral flavor symmetry. Its bulk description consists of five-dimensional gravity coupled to flavor gauge fields in an asymptotically AdS\(_5\) spacetime. The field content of the bulk theory is dictated by the types of operators that we want the dual (boundary) quantum field theory to include. In the present case, in addition to the metric dual to the stress-energy tensor, the operators of interest are the chiral flavor currents $\mathcal{J}^{(L/R)}_\mu$, which are dual to chiral gauge fields in the five-dimensional bulk $\mathbf{L}_M$ and $\mathbf{R}_M$, where \(\mu\) and \(M\) denote boundary and bulk spacetime indices, respectively

The bulk action describing the dynamics of these fields is given by
\begin{align}\nonumber
S = &M^3N_c^2 \int\mathrm{d}^5x\sqrt{-g}\, \left(R + \frac{12}{\ell^2}\right) +\\ \label{eq:Sb}
& - \frac{1}{8\ell}(M\ell)^3w_0^2 N_c \int\mathrm{d}^5x\sqrt{-g}\,\text{Tr}\left(\mathbf{F}_{MN}^{(L)}\mathbf{F}^{MN,(L)} + \mathbf{F}_{MN}^{(R)}\mathbf{F}^{MN,(R)} \right) \, .
\end{align}
The first term in \eqref{eq:Sb}, i.e.,  the five-dimensional Einstein-Hilbert action, is  associated with the color sector. In this term $R$ is the five-dimensional Ricci scalar, $M$ the five-dimensional Planck mass, $\ell$ the AdS$_5$ radius and $N_c$ the number of colors. The second line in \eqref{eq:Sb} is a Yang-Mills action, quadratic in the field strengths, describing the flavor sector where $w_0$ controls the (inverse) flavor Yang-Mills coupling, and $\mathbf{F}^{(L/R)}$ are the field strengths of the gauge fields $\mathbf{L}$ and $\mathbf{R}$ defined as
\begin{equation}
\label{eq:defF} \mathbf{F}^{(L)} \equiv \mathrm{d}\mathbf{L} - i\mathbf{L}\wedge \mathbf{L} \,,\qquad\qquad \mathbf{F}^{(R)} \equiv \mathrm{d}\mathbf{R} - i\mathbf{R}\wedge \mathbf{R} \, .
\end{equation}

In order to treat the gravitational theory at a semi-classical level, we assume a large number of colors $N_c$. Moreover, we consider the back-reaction of the flavor sector on the color one in order to describe strongly interacting dense matter. These conditions are accomplished either if the non-abelian gauge bosons come from the bulk, or in the Veneziano large-$N_c$ limit~\cite{Veneziano:1976wm}
\begin{equation}
\label{eq:V1}
N_c \to \infty \,, \qquad N_f \to \infty \,, \qquad \frac{N_f}{N_c} \ \text{fixed} \, .
\end{equation}
if they come from bulk flavor branes.

The range of parameters $M\ell$ and $w_0$ can be  determined if we treat this as a toy model for QCD and compare orders of magnitude with the real world QCD, specifically the equation of state at high $T$ and $\mu_q$,
see \cite{Jarvinen:2023xrx,Apostolidis:2026efq} for details. Note that this means comparing the strong-coupling holographic model to a weak-coupling perturbative QCD result. However, such a comparison is enough to obtain a reasonable toy mode because the high temperature perturbative results give a rough description of true the equation of state QCD even down to low temperatures (of the order of $\Lambda_{\rm QCD}$).
For the numerical comparisons in Section \ref{sec:numerics}, we use the phenomenological values corresponding to $N_c=3$ and $N_f=2$:
\begin{align}
	&(M\ell)^3 = \dfrac{13}{6}\dfrac{1}{45\pi^2}\,,& &w_0 = \dfrac{180}{13}\,.&
\end{align}

For a chiral symmetry group $\mathrm{U}(2)_L \times \mathrm{U}(2)_R$, the chiral currents and their dual bulk gauge fields can be expanded in the Pauli basis $\{\sigma_a\}$, with $a=1,2,3$,
\begin{equation}\label{eq:JsP}
\mathcal{J}^{(L)}_\mu = \frac{1}{2}\hat{J}^{(L)}_\mu\mathbb{I}_2 + \frac{1}{2}\sum_{a=1}^3 J^{a,(L)}_\mu \sigma_a\,,\qquad\qquad \mathbf{L}_M = \frac{1}{2}\hat{L}_M\mathbb{I}_2 + \frac{1}{2}\sum_{a=1}^3 L^{a}_M \sigma_a \, ,
\end{equation}
\begin{equation}
\mathcal{J}^{(R)}_\mu = \frac{1}{2}\hat{J}^{(R)}_\mu\mathbb{I}_2 + \frac{1}{2}\sum_{a=1}^3 J^{a,(R)}_\mu \sigma_a \,,\qquad\qquad \mathbf{R}_M = \frac{1}{2}\hat{R}_M\mathbb{I}_2 + \frac{1}{2}\sum_{a=1}^3 R^{a}_M \sigma_a \, .
\end{equation}

From now on, we only consider the vector currents, $\mathcal{J}_{\mu}^{(V)}$, and their dual bulk vector gauge fields, $\mathbf{V}_M$. They are related to the left- and right-handed components by
\begin{equation}
\mathcal{J}^{(V)}_\mu = \frac{1}{2}(\mathcal{J}_{\mu}^{(L)}+\mathcal{J}_{\mu}^{(R)}) \,,\qquad\qquad \mathbf{V}_M =\dfrac{1}{2}(\mathbf{L}_M+\mathbf{R}_M)\, .
\end{equation}
They are associated with the vector symmetry group U(1)$_V$ $\times$ SU(2)$_V$, which is the unbroken symmetry group in real-world QCD.\footnote{We restrict to the vector sector by setting the axial combination \((\mathbf{L}_M-\mathbf{R}_M)/2\) to zero, so that \(\mathbf{L}_M=\mathbf{R}_M=\mathbf{V}_M\).}

\subsection{Background solution}
\label{sec:bkgsol}

We now present the background solution for the bulk action \eqref{eq:Sb}, at finite temperature and baryon and isospin density. The state of matter that it describes in the dual boundary theory, corresponds to a plasma of deconfined (generalized) quarks and gluons, with isospin asymmetry. This background solution has been found many times in the past in various multi-vector contexts, and recent discussion in the present context can be found in \cite{Apostolidis:2026efq}.

The metric takes the form of an asymptotically $\operatorname{AdS}_{5}$ Reissner-Nordström (RN) black brane
\begin{equation}
	\label{eq:ds2RN} \mathrm{ds}^2 =e^{2A(r)}\left( -f(r)\mathrm{dt}^2 + f(r)^{-1}\mathrm{d}r^2 + \vec{\mathrm{d x}}^2 \right) \, ,
\end{equation}
with $r$ the holographic coordinate defined such that $r=0$ is the UV boundary. The geometry enjoys an outer horizon at $r=r_H$, where the blackening function vanishes ($f(r_H)=0$). The vector gauge field takes the following form
\begin{equation}
	\label{eq:Ba5} \mathbf{V}_M = \frac{1}{2}\,\delta^{0}\,_M\, \left(\Phi(r)\mathbb{I}_2+\Phi_3(r)\sigma_3\right) \, .
\end{equation}

The analytic expression for the functions $A(r)$, $f(r)$, $\Phi(r)$ and $\Phi_3(r)$ are given by
\begin{equation}\label{bmet}
f(r) = 1 - \left(\frac{r}{r_H}\right)^4\left(1 + 2\left(1 - \pi T r_H\right)\right) + 2\left(1 - \pi T r_H\right)\left(\frac{r}{r_H}\right)^6  \, ,
\end{equation}
\begin{equation}
\label{eq3:Phiasrh}
e^{A(r)} = \frac{\ell}{r} \,,\qquad\Phi(r)=\mu_q\left(1-\left(\frac{r}{r_H}\right)^2\right)\,, \qquad \Phi_3(r)=\mu_3\left(1-\left(\frac{r}{r_H}\right)^2\right)\, .
\end{equation}
The presence of a non-zero $\mu_3$ enters in the expression of the horizon radius $r_H = r_H(T,\mu_q,\mu_3)$, with $T$ the temperature. For convenience we can define a total chemical potential $\mu$ as
\begin{equation}
\mu \equiv \sqrt{\mu_q^2+\mu_3^2}\, ,
\end{equation}
so that the horizon radius reads
\begin{equation}
\label{eq:rHRN} r_H = \frac{2}{\pi T} \left( 1 + \sqrt{1 + \frac{w_0^2}{3N_c}\frac{\mu^2}{\pi^2T^2}} \right)^{-1}\, .
\end{equation}
For a near-extremal gravitational solution, $\mu\gg T$, we can expand the horizon radius around its extremal value $r_e$:
\begin{equation} \label{eq:re}
	r_H = r_e\left(1-\dfrac{\pi\sqrt{3N_c}}{w_0}\dfrac{T}{\mu}+\mathcal{O}\left(\dfrac{T^2}{\mu^2}\right)\right)\,,\qquad 	r_e \equiv \dfrac{2\sqrt{3N_c}}{w_0\,\mu}\,.
\end{equation}

\subsection{Fluctuation equations}

We report in this section the equations of motion for the so-called \textit{charged} gauge field fluctuations. As it will be clear in a few lines, the terminology charged refers to charge under the residual Cartan subgroup generated by $\sigma_3$. We refer to \cite{Apostolidis:2026efq} for a detailed derivation of the fluctuation equations.

The linearized field equations for small perturbations $\delta V^{1,2}$ of the bulk gauge fields dual to the flavor vector currents $J_{(V)}^{1,2}$ are
\begin{equation}
\label{eq:EoML} \partial_M\left(\sqrt{-g} \,\delta F_{(V)}^{M N}\right)-i\sqrt{-g}\left[\delta V_{M},F_{(V)}^{MN}\right]-i\sqrt{-g}\left[ V_{M},\delta F_{(V)}^{MN}\right] = 0 \, ,
\end{equation}
where
\begin{equation}
\label{eq:deltaF}
\delta F^{(V)}_{MN}=\partial_M\delta V_N -\partial_N \delta V_M-i\left[\delta V_{M},V_N\right]-i\left[V_{M},\delta V_N\right]\, .
\end{equation}
We work in the radial gauge
\begin{equation}
\label{eq:Ag} \delta V^{a}_r=0 \, ,
\end{equation}
and define the four-dimensional Fourier transform of the perturbation as
\begin{equation}
\label{eq:ansL} \delta V^a_\mu(r;t,\vec{x}) = \int \frac{\intd^4 k}{(2\pi)^4} e^{-i(\omega t - \vec{k}.\vec{x})} \mathcal{V}^a_{\mu,(\omega,\vec k)}(r) \, .
\end{equation}
From now on, we drop the Fourier labels $\omega,\vec k$ on the gauge field fluctuation to simplify the notation. 

We can decompose the spatial components of the vector fluctuation into a transverse $\mathcal{V}_i^\perp$, and longitudinal part $\mathcal{V}_i^\parallel$ to the 3-momentum $\vec k$ as follows
\begin{equation}
	\mathcal{V}_i^{\parallel} = \dfrac{k_ik_j}{k^2}\mathcal{V}_j \quad,\quad \mathcal{V}_i^{\perp} = \mathcal{V}_i -\dfrac{k_ik_j}{ k^2}\mathcal{V}_j\,,
\end{equation}
where we have indicated the modulus of the spatial momentum with $k \equiv |\vec k|$.

Now, we perform a change of basis involving the generators of the SU(2) flavor group. We define the \textit{charged} Pauli matrices $\sigma_\pm$ as a linear combination of $\sigma_{1,2}$
\begin{equation}
	\sigma_\pm = \dfrac{\sigma_1\pm i\,\sigma_2}{2}\,,
\end{equation}
and use them instead of $\sigma_1$ and $\sigma_2$.
In this new basis, the flavor indices run over $a\in\{+,-,3\}$ and the charged components of the bulk gauge field can be written as
\begin{equation}\label{eq:chargedbasis}
	\mathcal{V}_M^\pm = \mathcal{V}^1_M\mp i\, \mathcal{V}^2_M\,.
\end{equation}
The terminology ``charged'' refers to charge under the residual Cartan subgroup U(1)$_{I_3}\subset$ SU(2) generated by $\sigma_3$. Indeed, the charged Pauli matrices are eigenvectors of the adjoint action of $\sigma_3$,
\begin{equation}
	[\sigma_3,\sigma_\pm]=\pm 2\sigma_\pm\,.
\end{equation}
Therefore the fluctuations along $\sigma_+$ and $\sigma_-$ carry charge $+1$ and $-1$, respectively, under U(1)$_{I_3}$ associated with the generator $T_3 = \sigma_3/2$.

Then, we can write equations of motion \eqref{eq:EoML} in the basis \eqref{eq:chargedbasis} where the charged components $\mathcal{V}_M^{\pm}$ receive contributions from commutators with the non-zero background isospin gauge field. At quadratic order, these fluctuations decouple from the other components (in the group manifold) of the vector fluctuations and the metric fluctuations.

We can define a gauge-invariant combination of the vector fluctuation components in the longitudinal sector which reads
\begin{equation}
	\label{defEpm} E^{\parallel,\pm} \equiv k\,\mathcal{V}_0^{\pm} + \frac{\omega\pm\Phi_3}{k}\,k_i\mathcal{V}_i^{\pm} \, .
\end{equation}

Then, the differential equations for the fluctuations $\mathcal{V}_i^{\perp,\pm}$ and $E^{\parallel,\pm}$ are
\begin{equation}
\label{EoMVhtpm} \partial_r\left(\frac{f(r)}{r}\partial_r\mathcal{V}_i^{\perp,\pm}\right) + \frac{1}{rf(r)}\left(\Omega_\pm(r)^2-k^2f(r)\right)\mathcal{V}_i^{\perp,\pm}=0\, ,
\end{equation}
\begin{equation}
	\label{EpEoM} \partial_r\left(\frac{ f(r) \partial_rE^{\parallel,\pm}}{r \big(\Omega_\pm(r)^2-f(r)k^2\big)}\right)+ \left(1 + f(r) \Phi_3'(r)\frac{2 f(r) \Phi_3'(r)\mp f'(r)\Omega_\pm(r)}{\big(\Omega_\pm(r)^2-f(r)k^2\big)^2}\right)\frac{E^{\parallel,\pm}}{r f(r)}=0\, ,
\end{equation}
where we have defined the \textit{shifted} frequency
\begin{equation}
	\label{defOpm} \Omega_\pm(r) \equiv \omega \pm \Phi_3(r) \, .
\end{equation}
The equation for $E^{\parallel,\pm}$ may be further simplified by considering the field redefinition
\begin{equation}
	\label{fr} E^{\parallel\pm}(r) \equiv \Omega_\pm(r) \varphi^\pm(r).
\end{equation}
We obtain
\begin{equation}
	\label{Eqvf} \partial_r\left(\frac{f(r) \Omega_\pm(r)^2}{r \big(\Omega_\pm(r)^2- f(r)k^2\big)}\,\partial_r\varphi^\pm\right) + \frac{\Omega_\pm(r)^2}{rf(r)}\varphi^\pm = 0 \, .
\end{equation}
The equations in the two charged sectors are related by \(\mu_3\to-\mu_3\). It therefore suffices to solve one sector.

\subsection{Holographic retarded correlators for the charged currents}

The general form of the two-point function for the charged currents has been derived in \cite{Apostolidis:2026efq}. Here we report its expression and we refer to \cite{Apostolidis:2026efq} for more details.
The general tensor structure of the charged current two-point function can be inferred from the symmetries of the background. Spatial rotational invariance and the non-Abelian Ward identities constrain the charged-current correlator to the following form
\begin{align}\nonumber
	\left<J^{\pm}_\lambda J^{\mp}_\sigma\right>^R(\omega,\vec k\,) =& \left(P_{\lambda\sigma}^{\perp,\pm}(\omega,\vec k\,) i\Pi^{\perp,\pm}(\omega,\vec k\,) + P_{\lambda\sigma}^{\parallel,\pm}(\omega,\vec k\,) i\Pi^{\parallel,\pm}(\omega,\vec k\,) \right) + \\ \label{ltcR}
	&\mp i\delta_\lambda^0\delta_\sigma^0\dfrac{n_3}{\omega\pm\mu_3} \, .
\end{align}
where $n_3$ is the background isospin density. It is the last term here which is fixed by the Ward identity.

The transverse and longitudinal projectors are defined with respect to the shifted momentum $k_\mu^\pm$, defined as
\begin{equation}\label{eq:shiftedk}
	k_{\mu}^{\pm}=k_{\mu}\mp\mu_3\delta^{0}\,_{\mu}\,,\qquad\qquad k_\mu = (-\omega,\vec k)\, .
\end{equation}
The non-zero components of $P^{\perp,\pm}_{\lambda\sigma}$ and $P^ {\parallel,\pm}_{\lambda\sigma}$ are
\begin{equation}\label{Po} P^{\perp,\pm}_{ij}=P^\perp_{ij} = \delta_{ij} - \frac{k_ik_j}{k^2} \, ,
\end{equation}
\begin{equation}
\label{Pp} P^{\parallel,\pm}_{00} = \frac{k^2}{\left(\omega\pm\mu_3\right)^2-k^2} \,\, , \,\, \,\,\,\,P^{\parallel,\pm}_{0i}= P^{\parallel,\pm}_{i0} =- \frac{\left(\omega\pm\mu_3\right) k_i}{\left(\omega\pm\mu_3\right)^2-k^2} \,\, , \,\,
\end{equation}
\begin{equation}\label{Pq}
P^{\parallel,\pm}_{ij} = \frac{k_ik_j}{k^2}\frac{\left(\omega\pm\mu_3\right)^2}{\left(\omega\pm\mu_3\right)^2-k^2} \, .
\end{equation}
The other components are zero,
\begin{align}
	P^{\perp,\pm}_{00} = P^{\perp,\pm}_{0i}=P^{\perp,\pm}_{i0}=0.
\end{align}

Following \cite{Son:2002sd}, we can derive, within the holographic model, the charged current retarded two-point function. We focus on the vector gauge fields, for which the Yang-Mills action to quadratic order in the perturbations takes the form
\begin{equation}
	\label{pert_quad_action}
	\delta^{(2)}S_f = - \frac{1}{16\ell}(M\ell)^3w_0^2 N_c \int\mathrm{d}^5x\sqrt{-g} \left(\delta F^{(V)}_{MN,a}\delta F^{MN,a}_{(V)} \right) \, ,
\end{equation}
where $\delta F^{MN}_{(V)}= g^{MP}g^{NQ}\delta F^{(V)}_{PQ}$ and the linearized Yang-Mills field strength with explicit group indices is given by
\begin{equation}
	\delta F^{(V)\,a}_{MN}=\partial_M\delta V^{a}_{N}-\partial_N\delta V_{M}^{a}+\epsilon^{abc}\left(\delta V^{b}_{M}V^{c}_{N}+V^{b}_{M}\delta V^{c}_{N}\right)\, .
\end{equation}
After some algebra, according to the prescription of \cite{Son:2002sd}, the transverse and longitudinal polarization functions introduced in \eqref{ltcR} for the flavor currents are given by
\begin{equation}
\label{PLt} \Pi^{\perp,\pm}(\omega,\vec{k}\,) = -\frac{1}{8}(M\ell)^3w_0^2 N_c \frac{1}{r} \left.\frac{\partial_r\mathcal{L}_i^{\perp,\pm}}{\mathcal{L}_i^{\perp,\pm}} \right|_{r=\delta} \, ,
\end{equation}
\begin{align}
	\label{PLl} \Pi^{\parallel,\pm}(\omega,\vec{k}\,) &= -\frac{1}{8}(M\ell)^3w_0^2 N_c \frac{1}{r} \left.\left(\frac{\partial_rE^{\parallel,\pm}}{E^{\parallel,\pm}}\mp\frac{\partial_r\Phi_3}{\omega\pm\mu_3}\right) \right|_{r=\delta} \\ \nonumber
	&=-\frac{1}{8}(M\ell)^3w_0^2 N_c \frac{1}{r} \left.\frac{\partial_r\varphi^{\pm}}{\varphi^{\pm}}\right|_{r=\delta}\,.
\end{align}
In the last line,  we used the field $\varphi^\pm$ defined in \eqref{fr} to absorb the contact term and rewrite the polarization function in a more compact way. The above expressions are evaluated at $r=\delta$ with $\delta$ the UV cutoff.

The polarization functions \eqref{PLt} and \eqref{PLl} require a renormalization procedure to remove the logarithmic UV divergence which affects the real part of the polarization functions. In \cite{Apostolidis:2026efq}, a consistent renormalization procedure has been carried out for these particular polarization functions. In this work, we shall be mainly interested in
the imaginary part of the longitudinal polarization function which is
finite 
and hence does not require renormalization.

\subsubsection{The matching procedure for the longitudinal retarded correlator}\label{sec:matchingprocedure}
In this section, we present the so-called matching procedure (see \cite{Faulkner:2009wj,Davison:2013bxa,
	Arean:2020eus,Gouteraux:2025kta}) to compute analytically thermal correlators at finite density. As mentioned in the introduction, we focus on the longitudinal correlator which displays the most interesting features, in particular the interplay between the hydro-like pole and the IR poles associated with the emergent AdS$_2$ region. The procedure for computing the thermal correlator in the transverse sector is conceptually the same, and we do not report it here.

It is useful to define the following quantity
\begin{equation}
	\tau = \pi\,T\,r_H = \dfrac{2\pi\sqrt{3N_c}}{w_0}\dfrac{T}{\mu}+\mathcal{O}(T^2/\mu^2)\,,
\end{equation}
where in the second equality we expanded the horizon radius in the near-extremal limit $\mu\gg T$, where it is linearly suppressed. 
Then, we define a dimensionless holographic coordinate
\begin{equation}
	u = \dfrac{r}{r_H}\,,
\end{equation}
in terms of which the AdS$_5$ boundary is placed at $u = 0$ while the horizon radius location is $u = 1$. Using the above definitions of $u$ and $\tau$, the blackening factor reads
\begin{equation}\label{eq:blackresc}
	f(u) = 1-(3-2\tau)u^4+ 2(1-\tau)u^6\,.
\end{equation}
The extremal limit corresponds to $\tau =0$.

We want to derive an analytic solution to the fluctuation equations in the limit where  frequency, momentum and isospin chemical potential are all small with respect to the total chemical potential $\mu$. Therefore, we rescale them with $r_H\sim\mu^{-1}$ as\footnote{In the formula \eqref{eq:exp2}, we have rescaled the modulus of the spatial momentum since it is only the modulus $k$ that enters the equation of motion. Such a rescaling has to be intended as we have rescaled with $r_H$ every component of the vector $\vec k$.}
\begin{align}\label{eq:exp2}
	&\boldsymbol{k} = k\,r_H\,,& &\mathfrak{m}=\mu_3\,r_H\,,& &\boldsymbol{\omega} = \omega\,r_H\,.&
\end{align}
For later convenience, it is useful to define the frequency rescaled with the temperature as
\begin{equation}\label{eq:hatomega}
	\hat\omega= \dfrac{\omega}{2\pi T}=\dfrac{\boldsymbol{\omega}}{2\tau}\,.
\end{equation}
We stress that, in the near-extremal limit, $\omega$ can be smaller, bigger or of the same order of $T$, therefore $\hat \omega$ is not parametrically small.

Then, the longitudinal fluctuation equation can be written in dimensionless coordinates as
\begin{equation}\label{eq:EOM}
	\partial_u\left(\frac{f(u) \boldsymbol{\Omega}_\pm(u)^2 \partial_u\varphi^\pm(u)}{u \big(\boldsymbol{\Omega}_\pm(u)^2- f(u)\boldsymbol{k}^2\big)}\right) + \frac{\boldsymbol{\Omega}_\pm(u)^2}{uf(u)}\varphi^\pm(u) = 0 \,,
\end{equation}
where we have also introduced the rescaled shifted frequency $\boldsymbol{\Omega}_\pm$ which takes the following form
\begin{equation}\label{eq:barOmega}
	\boldsymbol{\Omega}_\pm = r_H\,\Omega_\pm= \boldsymbol{\omega} \pm\mathfrak{m}\,(1-u^2) = 2\tau\,\hat\omega \pm\mathfrak{m}\,(1-u^2)\,.
\end{equation}

As a bookkeeping parameter, we introduce $\epsilon\ll 1$, which we use to rescale the dimensionless parameters $\tau$, $\boldsymbol{k}^2$, $\mathfrak{m}$, and $\boldsymbol{\omega}$ as
\begin{align}
	&\tau\to\epsilon\,\tau\,,&  &\boldsymbol{k}\to\sqrt{\epsilon}\,\boldsymbol{k}\,,&  &\mathfrak{m}\to\epsilon\,\mathfrak{m}\,,&  &\boldsymbol{\omega}\to\epsilon\,\boldsymbol{\omega}\,,&
	\label{om}\end{align}
The presence of the small parameter $\epsilon$ implements the near-extremal limit $\omega,k,T,\mu_3\ll\mu$. Note that the spatial momentum 
has a different scaling than the frequency
in $\epsilon$.
This is because in the longitudinal sector we expect a diffusive behavior of the hydrodynamic pole $\omega \sim k^2$ and thus we adapt the scaling to this behavior.

We stress that equation \eqref{om} is one particular choice of scaling with $\epsilon$ which is useful to zoom in the diffusive region of the longitudinal correlator. This scaling implements a near-extremal, low-energy limit adapted to the longitudinal hydro-like pole, for which \(\omega\pm\mu_3\) scales as \(k^2\). Other scalings probe different regimes and are not considered here. Since frequency and temperature carry the same power of \(\epsilon\), their ratio \(\hat\omega\) is held fixed as \(\epsilon\to0\).\footnote{In \cite{Faulkner:2009wj}, this is a consequence of how the time coordinate is rescaled to zoom in the AdS$_2$ throat in the near extremal limit.}

The computation of the dispersion relation and retarded correlator is performed using the method of matched asymptotic expansions, following \cite{Faulkner:2009wj}. In the near-extremal limit and near the horizon, the black-brane geometry develops an AdS$_2$ throat. We therefore divide the spacetime into an \textit{inner region}, which resolves the near-horizon AdS$_2$ throat, and an \textit{outer region}, which extends from the vicinity of the horizon to the asymptotic AdS$_5$ boundary.

In the inner region, the limit $\epsilon\to0$ is taken while keeping
\begin{equation}
	\zeta=\frac{1-u}{\epsilon\,\tau}
\end{equation}
fixed. The coordinate $\zeta$ is the holographic radial coordinate in the AdS$_2$ throat of the near-horizon geometry. More precisely, the coordinate transformation above together with a rescaling of the time coordinate transforms the background AdS$_5$ -RN metric into an AdS$_2\times \mathbb{R}^3$ metric, at leading order in the $\epsilon$ expansion. The horizon in this new radial coordinate is located at $\zeta=0$, which is where we impose the ingoing boundary condition appropriate for the retarded correlator. This condition selects a unique inner solution up to an overall normalization. The inner solution is then computed perturbatively in $\epsilon$, and then, it is expanded for large values of $\zeta$, namely towards the boundary of the AdS$_2$ throat.

In the outer region, the coordinate $u$ is instead kept fixed as $\epsilon\to0$. The fluctuation equation is solved perturbatively in $\epsilon$. At each order, because the equation is linear, its general solution contains two independent 
integration constants. Except for the leading order solution, there is also in addition a particular solution generated by the lower-order terms. The relative coefficient of the two independent outer solutions is determined by matching with the ingoing inner solution, while the overall normalization remains arbitrary.

The inner and outer expansions are then matched in an intermediate holographic
radial domain, whose parametric extent can be determined directly from the
fluctuation equation.  To see this, we define
\begin{equation}
	\delta u\equiv 1-u\,.
\end{equation}
Close to extremality, the blackening factor behaves as
\begin{equation}
	f(u)
	=12\,\delta u^2+4\epsilon\,\tau\,\delta u+\mathcal{O}(\delta u^3, \epsilon\, \delta u^2)\,.
\end{equation}
Using the equation above, in the outer region, the expansion in the equation of motion \eqref{eq:EOM} of the factor
$\big(\boldsymbol{\Omega}_\pm^2-f(u)\boldsymbol{k}^2\big)^{-1}$, after the rescaling \eqref{om} requires
\begin{equation}
	\delta u\gg\sqrt{\epsilon}\,.
	\label{eq:outer-validity-region}
\end{equation}
The inner solution, on the other hand, is obtained by keeping
\begin{equation}
	\zeta=\frac{1-u}{\epsilon\,\tau}
\end{equation}
fixed as $\epsilon\to0$ and therefore $\delta u\sim \epsilon \ll 1$. Combining
this requirement with \eqref{eq:outer-validity-region}, we obtain the
overlap region
\begin{equation}
		\sqrt{\epsilon}\ll1-u\ll1\,.
	\label{eq:matching-overlap-region}
\end{equation}
Therefore, for sufficiently small $\epsilon$, this is the interval where the matching can be performed.

The large-$\zeta$ expansion of the inner solution and the near-horizon
expansion of the outer solution can therefore be compared in this domain.
Rewriting both expansions in terms of the same variable through
$1-u=\epsilon\,\tau\,\zeta$ and equating the coefficients of the independent
powers and logarithms fixes the relative outer integration constants in terms
of the normalization of the ingoing inner solution.  Since the fluctuation
equation is linear, the remaining common normalization cancels from the ratio
of response to source that determines the retarded correlator.

Once the matching coefficients have been determined, the outer solution can be expanded close to the AdS$_5$ boundary. The near-boundary expansion of the longitudinal fluctuation in the $u$-coordinate takes the form
\begin{align}\label{eq:NBexp}
	&\varphi^{\pm} = \mathcal{A}^\pm+ u^2\left(\mathcal{B}^\pm - \frac{r^2_H\,\mathcal{A}^\pm}{2}\left((\omega\pm\mu_3)^2-k^2\right) \log(r_Hu/\ell)\right)\Big(1 + \mathcal{O}\big(r^2\big)\Big) \, ,
\end{align}
where we have denoted the source and the response respectively as $\mathcal{A}^\pm(\omega,k,\mu_3,T)$ and $\mathcal{B}^\pm(\omega,k,\mu_3,T)$.

The longitudinal retarded polarization function is therefore
\begin{align}\label{eq:long:matchedcorrelator}
	&\Pi^{\parallel,\pm}(\omega,k)
	=
	-\frac{1}{4r_H^2}(M\ell)^3w_0^2N_c\,
	\frac{\mathcal{B}^\pm(\omega,k,\mu_3,T)}
	{\mathcal{A}^\pm(\omega,k,\mu_3,T)}\,.
\end{align}

The poles of the retarded correlators, \textit{i.e.} the quasinormal modes of the black hole geometry, are obtained by imposing the absence of a source at the asymptotic boundary while maintaining the ingoing condition at the horizon. In terms of the matched coefficients, the location of the QNMs for the longitudinal correlator, is therefore determined by solving
\begin{align}\label{eq:dispersioncondition}
	\mathcal{A}^\pm(\omega,k,\mu_3,T)=0\,.
\end{align}
Expanding the denominator according to
\begin{equation}
	\mathcal{A}^\pm
	=
	\mathcal{A}^{\pm,(0)}
	+\epsilon\,\mathcal{A}^{\pm,(1)}
	+\mathcal{O}(\epsilon^2)\,,
\end{equation}
and solving \eqref{eq:dispersioncondition} order by order yields the leading- and next-to-leading-order contributions to the dispersion relation. For a simple root, writing
\begin{equation}
	\omega_\text{QNM}^{\pm}
	=
	\omega_\text{QNM}^{\pm,(0)}
	+\epsilon\,\omega_\text{QNM}^{\pm,(1)}
	+\mathcal{O}(\epsilon^2)\,,
\end{equation}
one obtains\footnote{The formulas follow from Taylor expanding the source coefficient around the leading-order quasinormal frequency $\omega_\text{QNM}^{\pm,(0)}$ and setting it to zero order by order in $\epsilon$
	\begin{align}\nonumber
		\mathcal{A}^\pm\left(\omega_\text{QNM}^{\pm,(0)}
		+\epsilon\,\omega_\text{QNM}^{\pm,(1)}\right)  = &\mathcal{A}^{\pm,(0)}\left(\omega_\text{QNM}^{\pm,(0)}\right) +\epsilon\,\omega_\text{QNM}^{\pm,(1)}\,\partial_{\omega}\mathcal{A}^{\pm,(0)}\left(\omega_\text{QNM}^{\pm,(0)}\right) +\\
		&\epsilon\,\mathcal{A}^{\pm,(1)}\left(\omega_\text{QNM}^{\pm,(0)}\right) +\mathcal{O}(\epsilon^2) = 0\,.
\end{align}}
\begin{align}
	&\mathcal{A}^{\pm,(0)}
	\left(\omega_\text{QNM}^{\pm,(0)},k,\mu_3,T\right)=0\,,
	\\
	&\omega^{\pm,(1)}_\text{QNM}
	=
	-\left.
	\frac{\mathcal{A}^{\pm,(1)}(\omega,k,\mu_3,T)}
	{\partial_{\omega}\mathcal{A}^{\pm,(0)}(\omega,k,\mu_3,T)}
	\right|_{\omega=\omega_\text{QNM}^{\pm,(0)}}\,.
\end{align}

When two QNMs approach one another, the expansion around either isolated root can cease to be reliable. We therefore develop a local approximation that treats the two nearby modes together, as discussed in section \ref{sec:QNMs}.

\section{The NLO retarded correlator}\label{sec:NLOlong}
In this section, we derive the formula for the longitudinal retarded polarization function at NLO in the hydrodynamical expansion. The section is organized as follows: we first derive the leading- and next-to-leading-order solutions of the differential equation \eqref{eq:EOM} in the inner region (section \ref{sec:inner}) and in the outer region (section \ref{sec:outer}). As we shall explain, in the outer region,  we need also the next-to-next-to-leading order solution. We then perform the matching of the inner and outer solutions in section \ref{sec:matching}, from which the source $\mathcal{A}^\pm$ and response $\mathcal{B}^\pm$ are computed up to NLO. The NLO longitudinal polarization function is finally obtained as the ratio of these two quantities. The quasinormal modes of this correlator, $\omega^{\pm}_\text{QNM} = \omega^{\pm}_\text{QNM}(k,\mu_3,T)$, are studied in section~\ref{sec:QNMs}, where we also analyse the interplay between the two different kinds of poles seen by the NLO correlator: the hydro-like pole and the IR AdS$_2$ poles.

\subsection{Inner region}\label{sec:inner}

In this  subsection, we present the solution to the equation of motion \eqref{eq:EOM} in the inner region, where we define
\begin{equation}\label{eq:zetadef}
    u = 1-\epsilon\,\tau\,\zeta\,\qquad \qquad \zeta\geq 0\,,
\end{equation}
with $\zeta$ the holographic coordinate of the AdS$_2$ throat. The horizon is placed at $\zeta=0$, while the boundary of the AdS$_2$ region is at $\zeta =\infty$. Using this coordinate, the blackening factor displays the following structure
\begin{equation}\label{eq:fzeta}
    f(\zeta) = 4h(\zeta)\,\tau^2\epsilon^2 + g(\zeta)\,\tau^3\epsilon^3+\mathcal{O}(\epsilon^4)\,,
\end{equation}
where we used the rescaling \eqref{om}. The functions $h(\zeta)$ and $g(\zeta)$ are defined as
\begin{align}\label{eq:hg}
    &h(\zeta) = \zeta(1+3\zeta)\,,& &g(\zeta)=-2\zeta^2(9+14\zeta)\,.&
\end{align}

The rescaled shifted frequency using the rescalings \eqref{om} reads
\begin{align}\nonumber
    \boldsymbol{\Omega}_\pm(\zeta) &= 2\tau\,\hat\omega\,\epsilon \pm2\zeta\,\mathfrak{m}\,\tau\,\epsilon^2 \mp \zeta^2\,\mathfrak{m}\,\tau^2\epsilon^3=\\ \label{eq:Omegazeta}
    &=\boldsymbol{\omega}\,\epsilon \pm2\zeta\,\mathfrak{m}\,\tau\,\epsilon^2 \mp \zeta^2\,\mathfrak{m}\,\tau^2\epsilon^3\,.
\end{align}

We now  write the general solution of the equation of motion \eqref{eq:EOM} in the inner region up to NLO terms. The linearity of the differential equation allows as to write the NLO solution as the sum of three independent contributions proportional to $\boldsymbol{k}^2,\mathfrak{m},\tau$ respectively.
Therefore, the leading and next-to-leading contributions in the $\epsilon$ expansion may be written as
\begin{equation}\label{eq:expvarphi}
    \varphi^\pm_I(\zeta) = \varphi^{\pm,(0)}_{I}(\zeta) + \epsilon\,\boldsymbol{k}^2\varphi^{\pm,(1)}_{I,\boldsymbol{k}}(\zeta)\pm \epsilon\,\mathfrak{m}\,\varphi^{\pm,(1)}_{I,\,\mathfrak{m}}(\zeta)+\epsilon\,\tau\,\varphi^{\pm,(1)}_{I,\tau}(\zeta) +\mathcal{O}(\epsilon^2)\,,
\end{equation}
We recall that, after extracting the explicit powers of $\epsilon$ through
\eqref{om}, the parameters
$\boldsymbol{k}^2$, $\mathfrak m$, $\tau$, and $\boldsymbol\omega$
are held fixed and are of order $\epsilon^0$. We are going to show that, order by order, the fluctuation equation in the inner region depends on the frequency through $\hat\omega$ which we remind is of order $\epsilon^0$, so there will be no expansion in $\hat\omega$.

\subsubsection*{Leading order solution}
In this section, we derive the leading order solution in the inner region $\varphi^{\pm,(0)}_{I}(\zeta)$. The equation of motion at leading order for $\varphi^{\pm,(0)}_{I}(\zeta)$ reduces to
\begin{equation}
    \partial_\zeta\left(h(\zeta)\partial_\zeta\varphi^{\pm,(0)}_{I}(\zeta)\right) +\dfrac{\hat\omega^2}{4h(\zeta)}\varphi^{\pm,(0)}_{I}(\zeta) =0\,,
\end{equation}
where the function $h(\zeta)$ is defined in \eqref{eq:hg}. We stress that this is the leading order equation in $\boldsymbol{k}^2,\mathfrak{m},\tau$ while it is exact in $\hat\omega$ which is of order $\epsilon^0$. For later convenience, it is useful to define the differential operator applied to $\varphi^\pm_{I,(0)}$:
\begin{equation}\label{eq:L0}
    \mathcal{L}_0[\chi]\equiv  \partial_\zeta\left(h(\zeta)\partial_\zeta \chi(\zeta)\right) +\dfrac{\hat\omega^2}{4h(\zeta)}\chi(\zeta)\,.
\end{equation}
This way, the leading order equation of motion in the inner region can be written as
\begin{equation}
    \mathcal{L}_0\left[\varphi^{\pm,(0)}_{I}\right] =0\,.
\label{L0eq}\end{equation}
Note that, the differential operator $\mathcal{L}_0$ does not depend on $\mathfrak{m}$ and therefore the solution will not distinguish between plus and minus charged fluctuations. However, we hall not drop the $\pm$ from the notation.

The ingoing solution of (\ref{L0eq}) reads
\begin{equation}
    \varphi^{\pm,(0)}_{I}= \mathcal{N}\,\left(\dfrac{\zeta}{1+3\zeta}\right)^{-i\,\hat\omega/2}\,.
\end{equation}
where $\NN$ is an arbitrary constant normalization. Near the horizon the solution behaves as
\begin{equation}
    \varphi^{\pm,(0)}_{I}=\mathcal{N}\,\zeta^{-i\,\hat\omega/2}\left(1+\mathcal{O}\left(\zeta\right)\right)= \NN\,\zeta^{-i\,\omega/(4\pi T)}\left(1+\mathcal{O}\left(\zeta\right)\right)\,.
\end{equation}
as expected for the ingoing solution. Note that to leading order, the near-horizon solution depends only on the frequency and not also on $\mu_3$, as it should, since the isospin  potential vanishes at the horizon $\Phi_3(r_H) = 0$.

We conclude by writing down the expansion of the present solution in the overlap region $\zeta\to \infty$
\begin{equation}
    \varphi^{\pm,(0)}_{I}= \mathcal{N}\,3^{i\,\hat\omega/2}\left(1+\dfrac{i\,\hat\omega}{6\zeta}+\mathcal{O}(\zeta^{-2})\right)\,.
\end{equation}

Since the fluctuation equations are linear, the normalization constant of the leading order solution is a trivial parameter.
It is useful to fix the normalization such that the solutions approaches one at the AdS$_2$ boundary $\zeta\to \infty$:
\begin{equation}\label{eq:INNleadinglon}
    \varphi^{\pm,(0)}_{I}= \left(\dfrac{3\zeta}{1+3\zeta}\right)^{-i\,\hat\omega/2}\,.
\end{equation}

\subsubsection*{Next-to-leading order solution}
In this section, we compute the next-to-leading order solution to the equation of motion in the inner region. 
We  solve three independent equations of motion for the three types of fluctuations in \eqref{eq:expvarphi}: $\varphi_{I,\boldsymbol{k}}^{\pm,(0)}(\zeta), \,\varphi_{I,\,\mathfrak{m}}^{\pm,(1)}(\zeta),\,\varphi_{I,\tau}^{\pm,(1)}(\zeta)$.

The next-to-leading equation of motion can be written in very compact way as follows
\begin{equation}
    \mathcal{L}_0\left[\boldsymbol{k}^2\varphi_{I,\boldsymbol{k}}^{\pm,(1)}\pm \mathfrak{m}\,\varphi_{I,\,\mathfrak{m}}^{\pm,(1)}+\tau\,\varphi_{I,\tau}^{\pm,(1)}\right] = \boldsymbol{k}^2\,S_{\boldsymbol{k}}\left[\varphi_I^{\pm,(0)}\right]\pm\mathfrak{m}\,S_\mathfrak{m}\left[\varphi_I^{\pm,(0)}\right]+\tau\,S_\tau\left[\varphi_I^{\pm,(0)}\right].
\end{equation}
where
\begin{align}
    &S_{\boldsymbol{k}}\left[\varphi_I^{\pm,(0)}\right] = \dfrac{1}{4}\varphi^{\pm,(0)}_I(\zeta)-\dfrac{h(\zeta)(1+6\zeta)}{\hat\omega^2}\partial_\zeta\varphi_I^{\pm,(0)}(\zeta)\,,\\
    &S_\mathfrak{m}\left[\varphi_I^{\pm,(0)}\right] = -\dfrac{\hat\omega}{2(1+3\zeta)}\varphi_I^{\pm,(0)}(\zeta)\,,\\
    &S_\tau\left[\varphi_I^{\pm,(0)}\right] = \dfrac{g(\zeta)}{8h(\zeta)^2}\varphi_I^{\pm,(0)}(\zeta)+\dfrac{\zeta^2(7+8\zeta)(2+3\zeta)}{2h(\zeta)}\partial_\zeta\varphi_I^{\pm,(0)}(\zeta)\,,
\end{align}
are the source terms and $\mathcal{L}_0$ was defined in~\eqref{eq:L0}.

We use the linearity of the differential operator $\mathcal{L}_0$, and solve three independent equations of motion since the parameters $\boldsymbol{k}^2,\mathfrak{m},\tau$ are independent:
\begin{align}
    &\mathcal{L}_0\left[\varphi_{I,\boldsymbol{k}}^{\pm,(1)}\right] = S_{\boldsymbol{k}}\left[\varphi_I^{\pm,(0)}\right]\,,& &\mathcal{L}_0\left[\varphi_{I,\,\mathfrak{m}}^{\pm,(1)}\right] = S_\mathfrak{m}\left[\varphi_I^{\pm,(0)}\right]\,,& &\mathcal{L}_0\left[\varphi_{I,\tau}^{\pm,(1)}\right] = S_\tau\left[\varphi_I^{\pm,(0)}\right]\,.&
\end{align}

The details about the procedure to solve the inner differential equations are given in appendix \ref{app:details}. As a critical step in this procedure, we impose that the fluctuation (the leading order plus the NLO fluctuation) must contain no outgoing factor $\zeta^{i\hat\omega/2}$. Note also that, we have the freedom to add a term $\varphi_{I}^{\pm,(0)}$ to the NLO solution, changing the ingoing amplitude of the full solution. We can remove this freedom by fixing the normalization of the full ingoing solution to be the one set by the leading order solution
\begin{equation}
	\underset{\zeta\to 0}{\lim}\dfrac{\varphi_{I,(\boldsymbol{k},\mathfrak{m},\tau)}^{\pm,(1)}}{\varphi_{I}^{\pm,(0)}}=0\,.
\end{equation}

These two conditions, namely no outgoing factor, and no changing in the overall normalization, fix uniquely the NLO solutions which read
\begin{align}
	\varphi_{I,\boldsymbol{k}}^{\pm,(1)}(\zeta)=& \left(\frac{3\zeta}{1+3\zeta}\right)^{-i\hat\omega/2}\dfrac{i\,\zeta}{4\hat\omega}\left(2+\hat\omega^2\dfrac{\Gamma(-i\,\hat\omega)}{\Gamma(2-i\,\hat\omega)}{}_2F_1(1,1,2-i\,\hat\omega;-3\zeta)\right)\,,\\
		\varphi_{I,\,\mathfrak{m}}^{\pm,(1)}(\zeta)
	=&
	\frac{i}{6} \left(\frac{3 \zeta }{3 \zeta +1}\right)^{-i\hat\omega/2 } \bigg(\frac{3\zeta}{(1+3\zeta)^{i\hat\omega}}\frac{1}{1-i\hat{\omega}}\times\\ \nonumber
	&\times{}_2F_1\!\left(
	1-i\hat{\omega},\,1-i\hat{\omega};
	\,2-i\hat{\omega};
	\,-3\zeta\right)-\log \left(1+3\zeta\right)\bigg),\\ \nonumber
	\varphi_{I,\tau}^{\pm,(1)}(\zeta)=&\dfrac{i\,\hat\omega}{6}\left(\frac{3\zeta}{1+3\zeta}\right)^{-i\hat\omega/2}\bigg[\frac{3\zeta}{(1+3\zeta)^{i\hat\omega}}\frac{1}{1-i\hat{\omega}}\,
	{}_2F_1\!\left(
	1-i\hat{\omega},\,1-i\hat{\omega};
	\,2-i\hat{\omega};
	\,-3\zeta
	\right)+\\ &-\dfrac{13}{2}\frac{\zeta}{1+3\zeta}-\dfrac{7}{3}\log(1+3\zeta)\bigg]\,,
\end{align}
where ${}_2F_1$ is the ordinary hypergeometric function.

\subsection{Outer region}\label{sec:outer}

In this section, we present the solution to the fluctuation equation in the outer region. We employ the same scaling of the frequency, the momentum and the isospin chemical potential as in \eqref{eq:exp2}. However, in the outer region the holographic coordinate\footnote{We recall that the AdS$_5$ boundary is at $u =0$, while the horizon position is $u =1$.} $u$ is not expanded in the small parameter $\epsilon$
\begin{equation}
    u\sim\mathcal{O}(\epsilon^0)\,.
\end{equation}
Then, the shifted frequency after the rescaling with $\epsilon$ reads
\begin{equation}
    \boldsymbol{\Omega}_\pm(u) = (\boldsymbol{\omega}\pm\mathfrak{m}\left(1-u^2\right))\,\epsilon\,.
\end{equation}
This result differs from that of the inner region, where the term proportional to $\mathfrak{m}$ obtained an additional suppression factor from being close to the horizon (see equation \eqref{eq:Omegazeta}). Since we are in the outer region, at leading order in $\epsilon$, the blackening factor is the one at extremality $T=0$. We can write it as
\begin{equation}
    f(u) = f_0(u) +\epsilon\, f_1(u)\,,
\end{equation}
where
\begin{align}
\label{eq:f0def}
    &f_0(u) = 1 - 3 u^4 + 2 u^6 = (1-u^2)^2(1+2u^2)\,,\\
\label{eq:f1def}
    &f_1(u) = 2\tau \,u^4(1-u^2)\,.
\end{align}

We write the general solution to the equation of motion \eqref{eq:EOM} in the outer region as
\begin{equation}
	\varphi_{O}^\pm(u) = \varphi^{\pm,(0)}_O(u)+\epsilon\,\varphi^{\pm,(1)}_O(u)+\epsilon^2\,\varphi^{\pm,(2)}_O(u)+\mathcal{O}(\epsilon^3)\,,
\end{equation}
where we keep also the NNLO fluctuation $\varphi^{\pm,(2)}_O$ in the outer region. We need this term to perform a consistent matching between the inner and the outer solution in the overlapping region.

\subsubsection*{Leading order solution}
At leading order, the equation of motion reduces to a simple form
\begin{equation}
    \partial_u\left(\dfrac{\boldsymbol{\Omega}_\pm(u)^2}{u}\partial_u\varphi^{\pm,(0)}_O\right) =0\,.
\end{equation}
The solution can be found by integrating the equation twice:
\begin{equation}
   \varphi^{\pm,(0)}_O(u) = a^{\pm,(0)}_O +b^{\pm,(0)}_O\int_0^udx\,\dfrac{x}{\boldsymbol{\Omega}_\pm(x)^2}\,.
\end{equation}
Since the above integral is zero for $u = 0$, the non-normalizable/source coefficient is simply $a^{\pm,(0)}_O$. The integral can be carried out explicitly to give
\begin{equation}\label{eq:long:LOout}
    \varphi^{\pm,(0)}_O(u) = a^{\pm,(0)}_O + b^{\pm,(0)}_O\,\dfrac{u^2}{2(\boldsymbol{\omega}\pm\mathfrak{m})(\boldsymbol{\omega}\pm \mathfrak{m}\,(1-u^2))}\,.
\end{equation}

It is instructive to write the expansion of this solution close to the boundary $u=0$:
\begin{align}
     \varphi^{\pm,(0)}_O (u) = a^{\pm,(0)}_O  + \dfrac{b^{\pm,(0)}_O }{2(\boldsymbol{\omega}\pm\mathfrak{m})^2}\,u^2 +\mathcal{O}(u^4)\,.
\end{align}
By confronting the expansion above with the equation \eqref{eq:NBexp}, we can identify the source and the response coefficient $\mathcal{A}^{\pm} $ and $\mathcal{B}^{\pm} $ as
\begin{align} \label{eq:sourceres}
&	\mathcal{A}^{\pm} = a^{\pm,(0)}_O+\mathcal{O}(\epsilon) \,,& &\mathcal{B}^{\pm} =\dfrac{b^{\pm,(0)}_O }{2(\boldsymbol{\omega}\pm\mathfrak{m})^2}+\mathcal{O}(\epsilon)\,.&
\end{align}

\subsubsection*{Next-to-leading order solution}
The next-to-leading equation of motion for $\varphi^{\pm,(1)}_O(u)$ reads
\begin{equation}
    \partial_u\left(\dfrac{\boldsymbol{\Omega}_\pm(u)^2}{u}\partial_u\varphi^{\pm,(1)}_O(u)\right) = \mathcal{S}^{\pm,(1)}(u)\,,
\end{equation}
where
\begin{equation}
    \mathcal{S}^{\pm,(1)}(u) =-\dfrac{1}{\boldsymbol{k}^2}\partial_u\left(\dfrac{\boldsymbol{\Omega}_\pm(u)^4}{u\,f_0(u)}\partial_u\varphi^{\pm,(0)}_O(u)\right)+\boldsymbol{k}^2\dfrac{\boldsymbol{\Omega}_\pm(u)^2}{u\,f_0(u)}\varphi^{\pm,(0)}_O(u)\,,
\end{equation}
is a source term which depends on the leading order solution. We can use the explicit solution of $\varphi^{\pm,(0)}_O(u)$ to simplify it as
\begin{equation}\label{eq:S1pm}
    \mathcal{S}^{\pm,(1)}(u) =-\dfrac{b^{\pm,(0)}_O }{\boldsymbol{k}^2}\partial_u\left(\dfrac{\boldsymbol{\Omega}_\pm(u)^2}{f_0(u)}\right)+\boldsymbol{k}^2\dfrac{\boldsymbol{\Omega}_\pm(u)^2}{u\,f_0(u)}\left(a^{\pm,(0)}_O  +b^{\pm,(0)}_O \int_0^udx
    \,\dfrac{x}{\boldsymbol{\Omega}_\pm(x)^2}\right)\,.
\end{equation}

Note that the homogeneous equation associated with $\varphi^{\pm,(1)}_O(u)$ is the same as the one associated with $\varphi^{\pm,(0)}_O(u)$, so we can write
\begin{equation}
    \varphi^{\pm,(1)}_{O,\text{hom.}}(u)= a^{\pm,(1)}_O + b^{\pm,(1)}_O\,\dfrac{u^2}{2(\boldsymbol{\omega}\pm\mathfrak{m})(\boldsymbol{\omega}\pm \mathfrak{m}\,(1-u^2))}\,.
\end{equation}
Including also the particular solution, the full NLO solution for $\varphi^{\pm,(1)}_O$ reads
\begin{align}\nonumber
	\varphi_{O}^{\pm,(1)}(u)
	={}&
	a^{\pm,(1)}_O + b^{\pm,(1)}_O\,\dfrac{u^2}{2(\boldsymbol{\omega}\pm\mathfrak{m})(\boldsymbol{\omega}\pm \mathfrak{m}\,(1-u^2))}+ a_O^{\pm,(1)}\bigg\{\dfrac{\boldsymbol{k}^2(\boldsymbol{\omega}\pm\mathfrak{m})u^2}
	{2\boldsymbol{\Omega}_\pm(u)}
	\log u+\\ 	\nonumber
	&- 	\dfrac{\boldsymbol{k}^2
		\left[
		\pm3\mathfrak{m}(u^2-1)
		+\boldsymbol{\omega}(5u^2-2)
		\right]}
	{36\boldsymbol{\Omega}_\pm(u)}
	\log(1-u^2)+\\ \nonumber
	&-
	\dfrac{\boldsymbol{k}^2
		(2\boldsymbol{\omega}\pm3\mathfrak{m})(1+2u^2)}
	{36\boldsymbol{\Omega}_\pm(u)}
	\log(1+2u^2)\mp
	\dfrac{\boldsymbol{k}^2\mathfrak{m}\,\boldsymbol{\omega}\,u^2}
	{12(\boldsymbol{\omega}\pm\mathfrak{m})\boldsymbol{\Omega}_\pm(u)}\bigg\}+\\  \nonumber
	&+ b_O^{\pm,(1)}\bigg\{		\dfrac{
		u^2
		\left[
		\boldsymbol{k}^4\boldsymbol{\omega}
		-
		\boldsymbol{\Omega}_\pm(u)
		\left(
		\boldsymbol{k}^4+4(\boldsymbol{\omega}\pm\mathfrak{m})^2
		\right)
		\right]}
	{24\boldsymbol{k}^2(\boldsymbol{\omega}\pm\mathfrak{m})^2
		(1-u^2)\boldsymbol{\Omega}_\pm(u)} +\\
	&+	\dfrac{
		\boldsymbol{k}^4(1+2u^2)+8\left[
	\mathfrak{m}^2(u^2-1)
	\pm\mathfrak{m}\boldsymbol{\omega}(u^2-2)
	-\boldsymbol{\omega}^2
	\right]}
		{72\boldsymbol{k}^2(\boldsymbol{\omega}\pm\mathfrak{m})\boldsymbol{\Omega}_\pm(u)}\log\left(\dfrac{1+2u^2}{1-u^2}\right)\bigg\} \,. \label{NLOouterexplicit}
\end{align}
The solution as written above implies
\begin{equation}
	\underset{u\to0}{\lim}\varphi_O^{\pm,(1)}(u)
	=
	a_O^{\pm,(1)}\,,
\end{equation}
so the source term of the NLO solution is directly identified with $a_O^{\pm,(1)}$.

We can now look to the near-boundary expansion of the outer solution up to NLO. It reads
\begin{align}\nonumber
	&\varphi_O^{\pm,(0)}+\epsilon \,	\varphi_O^{\pm,(1)}  =		a_O^{\pm,(0)}
	+
	\epsilon\,a_O^{\pm,(1)}
	+  u^2\bigg\{\dfrac{1}{2(\boldsymbol{\omega}\pm\mathfrak{m})^2}\left(b_O^{\pm,(0)}
	+
	\epsilon\,b_O^{\pm,(1)}\right)+\\ \nonumber
	&+ \epsilon\,	a_O^{\pm,(0)} \boldsymbol{k}^2\dfrac{6(\boldsymbol{\omega}\pm\mathfrak{m})^2\log(u) - 3\mathfrak{m}^2\mp6\mathfrak{m}\,\boldsymbol{\omega}-2\boldsymbol{\omega}^2}{12(\boldsymbol{\omega}\pm\mathfrak{m})^2}+ \epsilon\,b_O^{\pm,(0)}\dfrac{\boldsymbol{k}^4\boldsymbol{\omega} -12(\boldsymbol{\omega}\pm\mathfrak{m})^3}{24\boldsymbol{k}^2(\boldsymbol{\omega}\pm\mathfrak{m})^3}\bigg\}\\ &+\mathcal{O}(u^4\log(u))\,.
\end{align}
Therefore, we can identify the source and the response coefficients at NLO
\begin{align}\label{eq:TT}
	&	\mathcal{A}^{\pm} = 	a_O^{\pm,(0)}
	+
	\epsilon\,a_O^{\pm,(1)}+\mathcal{O}(\epsilon^2) \,,\\ \nonumber
	&\mathcal{B}^{\pm} =\dfrac{1}{2(\boldsymbol{\omega}\pm\mathfrak{m})^2}\left(b_O^{\pm,(0)}
	+
	\epsilon\,b_O^{\pm,(1)}\right)+\epsilon\,	a_O^{\pm,(0)} \boldsymbol{k}^2\dfrac{- 3\mathfrak{m}^2\mp6\mathfrak{m}\,\boldsymbol{\omega}-2\boldsymbol{\omega}^2}{12(\boldsymbol{\omega}\pm\mathfrak{m})^2}+\\ \label{eq:PP}
	&\qquad+\epsilon\,b_O^{\pm,(0)}\dfrac{\boldsymbol{k}^4\boldsymbol{\omega} -12(\boldsymbol{\omega}\pm\mathfrak{m})^3}{24\boldsymbol{k}^2(\boldsymbol{\omega}\pm\mathfrak{m})^3}+\mathcal{O}(\epsilon^2)\,.
\end{align}

\subsubsection*{Next-to-next-to-leading order}

Now we go to the NNLO order, where the equation of motion reads
\begin{equation}
    \partial_u\left(\dfrac{\boldsymbol{\Omega}_\pm(u)^2}{u}\partial_u\varphi^{\pm,(2)}_O(u)\right) = \mathcal{S}^{\pm,(2)}(u)\,,
\end{equation}
where
\begin{align}\nonumber
    \mathcal{S}^{\pm,(2)}(u) =&-\partial_u\left(\dfrac{\boldsymbol{\Omega}_\pm(u)^4}{u\,f_0(u)\boldsymbol{k}^2}\partial_u\varphi^{\pm,(1)}_O(u)+\left(\dfrac{\boldsymbol{\Omega}_\pm(u)^4f_1(u)}{u\,f_0(u)^2\boldsymbol{k}^2}-\dfrac{\boldsymbol{\Omega}_\pm(u)^6}{u\,f_0(u)^2\boldsymbol{k}^4}\right)\partial_u\varphi^{\pm,(0)}_O(u)\right)
    \\&+\boldsymbol{k}^2\dfrac{\boldsymbol{\Omega}_\pm(u)^2}{u\,f_0(u)}\varphi^{\pm,(1)}_O(u)-\boldsymbol{k}^2\dfrac{\boldsymbol{\Omega}_\pm(u)^2f_1(u)}{u\,f_0(u)^2}\varphi^{\pm,(0)}_O(u)\,.
\end{align}
with $f_0$ and $f_1$ defined in~\eqref{eq:f0def} and~\eqref{eq:f1def}, respectively.
Using the equation of motion for the leading order fluctuation $\varphi_{O,(0)}^\pm$ we can rewrite it as
\begin{align}\nonumber
	\mathcal{S}^{\pm,(2)}(u) =&-\dfrac{1}{\boldsymbol{k}^2}\partial_u\left(\dfrac{\boldsymbol{\Omega}_\pm(u)^4}{u\,f_0(u)}\partial_u\varphi^{\pm,(1)}_O(u)\right)+\partial_u\left(\dfrac{\boldsymbol{\Omega}_\pm(u)^2}{f_0(u)^2}\left(\boldsymbol{k}^2f_1(u)-\boldsymbol{\Omega}_\pm(u)^2\right)\right)\dfrac{b^{\pm,(0)}_O}{\boldsymbol{k}^4}
	\\&+\boldsymbol{k}^2\dfrac{\boldsymbol{\Omega}_\pm(u)^2}{u\,f_0(u)}\varphi^{\pm,(1)}_O(u)-\boldsymbol{k}^2\dfrac{\boldsymbol{\Omega}_\pm(u)^2f_1(u)}{u\,f_0(u)^2}\varphi^{\pm,(0)}_O(u)\,.
\end{align}

The NNLO outer solution can then be written as
\begin{align}
	\varphi^{\pm,(2)}_O(u)
	={}&
	a^{\pm,(2)}_O
	+
	b^{\pm,(2)}_O
	\frac{u^2}{2(\boldsymbol{\omega}\pm\mathfrak{m})\boldsymbol{\Omega}_\pm(u)}+
	\nonumber\\
	&+
	a^{\pm,(1)}_O\,\mathcal F_{a_1}^\pm(u)
	+
	b^{\pm,(1)}_O\,\mathcal F_{b_1}^\pm(u)+
	b^{\pm,(0)}_O\,\mathcal F_{b_0}^\pm(u)
	+
	a^{\pm,(0)}_O\,\mathcal F_{a_0}^\pm(u)\,.
	\label{eq:NNLOoutersol}
\end{align}
where the functions $\mathcal F_{a_1}^\pm\,,\mathcal F_{b_1}^\pm\,,\mathcal F_{b_0}^\pm\,,\mathcal F_{a_0}^\pm$ have a rather complicated expressions. They are given in appendix \ref{app:details}.

\subsection{The matching and the NLO expression of the retarded correlator}\label{sec:matching}

In this section, we provide the NLO formula for the retarded correlator. We start by matching the inner and outer solution at leading order (section \ref{sec:LOmatch}) and then move to the NLO matching (section \ref{sec:NLOmatch}).

The key formulas to perform the matching are the expansion of the inner and outer solution in the overlapping region. All matching equations below are understood in the common domain
\eqref{eq:matching-overlap-region}, where both asymptotic expansions are
parametrically controlled. These formulas are rather long and discussed in detail in appendix \ref{app:details}. Here, we report only the necessary formulas.

\subsubsection{The matching at leading order}\label{sec:LOmatch}
The series expansion of the leading order inner solution~\eqref{eq:INNleadinglon} in the overlapping region ($\zeta\to\infty$) reads
\begin{equation}\label{eq:leadinINNERexp}
	\varphi_I^{\pm,(0)}(\zeta) = 1+\frac{i \,\hat\omega }{6 \zeta }-\frac{\hat\omega  (\hat\omega +2 i)}{72 \zeta ^2}+\mathcal{O}\left(\left(\frac{1}{\zeta }\right)^3\right)\,.
\end{equation}

In order to match this expansion with the one of the outer solution we need both the leading and the next-to-leading solutions. After some algebra, the result of the matching yields the fixing of the integration constants $a_O^{\pm,(0)}$, $b_O^{\pm,(0)}$ of the solution \eqref{eq:long:LOout}:

\begin{align}\label{eq:long:LOmatch}
    &a_O^{\pm,(0)}= 1 + \dfrac{i\,\boldsymbol{k}^2}{2(\boldsymbol{\omega}\pm \mathfrak{m})}\,,& & b_O^{\pm,(0)} = -i\boldsymbol{\omega}\,\boldsymbol{k}^2\,.&
\end{align}

These connection coefficients have to be translated into the boundary data encoded into the non-normalizable (source) and normalizable coefficients in~\eqref{eq:sourceres},
\begin{align}
    &\mathcal{A}^\pm =  a_O^{\pm,(0)}+ \mathcal{O}(\epsilon)\,, & &\mathcal{B}^\pm =\dfrac{b_O^{\pm,(0)} }{2(\boldsymbol{\omega}\pm \mathfrak{m})^2}+\mathcal{O}(\epsilon)\,.&
\end{align}
Then, we can substitute our results for $a_O^{\pm,(0)}$ and $b_O^{\pm,(0)}$ to obtain
\begin{align}
    &\mathcal{A}^\pm =  1 + \dfrac{i\,\boldsymbol{k}^2}{2(\boldsymbol{\omega}\pm \mathfrak{m})}+\mathcal{O}(\epsilon)\,, & &\mathcal{B}^\pm =-\dfrac{i\,\boldsymbol{\omega} \boldsymbol{k}^2}{2(\boldsymbol{\omega}\pm \mathfrak{m})^2}+\mathcal{O}(\epsilon)\,.&
\end{align}

The retarded correlator associated with the longitudinal current fluctuation is proportional to the ratio of the normalizable and non-normalizable coefficients
\begin{align}\label{eq:BA_LO}
    \dfrac{\mathcal{B}^\pm}{\mathcal{A}^\pm} = -\dfrac{1}{2}\dfrac{i\,\boldsymbol{\omega} \,\boldsymbol{k}^2}{(\boldsymbol{\omega} \pm \mathfrak{m})\left(\boldsymbol{\omega}\pm\mathfrak{m}+\frac{i}{2}\boldsymbol{k}^2\right)}+\mathcal{O}(\epsilon)\,.
\end{align}

We now make an observation. The exact longitudinal polarization function obeys an exact
kinematic constraint. Indeed, the longitudinal tensor projector $P_{\mu\nu}^{\parallel,\pm}$ (see equations \eqref{Pp}-\eqref{Pq}) contains the
denominator
$\big[(\omega\pm\mu_3)^2-k^2\big]^{-1}$.  In the absence of an independent
physical singularity at this locus, regularity of the exact current-current
correlator \eqref{ltcR} requires the longitudinal polarization function to vanish
when
\begin{equation}
	k^2=(\omega\pm\mu_3)^2\,.
	\label{eq:longitudinal-kinematic-zero}
\end{equation}
Equivalently, the exact polarization function contains the kinematic factor
$k^2-(\omega\pm\mu_3)^2$.  This property also follows directly from the Ward
identity and from the exact near-boundary solution discussed in
\cite{Apostolidis:2026efq}.

However, the ratio \eqref{eq:BA_LO} was derived using the diffusive scaling
\begin{equation}
	\boldsymbol{\omega},\mathfrak{m},\tau
	\sim\boldsymbol{k}^2\sim\mathcal{O}(\epsilon)\,.
\end{equation}
Within this scaling, the term
$(\boldsymbol{\omega}\pm\mathfrak{m})^2$ is suppressed by one additional
power of $\epsilon$ relative to $\boldsymbol{k}^2$.  It is therefore absent
from the strict leading-order expression \eqref{eq:BA_LO}.

It is useful to restore the exact zero \eqref{eq:longitudinal-kinematic-zero} while retaining the information
contained in the leading-order matching calculation.  We therefore define the LO
\emph{kinematically completed} ratio
\begin{align}
	\left.
	\frac{\mathcal B^\pm}{\mathcal A^\pm}
	\right|_{\text{LO},\textit{compl.}}
	\equiv
	-\frac{1}{2}
	\frac{
		i\boldsymbol{\omega}
		\left[
		\boldsymbol{k}^2
		-\epsilon(\boldsymbol{\omega}\pm\mathfrak m)^2
		\right]
	}{
		(\boldsymbol{\omega}\pm\mathfrak m)
		\left(
		\boldsymbol{\omega}\pm\mathfrak m
		+\frac{i}{2}\boldsymbol{k}^2
		\right)
	}\,.
	\label{eq:BA_LO_completed}
\end{align}
The added term is subleading in the diffusive expansion, so
\eqref{eq:BA_LO_completed} agrees with \eqref{eq:BA_LO} at leading order.  It
should not be interpreted as a complete determination of all NLO terms.
Rather, it is a kinematically constrained completion of the LO result.  The
coefficient of the restored term is independently reproduced by the
near-boundary expansion of the NLO outer solution.

We can substitute this formula in the formula for the polarization function \eqref{eq:long:matchedcorrelator}, restore the dimensionful parameters and remove $\epsilon$ to obtain
\begin{equation}\label{eq:Pi_long_LO}
	\Pi_\text{LO}^{\parallel,\pm}(\omega,k)
	=
	\frac{1}{8r_H}(M\ell)^3w_0^2N_c\,\dfrac{i\,\omega \,\left(k^2-(\omega\pm\mu_3)^2\right)}{(\omega \pm \mu_3)\left(\omega\pm\mu_3+\frac{i\,r_H}{2}k^2\right)}\,.
\end{equation}

We comment on the LO pole's structure. For a QNM excitation, the boundary source must vanish
\begin{align}
	\mathcal{A}^\pm(\omega_\text{QNM}^{\pm},k,\mu_3,T)=0\,,
\end{align}
or in other words, the denominator of the polarization function has to vanish.\footnote{From the formula for the polarization function \eqref{eq:Pi_long_LO}, it seems that $\omega = \mp \mu_3$ is a pole of the correlator. However, such a value for the frequency does not correspond to a QNM since it does not set to zero the source $\mathcal{A}^\pm$. Therefore $\omega = \mp \mu_3$  has to be regarded as an apparent singularity. See \cite{Preau:2026hge} for more details about such apparent singularities.}
The above equation admits one solution for each charged channel and this leads to the leading order dispersion relation for the hydro-like pole that we write as
\begin{equation}\label{eq:long:QNMLO}
	\boldsymbol{\omega}_\text{hydro,LO}^{\pm} =\mp\mathfrak{m} -\dfrac{i}{2}\boldsymbol{k}^2\,.
\end{equation}
We can restore the original dimensionful parameters 
to obtain
\begin{equation}
	\omega_\text{hydro,LO}^{\pm} =\mp\mu_3 -i\dfrac{r_H}{2}k^2\,.
\end{equation}
From the above relation, we can identify the diffusion constant $D$ as
\begin{equation}
	D = \dfrac{r_H}{2}\,.
\end{equation}

\subsubsection{The matching at next-to-leading order}\label{sec:NLOmatch}
The matching procedure at NLO requires to match the NLO inner solution with the NNLO outer solution.  For clarity, we report the series expansion of the inner solution (up to NLO terms) in the overlapping region $\zeta\to \infty$
\begin{align}\nonumber
    \varphi_{I}(\zeta) = &1+\frac{i \,\hat\omega }{6}\dfrac{1}{\zeta} -\frac{\hat\omega  (\hat\omega +2 i)}{72}\dfrac{1}{\zeta^2}+\dfrac{i\boldsymbol{k}^2}{2 \hat\omega}\epsilon\,\zeta+ \left(-\frac{\boldsymbol{k}^2}{12}-\dfrac{2i\hat\omega\,\tau}{9}\right)\epsilon\,\log\zeta+\\ \nonumber
    &+\bigg[\frac{\boldsymbol{k}^2}{12}\left( \psi(1-i\,\hat\omega)+\gamma_E-1-\log (3)\right)\mp\frac{\mathfrak{m}}{6} i ( \psi(1-i\,\hat\omega)+\gamma_E)+\\ \nonumber
    &-\frac{i \hat\omega\,\tau}{36}   \left(6( \psi(1-i\,\hat\omega)+\gamma_E)+13+8 \log (3)\right)\bigg]\epsilon+\\ \nonumber
    &+\dfrac{\epsilon\,\log\zeta}{\zeta}\left[\dfrac{i\hat\omega\,\boldsymbol{k}^2}{72}\pm\dfrac{\mathfrak{m}\,\hat\omega}{18}+\dfrac{5\hat\omega^2\,\tau}{54}\right]+\\ \nonumber
    &+\dfrac{\epsilon}{\zeta}\Bigg[\boldsymbol{k}^2\left(-\frac{i \hat\omega  \left(
    2(\psi(-i\,\hat\omega)+\gamma_E)-3-\log (9)\right)}{144}\right) +\\ \nonumber
    &\pm\mathfrak{m}\left( -\frac{\hat\omega  \left(\psi(1-i\,\hat\omega)+\gamma_E-2-\log (9)\right)}{36}\right)+\\ \nonumber
    &+\tau\left(-\frac{\hat\omega  \left(6 \hat\omega  (\psi(1-i\,\hat\omega)+\gamma_E)-25 \hat\omega -5 \hat\omega  \log (81)-10 i\right)}{216 }\right)\Bigg] +\\
    &+\mathcal{O}(\epsilon^2,\zeta^{-2}\log\zeta,\zeta^{-3})\,,
\end{align}
with $\psi(z) = \Gamma'(z)/\Gamma(z)$ the digamma function and $\gamma_E$ the Euler-Mascheroni constant.

In order to match all these terms, we  need the series expansion of the NNLO outer solution. This is analogous to what happened in the leading order matching where we needed the NLO outer solution.
The reason for this is that terms in the outer solution that contain singularities, when expressed in the near-horizon coordinate, acquire inverse powers of $\e$, and these cancel the powers of $\e$ already multiplying them.

 Below, we write the expansion of the outer solution in terms of the holographic coordinate $\zeta$
\begin{align}\nonumber
	\varphi_{O}^\pm(\zeta)&= Y^{\pm,(0)}_O
	-
	\dfrac{b_O^{\pm,(0)}}{\boldsymbol{\omega}^2}\epsilon\,\tau\,\zeta -\dfrac{b_O^{\pm,(0)}}{12\boldsymbol{k}^2}\dfrac{1}{\tau\,\zeta}
	+\\ \nonumber
	&+\left(
	\dfrac{b_O^{\pm,(0)}}{9\boldsymbol{k}^2}
	-
	\dfrac{\boldsymbol{k}^2}{12}Y^{\pm,(0)}_O
	\right)\,\epsilon\,\log(\epsilon\,\tau\,\zeta)
	+
	\widetilde Y^{\pm,(1)}_O\,\epsilon+\\ \nonumber
	&+\left[
	m^{\pm,(0)}_O\log(\epsilon\,\tau\,\zeta)
	+
	n^{\pm,(1)}_O
	\right]\,\epsilon^2\,\tau\,\zeta+ \frac{E_{-2}^{\pm,(0)}}{\tau^2\,\zeta^2}
	+\\
	&+
	\frac{
		E_{-1,\log}^{\pm,(0)}\log(\epsilon\,\tau\,\zeta)
		+
		E_{-1}^{\pm,(1)}
	}{\tau\,\zeta}\,\epsilon
	+\mathcal{O}(\epsilon^3,\zeta^{-2}\log\zeta,\zeta^{-3})\,.
\end{align}
The explicit expression of the various coefficients appearing in this expression is given in appendix \ref{app:details}.

In the previous section, we performed the matching of the constant and the $\zeta^{-1}$ terms, from which we obtained the following expressions for the leading order connection coefficients
\begin{align}
    &a^{\pm,(0)}_O = 1 + \dfrac{i\,\boldsymbol{k}^2}{2(\boldsymbol{\omega}\pm \mathfrak{m})}\,,& &b^{\pm,(0)}_O = -i\boldsymbol{\omega}\,\boldsymbol{k}^2\,.&
\end{align}
We observe that imposing these data, we automatically satisfy the matching of the terms proportional to $\epsilon\,\zeta$, $\epsilon\log\zeta$, $\zeta^{-2}$, and $\epsilon\,\zeta^{-1}\log\zeta$.

The non-trivial NLO matching comes from equating the terms proportional to $\epsilon$ and $\epsilon/\zeta$. Here, we report just the final result of the matching which is
\begin{align}\nonumber
	a^{\pm,(1)}_O
	={}&
	\frac{
		\boldsymbol{k}^2\boldsymbol{\omega}
		-i\left[
		2(\boldsymbol{\omega}\pm\mathfrak{m})(\boldsymbol{\omega}\pm2\mathfrak{m})
		+\boldsymbol{k}^4
		\right]
	}
	{24(\boldsymbol{\omega}\pm\mathfrak{m})}
	\,\left(\psi\left(1-i\,\frac{\boldsymbol{\omega}}{2\tau}\right)+\gamma_E\right)
	+\\ \nonumber
	&+	\left[
	-\frac{\boldsymbol{k}^2(2\boldsymbol{\omega}\pm3\mathfrak{m})}
	{36(\boldsymbol{\omega}\pm\mathfrak{m})}
	+i\left(
	\frac{\boldsymbol{\omega}}{9}
	-\frac{\boldsymbol{k}^4}{24(\boldsymbol{\omega}\pm\mathfrak{m})}
	\right)
	\right]\log\tau\\ 	\nonumber
	&
	+
	\frac{\boldsymbol{k}^2}
	{36(\boldsymbol{\omega}\pm\mathfrak{m})}
	\left[
	\pm3\mathfrak{m}\log\!\left(\frac{9}{2}\right)
	+\boldsymbol{\omega}\left(1+2\log\!\left(\frac{9}{2}\right)\right)
	\right]\\
	&	-
	\frac{i\boldsymbol{\omega}}{18}
	\left(1+2\log\!\left(\frac{9}{2}\right)\right)
	+
	\frac{i\boldsymbol{k}^4}{24(\boldsymbol{\omega}\pm\mathfrak{m})}\log\!\left(\frac{9}{2}\right)\,,
	\label{a1connection}
\end{align}
and
\begin{align}\nonumber
	b^{\pm,(1)}_O
	={}&
	\frac{\boldsymbol{k}^2\boldsymbol{\omega}\left(\boldsymbol{\omega}\pm2\mathfrak{m}+i\boldsymbol{k}^2\right)}{12}\,
	\left(\psi\left(1-i\,\frac{\boldsymbol{\omega}}{2\tau}\right)+\gamma_E\right)
	+
	\frac{\boldsymbol{k}^2\boldsymbol{\omega}}{36}\left[10\boldsymbol{\omega}\pm12\mathfrak{m}+3i\boldsymbol{k}^2
	\right]\log\tau
	\\ \nonumber
	&
	+
	\frac{\boldsymbol{k}^2}{18}
	\Big[
	9\mathfrak{m}^2\log3
	\pm6\mathfrak{m}\,\boldsymbol{\omega}\log6
	+\boldsymbol{\omega}^2\left(-4-\log3+5\log2\right)
	\Big]+\\
	&+\frac{i\boldsymbol{k}^4}{12}
	\Big[
	3\pm\mathfrak{m}\log3
	+\boldsymbol{\omega}\,\log  6
	\Big].
	\label{b1connection}
\end{align}

Now we have all the ingredients to derive the formula for the retarded correlator at NLO. In order to do so, we use formulae \eqref{eq:TT}, and \eqref{eq:PP}. In these formulae, we derived, up to order $\epsilon^2$, the non-normalizable $\mathcal{A}_\pm$ and normalizable $\mathcal{B}_\pm$ coefficients from the outer solution expansion around the boundary. Therefore, the ratio between the two coefficients can be written as
\begin{align}\nonumber
\dfrac{\mathcal{B}^\pm }{\mathcal{A}^\pm } = &\dfrac{b_O^{\pm,(0)} + \epsilon\left(b_O^{\pm,(1)}-a_O^{\pm,(0)}\boldsymbol{k}^2\,\dfrac{ 3\mathfrak{m}^2\pm6\mathfrak{m}\,\boldsymbol{\omega}+2\boldsymbol{\omega}^2}{6}+b_O^{\pm,(0)}\dfrac{\boldsymbol{k}^4\boldsymbol{\omega} -12(\boldsymbol{\omega}\pm\mathfrak{m})^3}{12\boldsymbol{k}^2(\boldsymbol{\omega}\pm\mathfrak{m})}\right)}{2(\boldsymbol{\omega}\pm\mathfrak{m})^2\left(a_O^{\pm,(0)}+\epsilon\,a_O^{\pm,(1)}\right)}\\
&+\mathcal{O}(\epsilon^2)\,.
\end{align}
We can substitute the explicit expressions for $a_O^{\pm,(0)} ,b_O^{\pm,(0)},a_O^{\pm,(1)} ,b_O^{\pm,(1)} $ to obtain
\begin{equation}
\frac{\mathcal B^\pm}{\mathcal A^\pm}
=
\frac{
-i\boldsymbol{k}^2\boldsymbol{\omega}+\epsilon\,\mathfrak N^\pm
}{
2(\boldsymbol{\omega}\pm\mathfrak{m})^2
\left[
1+\dfrac{i\boldsymbol{k}^2}{2(\boldsymbol{\omega}\pm\mathfrak{m})}
+\epsilon\,\mathfrak D^\pm
\right]
}+\mathcal{O}(\epsilon^2)
\,,
\label{BAepsilon}
\end{equation}
with
\begin{align}
\mathfrak N^\pm
={}&
i\boldsymbol{\omega}(\boldsymbol{\omega}\pm\mathfrak{m})^2
-\frac{\boldsymbol{k}^2}{2}(\boldsymbol{\omega}\pm\mathfrak{m})^2
-\frac{i\boldsymbol{k}^4}{4}(\boldsymbol{\omega}\pm\mathfrak{m})
-\frac{\boldsymbol{k}^2\boldsymbol{\omega}^2}{18}+
\nonumber\\
&+\frac{\boldsymbol{k}^2\boldsymbol{\omega}(\boldsymbol{\omega}\pm2\mathfrak{m}+i\boldsymbol{k}^2)}{12}\left(\psi\left(1-i\,\frac{\boldsymbol{\omega}}{2\tau}\right)+\gamma_E\right)+\nonumber\\
&+
\boldsymbol{k}^2\boldsymbol{\omega}\left[
\frac{6(\boldsymbol{\omega}\pm\mathfrak{m})-\boldsymbol{\omega}}{18}
+\frac{i\boldsymbol{k}^2}{12}
\right]
\log\left(\frac{2\tau}{9}\right)
+\frac{\boldsymbol{k}^2(\boldsymbol{\omega}\pm\mathfrak{m})(2(\boldsymbol{\omega}\pm\mathfrak{m})+i\boldsymbol{k}^2)}{4}\log3 ,
\label{Npm}
\\[2mm]
\mathfrak D^\pm
={}&
\left[
-\frac{i(\boldsymbol{\omega}\pm2\mathfrak{m})}{12}
+\frac{\boldsymbol{k}^2(\boldsymbol{\omega}-i\boldsymbol{k}^2)}{24(\boldsymbol{\omega}\pm\mathfrak{m})}
\right] \left(\psi\left(1-i\,\frac{\boldsymbol{\omega}}{2\tau}\right)+\gamma_E\right)\nonumber\\
&+
\left[
-\frac{\boldsymbol{k}^2}{12}
+\frac{i\boldsymbol{\omega}}{9}
+\frac{\boldsymbol{k}^2\boldsymbol{\omega}}{36(\boldsymbol{\omega}\pm\mathfrak{m})}
-\frac{i\boldsymbol{k}^4}{24(\boldsymbol{\omega}\pm\mathfrak{m})}
\right]
\log\left(\frac{2\tau}{9}\right)
-\frac{i\boldsymbol{\omega}}{18}
+\frac{\boldsymbol{k}^2\boldsymbol{\omega}}{36(\boldsymbol{\omega}\pm\mathfrak{m})}.
\label{Dpm}
\end{align}

The kinematic constraint discussed for the LO correlator below formula \eqref{eq:BA_LO} must also be satisfied by the
full NLO polarization function. To make this property manifest, we define the NLO correlator from \eqref{BAepsilon} by pulling out the appropriate factor:
\begin{equation}
\Pi_\text{NLO}^{\parallel,\pm}(\omega,k)
=
\frac{(M\ell)^3}{8r_H^2}w_0^2N_c\,
	\left(1-\dfrac{(\bm{\omega}\pm\mathfrak{m})^2}{\bm{k}^2}\right)\frac{
i\bm{k}^2\bm{\omega}-\left(\mathfrak N^\pm-i\bm{\omega}(\bm{\omega}\pm\mathfrak{m})^2\right)
}{(\bm{\omega}\pm\mathfrak{m})^2
\left[
1+\dfrac{i\bm{k}^2}{2(\bm{\omega}\pm\mathfrak{m})}
+\mathfrak D^\pm
\right]
}\,.
\label{BAdimensionful}
\end{equation}

Note that our result for the retarded correlator is the analogue of the current correlator computed in an axionic four-dimensional black hole discussed in \cite{Gouteraux:2025kta}, but generalized to a five-dimensional RN black hole with a non-zero isospin density.

The limit $\mu_3\to0$ of the retarded correlator \eqref{BAdimensionful} is straightforward. The limit $T\to0$ requires more care. The functions $\N^\pm$ and $\Dg^\pm$ contain the digamma function and $\log(2\tau/9)$, which are separately singular as $\tau\to0$, and the logarithms of the temperature must cancel against the $\tau\to0$ limit of the digamma function. Only the
combination
\begin{equation}
	\Lam \equiv \psi\!\left(1-\frac{i\bw}{2\tau}\right)+\gamma_E+\log\!\left(\frac{2\tau}{9}\right)
	\label{eq:Lambda}
\end{equation}
has a finite limit at fixed $\bw\neq0$. Since $\psi(1+z)=\log z+O(1/z)$ for
$|z|\to\infty$, $|\arg z|<\pi$,
\begin{equation}
	\lim_{\tau\to0}\Lam=\gamma_E+\log\!\left(-\frac{i\bw}{9}\right)\,.
	\label{eq:LambdaT0}
\end{equation}
The logarithm is defined by analytic continuation from
$\operatorname{Im}\boldsymbol{\omega}>0$. In particular, for real non-zero
frequency,
\begin{equation}
	\log\!\left[-\frac{i(\boldsymbol{\omega}+i0^+)}{9}\right]
	=
	\log\!\left(\frac{|\boldsymbol{\omega}|}{9}\right)
	-\frac{i\pi}{2}\operatorname{sgn}\boldsymbol{\omega} .
\end{equation}

Writing $\psi+\gamma_E=\Lam-\log(2\tau/9)$ in \eqref{Npm} and \eqref{Dpm}, the
remaining logarithms of $\tau$ turn out to be proportional to the
leading-order numerator and denominator, with a common coefficient:
\begin{equation}
	\begin{aligned}
		\Dg^\pm &= c^\pm_\psi\,\Lam + F^\pm\,A^\pm_0\,\log\!\left(\frac{2\tau}{9}\right)
		-\frac{i\bw}{18}\,A^\pm_0 ,\\
		\N^\pm &= \overline{\N}^\pm + n^\pm_\psi\,\Lam + F^\pm\,N_0\,\log\!\left(\frac{2\tau}{9}\right),
	\end{aligned}
	\label{eq:NDdecomp}
\end{equation}
where $A^\pm_0\equiv 1+\frac{i\bk^2}{2(\bw\pm\mathfrak{m})}$ and $N_0\equiv-i\bk^2\bw$ are the
leading-order denominator and numerator in \eqref{BAepsilon}, and\footnote{The first line of \eqref{eq:NDdecomp} is the decomposition of the denominator where we used $-\frac{i\bw}{18}+\frac{\bk^2\bw}{36(\bw\pm\mathfrak{m})}=-\frac{i\bw}{18}A_0^\pm$.}
\begin{align}
		F^\pm=&\frac{i}{36}\,(7\bw\pm6\mathfrak{m}), \quad
		c^\pm_\psi=-\frac{i(\bw\pm2\mathfrak{m})}{12}+\frac{\bk^2(\bw-i\bk^2)}{24(\bw\pm\mathfrak{m})},\quad
		n^\pm_\psi=\frac{\bk^2\bw(\bw\pm2\mathfrak{m}+i\bk^2)}{12},\\
		\nonumber
		\overline{\N}^\pm = &i\bw(\bw\pm\mathfrak{m})^2-\frac{\bk^2}{2}(\bw\pm\mathfrak{m})^2-\frac{i\bk^4}{4}(\bw\pm\mathfrak{m})
		-\frac{\bk^2\bw^2}{18}+\\
		&+\frac{\bk^2(\bw\pm\mathfrak{m})\left[2(\bw\pm\mathfrak{m})+i\bk^2\right]}{4}\log3 .
	\end{align}

At NLO the ratio \eqref{BAepsilon} is unchanged if its numerator and denominator are
both multiplied by the same factor $1+\epsilon\,G(\bw)$. Choosing
$G=-F^\pm\log(2\tau/9)$, i.e. multiplying by
$1-\epsilon\,\frac{i}{36}(7\bw\pm6\mathfrak{m})\log(2\tau/9)$, we obtain
\begin{equation}
	\frac{\mathcal B^\pm}{\mathcal A^\pm}
	=\frac{-i\bk^2\bw+\epsilon\,\NT^\pm}
	{2(\bw\pm\mathfrak{m})^2\left[1+\dfrac{i\bk^2}{2(\bw\pm\mathfrak{m})}+\epsilon\,\DT^\pm\right]}
	+\mathcal O(\epsilon^2),
	\label{eq:BAtilde}
\end{equation}
with
\begin{equation}
	\DT^\pm = c^\pm_\psi\,\Lam-\frac{i\bw}{18}+\frac{\bk^2\bw}{36(\bw\pm\mathfrak{m})},
	\qquad
	\NT^\pm = \overline{\N}^\pm+n^\pm_\psi\,\Lam .
	\label{eq:NDtilde}
\end{equation}
In this form the temperature enters only through $\Lam$, so numerator and
denominator have finite $\tau\to0$ limits separately, obtained by the
replacement \eqref{eq:LambdaT0}. The forms \eqref{BAepsilon} and \eqref{eq:BAtilde} differ only at
$\mathcal O(\epsilon^2)$, where both are beyond the accuracy of the matching. 

Proceeding as in \eqref{BAdimensionful}, the kinematically completed polarization function can be written as
\begin{equation}
	\widetilde\Pi^{\parallel,\pm}_{\rm NLO}(\omega,k)=\frac{1}{8r_H^4}(M\ell)^3w_0^2N_c
	\left(1-\frac{(\omega\pm\mu_3)^2}{k^2}\right)
	\frac{ir_H^3k^2\omega-\left(\NT^\pm_{\rm phys}-ir_H^3\omega(\omega\pm\mu_3)^2\right)}
	{(\omega\pm\mu_3)^2\left[1+\dfrac{ir_Hk^2}{2(\omega\pm\mu_3)}+\DT^\pm_{\rm phys}\right]},
	\label{eq:PiTilde}
\end{equation}
where, with
$\Lam_{\rm phys}\equiv\psi\!\left(1-\frac{i\omega}{2\pi T}\right)+\gamma_E+\log\!\left(\frac{2\pi Tr_H}{9}\right)$,
\begin{align}
	\DT^\pm_{\rm phys}&=\left[-\frac{ir_H(\omega\pm2\mu_3)}{12}+\frac{r_H^2k^2(\omega-ir_Hk^2)}{24(\omega\pm\mu_3)}\right]\Lam_{\rm phys}
	-\frac{ir_H\omega}{18}+\frac{r_H^2k^2\omega}{36(\omega\pm\mu_3)},
	\label{eq:DtildePhys}\\
	\NT^\pm_{\rm phys}&=ir_H^3\omega(\omega\pm\mu_3)^2-\frac{r_H^4k^2}{2}(\omega\pm\mu_3)^2
	-\frac{ir_H^5k^4}{4}(\omega\pm\mu_3)-\frac{r_H^4k^2\omega^2}{18}
+\nonumber\\
	&\quad	+\frac{r_H^4k^2\omega(\omega\pm2\mu_3+ir_Hk^2)}{12}\,\Lam_{\rm phys}+\frac{r_H^4k^2(\omega\pm\mu_3)\left[2(\omega\pm\mu_3)+ir_Hk^2\right]}{4}\log3 .
	\label{eq:NtildePhys}
\end{align}
We stress that \eqref{BAdimensionful} and \eqref{eq:PiTilde} agree at NLO and differ at $\mathcal{O}(\epsilon^2)$. To keep track of this difference, we denote the polarization function in the form \eqref{eq:PiTilde} by $\widetilde\Pi$. The numerical comparison of section~\ref{sec:numpol} uses \eqref{BAdimensionful}.

In the form \eqref{eq:PiTilde}, the $T\to0$ limit is immediate:  the temperature appears only through $\Lam_{\rm phys}$, whose limit is given by \eqref{eq:LambdaT0}.

\section{Analytic expressions for the NLO quasinormal mode frequencies}\label{sec:QNMs}
In this section, we derive approximate analytic formulas for the quasinormal modes of the NLO longitudinal retarded correlator, i.e.\ for the zeros of the NLO source coefficient. We first compute the NLO correction to the hydro-like pole (section~\ref{sec:QNMhydro}). We then discuss its zero-temperature limit which, at non-zero isospin, reveals a second hydro-like pole above a threshold momentum (section~\ref{sec:QNMT0}). Next, we derive the NLO corrections to the IR poles (section~\ref{sec:QNMIR}) and, finally, a local approximation describing the interaction between the hydro-like pole and one IR pole (section~\ref{sec:QNMint}). All the analytic formulas of this section are the same whether they are derived from \eqref{BAepsilon} or from \eqref{eq:BAtilde}, see below \eqref{eq:long:QNMNLO}.

\subsection{The hydro-like pole}\label{sec:QNMhydro}
In section~\ref{sec:LOmatch}, we computed the dispersion relation of the only dynamical pole of the LO correlator. This is a hydro pole whose dispersion relation reads 
\begin{equation}\label{eq:hydroLO}
	\boldsymbol{\omega}_\text{hydro,LO}^{\pm} =\mp\mathfrak{m} -\dfrac{i}{2}\boldsymbol{k}^2\,.
\end{equation}

The formula for the NLO correlator allows us to compute the NLO correction to the hydro-like pole. We remind the reader that the source coefficient of the near-boundary expansion of the longitudinal fluctuation, can be written in terms of the connection coefficients as (see equation \eqref{eq:TT})
\begin{equation}
	\label{NLOA}    \mathcal{A}^{\pm} = 	a_O^{\pm,(0)}
	+
	\epsilon\,a_O^{\pm,(1)}+\mathcal{O}(\epsilon^2) \,.
\end{equation}
In order to derive the NLO correction to the leading-order dispersion relation \eqref{eq:hydroLO}, we write the frequency of a QNM as the LO dispersion relation for the hydro-like pole plus a correction linear in $\epsilon$ (and subleading terms of higher order in $\epsilon$)
\begin{equation}
	\boldsymbol{\omega}_\text{QNM}^{\pm}  =  \boldsymbol{\omega}_\text{hydro,LO}^{\pm} + \epsilon\,\delta\boldsymbol{\omega}_\text{hydro}^{\pm}   +\mathcal{O}(\epsilon^2)\,.
\end{equation}
Then, we set the source coefficient to zero, inserting the above ansatz for the QNM frequency. After some algebra, we can derive the NLO correction $\delta\boldsymbol{\omega}_\text{hydro}^{\pm}$ for the hydro-like pole
\begin{align}
	\label{eq:long:QNMNLO}
	\delta\boldsymbol{\omega}_\text{hydro}^{\pm} =  \dfrac{\boldsymbol{k}^2(i\,\boldsymbol{k}^2\pm2\mathfrak{m})}{24}\left(\gamma_E+\psi\left(1+\dfrac{1}{4\tau}(\pm 2i\,\mathfrak{m}-\boldsymbol{k}^2)\right)+\log\!\left(\frac{2\tau}{9}\right)\right)\,.
\end{align}
The expression in parentheses in \eqref{eq:long:QNMNLO} is precisely the combination \eqref{eq:Lambda} evaluated on the leading-order pole, $\delta\boldsymbol{\omega}_\text{hydro}^{\pm}=\frac{\boldsymbol{k}^2(i\boldsymbol{k}^2\pm2\mathfrak{m})}{24}\,\Lam\big|_{\boldsymbol{\omega}=\boldsymbol{\omega}^\pm_\text{hydro,LO}}$. This follows from \eqref{eq:NDdecomp}: the terms proportional to $A_0^\pm$ vanish on the leading-order pole, where $A_0^\pm=0$. For the same reason, the analytic formulas \eqref{QNMphysical}, \eqref{eq:NLOIR} and \eqref{int:roots} derived in this section for the various types of poles are identical whether they are obtained from \eqref{BAepsilon} or from \eqref{eq:BAtilde}.

We group the LO hydro-pole dispersion relation and its NLO correction together to define the NLO hydro-pole dispersion relation
\begin{equation}
 \boldsymbol{\omega}_\text{hydro,NLO}^{\pm}  \equiv  \boldsymbol{\omega}_\text{hydro,LO}^{\pm}  +\epsilon\,\delta  \boldsymbol{\omega}_\text{hydro}^{\pm} \,.
\end{equation}
Then, we restore the original units to obtain
\begin{align}\nonumber
	&\omega_\text{hydro,NLO}^{\pm}
	={}
	\mp\mu_3
	-\frac{i r_H}{2}k^2
	+\\
	&+
	\frac{r_H^2k^2}{24}
	\left(
	\pm2\mu_3+i r_Hk^2
	\right)
	\Bigg[
	\gamma_E+\psi\left(1+\dfrac{1}{4\pi\,T}(\pm 2i\,\mu_3-r_Hk^2)\right)
	+
	\log\!\left(\frac{2\pi T r_H}{9}\right)
	\Bigg]\,.
	\label{QNMphysical}
\end{align}

We now  comment on this result. At zero spatial momentum, the NLO correction vanishes identically
\begin{equation}
	\omega_\text{hydro,NLO}^{\pm}(k=0) =\mp\mu_3\,.
\end{equation}
Temperature and isospin chemical potential,  at NLO, can modify the momentum-dependent coefficients, but they do not shift the $k=0$ precession frequency.

In our final formula for the dispersion relation \eqref{QNMphysical}, we restored the dimensionful quantities by re-introducing $r_H$. However, $r_H$ itself receives corrections in $T/\mu$. Therefore, we should expand also $r_H$ consistently:
\begin{equation}
	r_H = r_e\left(1-\dfrac{\pi\sqrt{3N_c}}{w_0}\dfrac{T}{\mu}+\mathcal{O}(T^2/\mu^2)\right)\,,
\end{equation}
where the extremal horizon radius $r_e$ has been defined in \eqref{eq:re} as
\begin{equation}
	r_e \equiv \dfrac{2\sqrt{3N_c}}{w_0\,\mu}\,.
\end{equation}
Then, the result for the dispersion relation, taking this into account, reads
\begin{align}\nonumber
	&\omega_\text{hydro,NLO}^{\pm}
	={}
	\mp\mu_3
	-\frac{i}{2}k^2r_e\left(1-\dfrac{\pi\sqrt{3N_c}}{w_0}\dfrac{T}{\mu}+\mathcal{O}\left(\dfrac{T^2}{\mu^2}\right)\right)
	+\\ \nonumber
	&+
	\frac{r_e^2k^2}{24}
	\left(
	\pm2\mu_3+i r_ek^2
	\right)
	\Bigg[
	\gamma_E+\psi\left(1+\dfrac{1}{4\pi\,T}(\pm 2i\,\mu_3-r_ek^2)\right)
	+
	\log\!\left(\frac{2\pi T r_e}{9}\right)
	\Bigg]\\
	&\times \left(1+\mathcal{O}\left(\dfrac{T}{\mu}\right)\right) \,.
	\label{QNMphysical:re}
\end{align}
Corrections generated by replacing $r_H$ with its near-extremal
expansion inside the term that is already NLO contribute only beyond
the accuracy considered here.

Now, we can consider the vanishing limit of the other two parameters that appear explicitly in the NLO dispersion relation: the chemical potential $\mu_3$ and the temperature $T$.  The limit of vanishing isospin chemical potential is trivial and it yields
\begin{align}\nonumber
	&\omega_\text{hydro,NLO}^{\pm}\bigg|_{\mu_3=0}
	={}
	-\frac{i}{2}k^2r_e\left(1-\dfrac{\pi\sqrt{3N_c}}{w_0}\dfrac{T}{\mu}+\mathcal{O}\left(\dfrac{T^2}{\mu^2}\right)\right)
	+\\
	&+
	\frac{ir_e^3k^4}{24}
	\Bigg[
	\gamma_E+\psi\left(1-\dfrac{r_ek^2}{4\pi\,T}\right)
	+
	\log\!\left(\frac{2\pi T r_e}{9}\right)
	\Bigg]\left(1+\mathcal{O}\left(\dfrac{T}{\mu}\right)\right)\,.
\end{align}

We comment now on the validity of the NLO correction that we computed. First of all, we are neglecting higher order corrections in $\epsilon$ and therefore in $k/\mu,T/\mu,\mu_3/\mu$. Consequently, our formula is valid up to the regime where these corrections become important. Moreover one could be worried about the presence of the digamma function in formula \eqref{QNMphysical} because it can be divergent. We write it below for the reader's convenience
\begin{equation}
\psi\left(1+\dfrac{1}{4\pi\,T}(\pm 2i\,\mu_3-r_Hk^2)\right)\,,
\end{equation}
which has poles when its argument is a non-positive integer, $1-n$ with $n=1,2,\ldots$. Equating the argument to $1-n$, we obtain
\begin{equation}\label{eq:psidiv}
	r_H^2k^2 =  4\pi\,n\,r_HT\pm2i\,r_H\mu_3\,.
\end{equation}

At $\mu_3=0$, this does have real solutions
\begin{equation}
	r_Hk = \pm 2\sqrt{\pi\,n\,r_HT}\,,
\end{equation}
and the digamma function in the NLO dispersion relation of the hydro pole diverges. This corresponds to the breakdown of the hydrodynamic regime corresponding to the collision of the hydro-like pole with the IR poles. The resulting scale $k\sim\sqrt{T\mu}$ is consistent with the
near-extremal breakdown of diffusive hydrodynamics studied
in~\cite{Arean:2020eus} and with the analytic charge-correlator
calculation of~\cite{Gouteraux:2025kta}.

However, for non-zero real $\mu_3$, equation \eqref{eq:psidiv} has no
solution at real momentum. Therefore, the exact collision between the hydrodynamic pole and the AdS$_2$ pole is displaced
away from the real $k$ axis. The expression \eqref{QNMphysical} has no breakdown region, apart from where higher order corrections become important. This observation reflects the results found in \cite{Apostolidis:2026efq}: at finite isospin chemical potential the hydro-like pole of the charged current correlator does not collide with the tower of infrared poles associated with the AdS$_2$ throat of the black hole geometry for real $k$. We must say that, even if they do not collide for real $k$, they can collide for imaginary $k$.\footnote{We thank B. Gouteraux for useful insights on this matter.}

Since $\mu_3$ is a continuous parameter, as we decrease it towards zero, the formula \eqref{QNMphysical} becomes less and less accurate in the region where the digamma function approaches its poles. This is a consequence of the fact, that we derived our formula perturbing around the hydro-pole. Therefore, it cannot capture the interactions with the IR poles correctly. We have to consider our formula, up to higher order corrections, valid away from the collision region identified by the poles of the digamma function.

\subsubsection{Zero-temperature limit and the second hydro-like pole}\label{sec:QNMT0}
Both at finite and zero $\mu_3$, the dispersion relation \eqref{QNMphysical:re} admits
a formal $T\to0$ limit at fixed $k$ and $\mu_3$:
\begin{align}\label{eq:QNMphysicalT0}
	\omega_\text{hydro,NLO}^{\pm}\bigg|_{T=0} = \mp\mu_3-i\dfrac{r_e}{2}k^2 + \dfrac{r_e^2k^2}{24}(\pm2\mu_3+i\,r_ek^2)\left(\gamma_E+\log\left(\dfrac{r_e}{9}\left[\pm i\,\mu_3-\dfrac{r_e}{2}k^2\right]\right)\right).
\end{align}
Without loss of generality we take $\mu_3>0$ in the following discussion. Changing the sign of $\mu_3$ exchanges the two charge channels.

\paragraph*{Branch of the logarithm.}
The logarithm in \eqref{eq:QNMphysicalT0} arises from the $T\to0$ limit of the digamma function $\psi(1-i\omega/2\pi T)$ appearing in the denominator of the polarization function, which in \eqref{QNMphysical} is evaluated on the leading-order pole. As a function of $\omega$, the only non-analyticity of this denominator at $T>0$ is the series of poles of $\psi(1-i\omega/2\pi T)$ at $\omega=-2\pi iTn$, $n=1,2,\ldots$, which accumulate into a branch cut along the negative imaginary $\omega$ axis as $T\to0$. Consistently with \eqref{eq:LambdaT0}, the logarithm is therefore the principal branch of $\log(-i\omega r_e/9)$, defined by continuation from $\mathrm{Im}\,\omega>0$. On the principal sheet of the logarithm, on which the retarded correlator is defined, its imaginary part tends to $+\pi$ just to the left of the cut ($\mathrm{Re}\,\omega\to0^-$) and to $-\pi$ just to the right of it ($\mathrm{Re}\,\omega\to0^+$).

In \eqref{eq:QNMphysicalT0} the logarithm is evaluated on the leading-order pole $\omega^\pm_{\rm hydro,LO}=\mp\mu_3-ir_ek^2/2$. For $\mu_3>0$ this point lies at a finite distance $\mu_3$ from the cut, so the branch is fixed unambiguously: the argument of the logarithm in \eqref{eq:QNMphysicalT0} lies in $(\pi/2,\pi)$ for the channel $+$ and in $(-\pi,-\pi/2)$ for the channel $-$. The two channels are related by $\omega\to-\omega^*$, as required by charge conjugation. Equivalently, \eqref{eq:QNMphysicalT0} is the uniform $T\to0$ limit of \eqref{QNMphysical}, because the argument of the digamma function stays at a distance $\mu_3/(2\pi T)\to\infty$ from its poles.

At $\mu_3=0$, instead, the point at which the logarithm is evaluated, $\omega_{\rm hydro,LO}=-ir_ek^2/2$, lies on the cut, and \eqref{eq:QNMphysicalT0} does not specify which determination of the logarithm to use. There, the argument of the logarithm in \eqref{eq:QNMphysicalT0}, $(r_e/9)(\pm i\mu_3-r_ek^2/2)$, reduces to the negative real number $-r_e^2k^2/18$, so that $\log=\log(r_e^2k^2/18)\pm i\pi$, and the determination has to be fixed by self-consistency: the resulting NLO pole must lie on the side of the cut on which the imaginary part of the logarithm takes the assumed value. With $+i\pi$ the NLO correction shifts the pole to $\mathrm{Re}\,\omega=-\pi r_e^3k^4/24<0$, i.e.\ to the left of the cut, where the imaginary part of the logarithm is indeed $+\pi$. Likewise, $-i\pi$ gives a pole to the right of the cut. Both determinations are therefore consistent, and one finds two hydro-like poles,
\begin{equation}
	\omega_{\rm hydro,NLO}\Big|_{T=0,\,\mu_3=0}=-\frac{ir_e}{2}k^2\mp\frac{\pi r_e^3k^4}{24}
	+\frac{ir_e^3k^4}{24}\left[\gamma_E+\log\!\left(\frac{r_e^2k^2}{18}\right)\right],
	\label{eq:twopoles}
\end{equation}
where the upper (lower) sign corresponds to the determination $+i\pi$ ($-i\pi$). The two poles are related by $\omega\to-\omega^*$ and are present in both (coincident) charge channels. For $\mu_3>0$, the pole with $\mathrm{Re}\,\omega<0$ (respectively $>0$) is continuously connected to the hydro-like pole of the channel $+$ (respectively $-$). The pair of hydro-like poles at $\mu_3=0$ is thus split between the two charge channels and, at small momentum, each channel has a single hydro-like pole, described by \eqref{eq:QNMphysicalT0}.

\paragraph*{Emergence of a second hydro-like pole.}
The last statement holds only below a threshold momentum. The reason is that the determination of the logarithm is not a free choice. A zero of the denominator of \eqref{BAdimensionful}, computed with a given determination, is a pole of the retarded correlator only if it lies on the side of the cut associated with that determination. Otherwise it is a zero of the continuation of the denominator through the cut, i.e.\ it lies on the second sheet of the logarithm, and it is not a pole of the retarded correlator.

The expansion leading to \eqref{eq:QNMphysicalT0} treats $\mu_3$ as a leading-order quantity, $r_e\mu_3\sim(r_ek)^2=\mathcal O(\epsilon)$ as in \eqref{intro:scaling}, parametrically larger than the NLO correction, which is of order $(r_ek)^4=\mathcal O(\epsilon^2)$. In that regime the NLO shift cannot move the pole across the cut, and the determination of the logarithm fixed at $\omega^\pm_{\rm hydro,LO}$ is the correct one. The hierarchy fails when $r_e\mu_3$ is as small as the NLO correction, i.e.\ for $r_e\mu_3=\mathcal O(\epsilon^2)$ at $r_ek=\mathcal O(\sqrt\epsilon)$. Equivalently, at fixed $r_e\mu_3=\mathcal O(\epsilon)$, it fails at the larger momenta $r_ek=\mathcal O(\epsilon^{1/4})$. In this regime the isospin shift and the NLO terms have to be treated together. We therefore expand around the point $\omega_0=-ir_ek^2/2$ on the cut, regarding $\mp\mu_3$ and the NLO terms as corrections of the same order, and we fix the determination of the logarithm only afterwards. Since $|\omega_{\rm hydro,NLO}-\omega_0|\ll|\omega_0|$, on either side of the cut we can write $\log(-i\omega r_e/9)=\log(r_e^2k^2/18)+i\eta\pi+\mathcal O(\mu_3/r_ek^2)$, with $\eta=+1$ to the left of the cut and $\eta=-1$ to the right of it. This gives
\begin{equation}
	\omega^{\pm}_{\eta}=\mp\mu_3-\frac{ir_e}{2}k^2+i\frac{r_e^3k^4}{24}
	\left[\gamma_E+\log\!\left(\frac{r_e^2k^2}{18}\right)+i\eta\pi\right],
	\label{eq:cutexp}
\end{equation}
where we have dropped the contribution of $\mu_3$ to the NLO correction, which in this regime is of order $\epsilon^3\log\epsilon$. \eqref{eq:cutexp} reduces to \eqref{eq:twopoles} at $\mu_3=0$. A root $\omega^\pm_\eta$ is a pole of the retarded correlator only if it lies on the side of the cut corresponding to its determination, i.e.\ if
\begin{equation}
	\eta\,\mathrm{Re}\,\omega^\pm_\eta<0 .
	\label{eq:selfcons}
\end{equation}

In the channel $+$, the root with $\eta=+1$ has $\mathrm{Re}\,\omega^+_{+}=-\mu_3-\pi r_e^3k^4/24<0$ for all $k$. It is the hydro-like pole \eqref{eq:QNMphysicalT0}, that \eqref{eq:cutexp} reproduces to the order considered. The root with $\eta=-1$ has
\begin{equation}
	\mathrm{Re}\,\omega^+_{-}=-\mu_3+\frac{\pi r_e^3k^4}{24} \,.
	\label{eq:Resecond}
\end{equation}
It satisfies \eqref{eq:selfcons} only when \eqref{eq:Resecond} is positive. The channel $-$ follows from $\omega\to-\omega^*$.

The meaning of \eqref{eq:Resecond} is simple. At $\mu_3=0$ the two hydro-like poles \eqref{eq:twopoles} sit symmetrically on the two sides of the cut. Their separation comes from the discontinuity of the logarithm across the cut, $2\pi i$, multiplied by the coefficient $ir_e^3k^4/24$ of the logarithm in \eqref{eq:twopoles}. Each pole is therefore at a distance $\pi r_e^3k^4/24$ from the cut. A non-zero $\mu_3$ shifts both poles of the channel $+$ to the left by $\mu_3$. The pole that was on the right of the cut remains a pole of the retarded correlator only as long as its distance from the cut exceeds this shift. The distance grows as $k^4$, while the shift does not depend on $k$. So there is a threshold momentum $k_c$, above which the second zero is also a pole of the retarded correlator. Setting \eqref{eq:Resecond} to zero gives
\begin{equation}
	r_ek_c=\left(\frac{24\,r_e\mu_3}{\pi}\right)^{1/4}.
	\label{eq:kc}
\end{equation}
For $k<k_c$, each charge channel has a single hydro-like pole, given by \eqref{eq:QNMphysicalT0}. For $k>k_c$ a second hydro-like pole appears in the retarded correlator, with $\mathrm{Re}\,\omega>0$ in the channel $+$ (and $\mathrm{Re}\,\omega<0$ in the channel $-$). It is given by \eqref{eq:cutexp} with $\eta=\mp1$. As $\mu_3\to0$, $k_c\to0$ and the two hydro-like poles of \eqref{eq:twopoles} are recovered at all momenta. Using $r_e=2\sqrt{3N_c}/(w_0\mu)$, the threshold scales as $k_c/\mu\propto(\mu_3/\mu)^{1/4}$.

For $\mu_3/\mu_q=1.2/13$ ($r_e\mu_3\simeq0.15$), the value at which we compute the QNMs numerically, \eqref{eq:kc} gives $r_ek_c\simeq1.0$. Since this value is of order one, the threshold lies at the edge of the regime of validity of the expansion, and \eqref{eq:kc} should be regarded as an estimate.

The second hydro-like pole differs from the first in that it is a pole of the retarded correlator only for $k>k_c$. At $k=k_c$ its real part vanishes, see \eqref{eq:Resecond}, so it emerges from the cut on the negative imaginary axis, at 
\begin{align}
\nn\omega_c&\simeq-\frac{ir_e}{2}k_c^2+i\frac{r_e^3k_c^4}{24}\left[\gamma_E+\log\!\left(\frac{r_e^2k_c^2}{18}\right)\right]\\ 
\label{eq:omegac} &= -\frac{2i}{r_e}\left(\frac{3\,r_e\mu_3}{2\pi}\right)^{1/2}+i\frac{\mu_3}{\pi}\left[\gamma_E + \log\!\left(\frac{2}{9}\right) + \frac{1}{2}\log\!\left(\frac{3\,r_e\mu_3}{2\pi}\right)\right] ,
\end{align}
that is with a finite damping rate. For $k<k_c$ the zero described by \eqref{eq:cutexp} with $\eta=-1$ still exists, but it lies on the second sheet of the logarithm and is not a pole of the retarded correlator. At $\mu_3=0$ (and $T=0$), instead, $k_c=0$ and the two hydro-like poles \eqref{eq:twopoles} are poles of the retarded correlator at all momenta.

\paragraph*{Finite temperature.}
At non-zero temperature the cut is replaced by the poles of $\psi(1-i\omega/2\pi T)$ at $\omega=-2\pi iTn$, i.e.\ by the tower of IR modes. The second hydro-like pole therefore does not emerge from a cut: it detaches from the IR tower, near the level whose frequency matches the imaginary part of the pole at threshold, $\omega_c$ in \eqref{eq:omegac}. Setting $-\mathrm{Im}\,\omega_c=2\pi Tn_c$ gives
	\begin{equation}
		n_c\simeq\frac{r_ek_c^2}{4\pi T}\left[1-\frac{r_e^2k_c^2}{12}\left(\gamma_E+\log\frac{r_e^2k_c^2}{18}\right)\right],
		\label{eq:nc}
	\end{equation}
	whose leading term, $n_c\simeq r_ek_c^2/(4\pi T)$, is the condition \eqref{eq:psidiv} at $\mu_3=0$. Since $n_c$ grows as $1/T$, the closer the background is to extremality, the higher the AdS$_2$ level from which the second pole emerges. For $\mu_3/\mu_q=1.2/13$, $r_ek_c\simeq1.0$, and the term in square brackets increases the leading-order value by about $20\%$: \eqref{eq:nc} gives $n_c\simeq4.1$ at $\mu_q/T=65$ and $n_c\simeq19$ at $\mu_q/T=300$.

The discreteness of the tower also smooths out the threshold. At $T=0$, whether the zero \eqref{eq:cutexp} with $\eta=-1$ is a pole of the retarded correlator is decided by the side of the cut on which it lies, i.e.\ by the sign of $\mathrm{Re}\,\omega$: across the cut the imaginary part of the logarithm jumps from $+\pi$ to $-\pi$. At non-zero temperature the logarithm is replaced by $\psi(1-i\omega/2\pi T)$, which has no cut, and its imaginary part changes continuously. Close to the negative imaginary axis, and away from the IR levels,
	\begin{equation}
		\mathrm{Im}\,\psi\!\left(1-\frac{i\omega}{2\pi T}\right)\simeq-\pi\tanh\frac{\mathrm{Re}\,\omega}{2T}\,,
	\end{equation}
	which approaches $\mp\pi$ only for $|\mathrm{Re}\,\omega|\gtrsim4T$. For smaller $|\mathrm{Re}\,\omega|$ the two determinations of the logarithm are not distinct, and neither is the second pole from the IR levels it passes close to. Near the threshold, \eqref{eq:Resecond} gives $\mathrm{Re}\,\omega^+_{-}=\mu_3[(k/k_c)^4-1]\simeq4\mu_3(k-k_c)/k_c$. The second pole is therefore a well-defined hydro-like pole only for $k\gtrsim k_c(1+\Delta k/k_c)$, with
	\begin{equation}
		\frac{\Delta k}{k_c}\simeq\frac{T}{\mu_3}\,.
	\end{equation}
In the window $[k_c-\Delta k,k_c+\Delta k]$, the hydro-like pole interacts with the IR levels instead of crossing a cut, and the threshold is sharp only for $\mu_3\gg T$. For $\mu_3/\mu_q=1.2/13$, $\Delta k/k_c\simeq0.04$ at $\mu_q/T=300$ and $\Delta k/k_c\simeq0.17$ at $\mu_q/T=65$.

The smoothing also shifts the level from which the second pole detaches. Since $-\mathrm{Im}\,\omega\propto k^2$ at leading order, the pole becomes distinct from the IR tower near the level $n_c(1+2\Delta k/k_c)$ rather than $n_c$. At $\mu_q/T=300$ the shift is small, about $7\%$. At $\mu_q/T=65$ it is about $30\%$, which moves the estimate from $n_c\simeq4.1$ to about $5.5$. Note that this is consistent with the numerical result, in which the second hydro-like mode detaches from the fifth IR level (figure~\ref{fq5}). Given that $r_ek_c\simeq1$, the agreement indicates that the perturbative treatment remains roughly valid for such values of momentum.

\subsection{The infrared poles}\label{sec:QNMIR}
Our discussion makes evident that the NLO correlator knows about the IR poles. Can we say something about their dispersion relation? If we set the denominator of the correlator to zero, we obtain a meromorphic equation due to the presence of the digamma function and it contains many roots. We would like to perform an analytic study, similar to what we did for the hydro-like pole. In order to do so, we look for the poles of the digamma function present in the correlator. The frequencies for which the digamma function diverges (see \eqref{Npm} and \eqref{Dpm}) are
\begin{equation}
	\boldsymbol{p}^{(n)}  \equiv -2i\,\tau\,n\,,\qquad n\in\mathbb{N}\,.
\end{equation}
Note that, the frequencies $\boldsymbol{p}^{(n)} $ are poles of the digamma function,
but they are not themselves quasinormal frequencies of the complete
NLO correlator. Indeed, both the numerator and the denominator of
the correlator contain singular contributions at these frequencies,
but their ratio remains generically finite. The values $\boldsymbol{p}^{(n)} $
should instead be regarded as the leading-order locations around
which the IR poles are organized. We can compute the NLO correction to the above formula by expanding a generic QNM frequency around $\boldsymbol{p}^{(n)} $
\begin{equation}
	\boldsymbol{\omega}^{\pm} _{\text{QNM}} = \boldsymbol{p}^{(n)}  +\epsilon\,\delta\boldsymbol{\omega}^{\pm,(n)}+\mathcal{O}(\epsilon^2)\,.
\end{equation}
We substitute the ansatz above in the denominator of the NLO correlator and we set it to zero. We solve the equation perturbatively in $\epsilon$ (like we did for the hydro-pole NLO correction) and we obtain the following formula for $\delta\boldsymbol{\omega}^{\pm,(n)} $:
\begin{align}
	\delta\boldsymbol{\omega}^{\pm,(n)}  = \tau\,\dfrac{\boldsymbol{k}^4+i\,\boldsymbol{p}^{(n)} \boldsymbol{k}^2+2(\boldsymbol{p}^{(n)} \pm\mathfrak{m})(\boldsymbol{p}^{(n)} \pm2\mathfrak{m})}{6\left[i\,\boldsymbol{k}^2+2(\boldsymbol{p}^{(n)} \pm\mathfrak{m})\right]}\,.
\end{align}

Now, we group the leading order dispersion relation for the IR poles $\boldsymbol{p}^{(n)} $ and its NLO correction together and define
\begin{equation}
	\boldsymbol{\omega}_\text{IR,NLO}^{\pm}  \equiv \boldsymbol{p}^{(n)}  +\epsilon\,\delta  \boldsymbol{\omega}^{\pm,(n)} \,.
\end{equation}
We restore the dimensions by using $r_H$ to obtain
\begin{align}\nonumber
	\omega_\text{IR,NLO}^{\pm,(n)} = &-2\pi i\,T\,n+\\ \label{eq:NLOIR}
	&+\pi\,T\,r_H\,\dfrac{r_H^2k^4+2\pi\,Tr_Hk^2n+2(-2\pi i\,T\,n\pm\mu_3)(-2\pi i\,T\,n\pm2\mu_3)}{6\left[i\,r_Hk^2+2(-2\pi i\,T\,n\pm\mu_3)\right]}\,.
\end{align}

As for the hydro-pole dispersion relation \eqref{QNMphysical}, the formula above is valid up to higher order corrections in the near-extremal limit. Moreover, we observe that the NLO correction diverges when its denominator vanishes. This happens for
\begin{equation}
	r_H^2k^2 =  4\pi\,n\,r_HT\pm2i\,r_H\mu_3\,.
\end{equation}
which is the same as \eqref{eq:psidiv}. Therefore, the NLO correction of both the hydro-like and IR pole diverges for the same value of parameters, that is when the two poles collide in the complex plane (with $k$ complex in general).
 As a consequence, it is not possible to say anything about the interacting region of the two poles by perturbing around a given LO pole. This is because we have assumed a non-degenerate expansion around an isolated root of the denominator.

\subsection{Interacting poles}\label{sec:QNMint}
The question is then whether we can retain analytic control in the region where the poles collide, or come close to each other. The answer is yes, as we now show. Since the formulas for isolated poles (hydro-like and IR) become singular and directly unusable, we shall instead keep the singular contributions, reduce the denominator to a second-order polynomial in $\boldsymbol{\omega}$ and then look for the two roots of the equation.

Given the function \eqref{Dpm} entering the NLO denominator, we can extract the divergent contribution coming from the digamma function as the frequency asymptotes to $\boldsymbol{p}^{(n)} $ and define the regular part of $\mathfrak D^\pm$
\begin{equation}
	\mathfrak D^{\pm,(n)} _{\mathrm{reg}}(\boldsymbol\omega)
	=
	\mathfrak D^\pm(\boldsymbol\omega)-	\frac{d_{-1}^{\pm,(n)} }{\boldsymbol\omega-\boldsymbol{p}^{(n)} }\,,
	\label{int:Ddecomposition}
\end{equation}
with
\begin{align}
	d_{-1}^{\pm,(n)}
	={}&
	-2i\tau
	\left[
	-\frac{i(\boldsymbol{p}^{(n)} \pm2\mathfrak m)}{12}
	+
	\frac{\boldsymbol k^2(\boldsymbol{p}^{(n)} -i\boldsymbol k^2)}
	{24(\boldsymbol{p}^{(n)} \pm\mathfrak m)}
	\right].
	\label{int:dminusone}
\end{align}
By construction $\mathfrak D_{\mathrm{reg}}^{\pm,(n)} $ is regular at $\boldsymbol\omega=\boldsymbol{p}^{(n)} $.

The QNM condition following from the denominator of the NLO correlator is
\begin{equation}
	\boldsymbol\omega
	\pm\mathfrak m
	+
	\frac{i\boldsymbol k^2}{2}
	+
	\epsilon
	(\boldsymbol\omega\pm\mathfrak m)
	\mathfrak D^\pm(\boldsymbol\omega)
	=0\,.
	\label{int:QNMcondition}
\end{equation}
Using \eqref{int:Ddecomposition} and multiplying by
$\boldsymbol\omega-\boldsymbol{p}^{(n)} $, this becomes
\begin{align}
	&
	\left(
	\boldsymbol\omega-\boldsymbol\omega_\text{hydro,LO}^{\pm}
	\right)
	\left(\boldsymbol\omega-\boldsymbol{p}^{(n)} \right)
	+
	\epsilon
	(\boldsymbol\omega\pm\mathfrak m)d_{-1}^{\pm,(n)}
	\nonumber\\
	&+
	\epsilon
	(\boldsymbol\omega\pm\mathfrak m)
	\left(\boldsymbol\omega-\boldsymbol{p}^{(n)} \right)
	\mathfrak D_{\mathrm{reg}}^{\pm,(n)} (\boldsymbol\omega)=0\,,
	\label{int:conditionmultiplied}
\end{align}
where 
\begin{equation}
	\boldsymbol\omega_\text{hydro,LO}^{\pm}
	=
	\mp\mathfrak m-\frac{i\boldsymbol k^2}{2}
	\label{int:hydroLO}
\end{equation}
is the leading hydro-like frequency.

As $\boldsymbol\omega$ varies, the leading hydrodynamic and infrared frequencies are close when
\begin{equation}
	\boldsymbol\omega_\text{hydro,LO}^{\pm}-\boldsymbol{p}^{(n)}
	=
	\mp\mathfrak m-\frac{i\boldsymbol k^2}{2} + 2i\,\tau\,n \ll 1\,.
\end{equation}
We call this region the interaction region. Since
$(\boldsymbol\omega\pm\mathfrak m)
\mathfrak D_{\mathrm{reg}}^{\pm,(n)}(\boldsymbol\omega)$
in the last term in \eqref{int:conditionmultiplied}  is analytic in this region, we can expand it around the leading
hydrodynamic frequency:
\begin{align}
	&(\boldsymbol\omega\pm\mathfrak m)
	\mathfrak D_{\mathrm{reg}}^{\pm,(n)}
	(\boldsymbol\omega)=
	\left(
	\boldsymbol\omega_\text{hydro,LO}^{\pm}
	\pm\mathfrak m
	\right)
	\mathfrak D_{\mathrm{reg}}^{\pm,(n)}
	\left(
	\boldsymbol\omega_\text{hydro,LO}^{\pm}
	\right)
	+
	\mathcal O(\epsilon)\,.
\end{align}
When this expression is inserted into the last term of
\eqref{int:conditionmultiplied}, the remainder contributes at higher order in
$\epsilon$ in the interaction region and can therefore be neglected
within the present accuracy.

It is then convenient to define the regularly corrected
hydrodynamic level
\begin{align}
	\overline{\boldsymbol\omega}_{\text{hydro}}^{\pm,(n)}
	\equiv{}&
	\boldsymbol\omega_\text{hydro,LO}^{\pm}
	-
	\epsilon
	\left(
	\boldsymbol\omega_\text{hydro,LO}^{\pm}
	\pm\mathfrak m
	\right)
	\mathfrak D_{\mathrm{reg}}^{\pm,(n)}
	\left(
	\boldsymbol\omega_\text{hydro,LO}^{\pm}
	\right).
	\label{int:regularhydro}
\end{align}
The bar indicates that this frequency includes the part of the NLO
correction that is regular near the selected infrared level
$\boldsymbol{p}^{(n)} $. It is not the complete NLO hydrodynamic frequency:
the singular contribution associated with the $n$-th infrared level
has been excluded and will be retained explicitly in the interaction
equation.

Using \eqref{int:regularhydro}, equation
\eqref{int:conditionmultiplied} reduces to
\begin{equation}
		\left(
		\boldsymbol\omega
		-
		\overline{\boldsymbol\omega}_{\text{hydro}}^{\pm,(n)}
		\right)
		\left(
		\boldsymbol\omega-\boldsymbol{p}^{(n)}
		\right)
		+
		\epsilon
		(\boldsymbol\omega\pm\mathfrak m)
		d_{-1}^{\pm,(n)}
		=0\,.
	\label{int:localequation}
\end{equation}

We now solve the local equation \eqref{int:localequation}. To
distinguish the sign labelling the charged channel from the sign
labelling the two solutions of the quadratic equation, we introduce
$\eta=\pm1$.  The two solutions are
\begin{align}
	\boldsymbol\omega_{\mathrm{int},\eta}^{\pm,(n)}
	={}&
	\frac{1}{2}
	\left[
	\boldsymbol{p}^{(n)}
	+
	\overline{\boldsymbol\omega}_{\text{hydro}}^{\pm,(n)}
	-
	\epsilon d_{-1}^{\pm,(n)}
	\right]
	\nonumber\\
	&+
	\frac{\eta}{2}
	\Bigg[
	\left(
	\boldsymbol{p}^{(n)}
	-
	\overline{\boldsymbol\omega}_{\text{hydro}}^{\pm,(n)}
	+
	\epsilon d_{-1}^{\pm,(n)}
	\right)^2
	-
	4\epsilon
	\left(
	\boldsymbol{p}^{(n)} \pm\mathfrak m
	\right)
	d_{-1}^{\pm,(n)}
	\Bigg]^{1/2}.
	\label{int:roots}
\end{align}
Equation \eqref{int:roots} should be understood within the accuracy
of the local NLO equation from which it was derived. We keep the
square root unexpanded because its two contributions become
comparable in the interaction region.

The square root in \eqref{int:roots} is defined by analytic
continuation from a region in which the two levels are well-separated. Its branch is chosen such that, when we neglect $d_{-1}^{\pm,(n)}$,
\begin{equation}
	\left[
	\left(
	\boldsymbol{p}^{(n)}
	-
	\overline{\boldsymbol\omega}_{\text{hydro}}^{\pm,(n)}
	+
	\epsilon d_{-1}^{\pm,(n)}
	\right)^2
	-
	4\epsilon
	\left(
	\boldsymbol{p}^{(n)} \pm\mathfrak m
	\right)
	d_{-1}^{\pm,(n)}
	\right]^{1/2}
	\longrightarrow
	\overline{\boldsymbol\omega}_{\text{hydro}}^{\pm,(n)}
	-
	\boldsymbol{p}^{(n)} .
\end{equation}

Away from the interaction region, the separation between the two
regular levels is large compared with the term proportional to $d_{-1}^{\pm,(n)}$. We define
\begin{equation}
	\Delta_n^\pm
	\equiv
	\overline{\boldsymbol\omega}_{\mathrm{hydro}}^{\pm,(n)}
	-\boldsymbol{p}^{(n)} .
\end{equation}
Choosing the root connected to
$\overline{\boldsymbol\omega}_{\mathrm{hydro}}^{\pm,(n)}$, we find the following expansion of the square root in \eqref{int:roots}
\begin{align}
	&
	\Big[
	\left(
	\boldsymbol{p}^{(n)}
	-
	\overline{\boldsymbol\omega}_{\mathrm{hydro}}^{\pm,(n)}
	+
	\epsilon d_{-1}^{\pm,(n)}
	\right)^2
	-
	4\epsilon
	\left(\boldsymbol{p}^{(n)} \pm\mathfrak m\right)
	d_{-1}^{\pm,(n)}
	\Big]^{1/2}
	\nonumber\\
	&\qquad =
	\Delta_n^\pm
	-
	\epsilon d_{-1}^{\pm,(n)}
	-
	\frac{
		2\epsilon
		\left(\boldsymbol{p}^{(n)} \pm\mathfrak m\right)
		d_{-1}^{\pm,(n)}
	}{
		\Delta_n^\pm
	}
	+
	\mathcal O(\epsilon^2).
	\label{int:sqrtexpansion}
\end{align}
This expansion 
is valid when
\begin{equation}
	\left|\Delta_n^\pm\right|
	\gg
	\epsilon\left|d_{-1}^{\pm,(n)}\right|,
	\qquad
	\left|\Delta_n^\pm\right|^2
	\gg
	\epsilon
	\left|
	\left(\boldsymbol{p}^{(n)} \pm\mathfrak m\right)
	d_{-1}^{\pm,(n)}
	\right|.
	\label{int:awaycondition}
\end{equation}

The hydrodynamic-like root becomes
\begin{align}
	\boldsymbol\omega_{\mathrm{int},+}^{\pm,(n)}
	={}&
	\overline{\boldsymbol\omega}_{\mathrm{hydro}}^{\pm,(n)}
	-
	\epsilon
	\frac{
		\left(
		\overline{\boldsymbol\omega}_{\mathrm{hydro}}^{\pm,(n)}
		\pm\mathfrak m
		\right)
		d_{-1}^{\pm,(n)}
	}{
		\Delta_n^\pm
	}
	+
	\mathcal O(\epsilon^2).
	\label{int:hydrobranch}
\end{align}
To NLO, the barred quantities in the second term can be replaced by
their leading-order values. Using the definition
\eqref{int:regularhydro}, we then recover
\begin{align}
	\boldsymbol\omega_{\mathrm{int},+}^{\pm,(n)}
	={}&
	\boldsymbol\omega_{\mathrm{hydro,LO}}^\pm
	-
	\epsilon
	\left(
	\boldsymbol\omega_{\mathrm{hydro,LO}}^\pm
	\pm\mathfrak m
	\right)
	\mathfrak D^\pm
	\left(
	\boldsymbol\omega_{\mathrm{hydro,LO}}^\pm
	\right)
	+
	\mathcal O(\epsilon^2)
	\nonumber\\
	={}&
	\boldsymbol\omega_{\mathrm{hydro,NLO}}^\pm
	+
	\mathcal O(\epsilon^2).
\end{align}
The infrared-like root becomes
\begin{align}
	\boldsymbol\omega_{\mathrm{int},-}^{\pm,(n)}
	={}&
	\boldsymbol{p}^{(n)}
	+
	\epsilon
	\frac{
		\left(\boldsymbol{p}^{(n)} \pm\mathfrak m\right)
		d_{-1}^{\pm,(n)}
	}{
		\Delta_n^\pm
	}
	+
	\mathcal O(\epsilon^2)
	\nonumber\\
	={}&
	\boldsymbol{p}^{(n)}
	+
	\epsilon\,
	\delta\boldsymbol\omega^{\pm,(n)}
	+
	\mathcal O(\epsilon^2)
	\nonumber\\
	={}&
	\boldsymbol\omega_{\mathrm{IR,NLO}}^{\pm,(n)}
	+
	\mathcal O(\epsilon^2).
\end{align}
Thus, away from the interaction region, the two roots recover the previously derived NLO hydro and IR poles. When condition \(\eqref{int:awaycondition}\) fails, the square root must remain unexpanded, and \(\eqref{int:roots}\) provides the local interacting-pole approximation.

Equation \eqref{int:roots} provides a uniform local
approximation to the QNM frequencies of the interacting poles in the region where the isolated-pole
expansions cease to be reliable. Away from this region, the two
roots reduce respectively to the NLO hydro-like dispersion
relation \eqref{QNMphysical} and to the NLO infrared
frequency \eqref{eq:NLOIR}. Close to the interaction region, the
square root must instead be retained unexpanded and describes the
redistribution of the two modes caused by their mixing.

The construction is local in the infrared level $n$: it describes
the interaction of the hydrodynamic-like pole with one selected
IR mode and assumes that all other infrared levels remain
sufficiently far away. Its accuracy is therefore controlled both
by the near-extremal expansion and by the separation from the
remaining QNMs.

\section{Comparison with numerical results}\label{sec:numerics}
In this section, we assess quantitatively the accuracy of the analytic results derived in sections~\ref{sec:NLOlong} and~\ref{sec:QNMs} by comparing them with the full numerical solution of the longitudinal fluctuation equation \eqref{EpEoM}. The main goal of this comparison is to determine to what extent the NLO corrections enlarge the regime in which the near-extremal analytic expansion provides an accurate description of the retarded correlator.

In section \ref{qnm}, we compare the numerical results for the QNMs derived with the Mathematica package of \cite{Jansen:2017oag} with the results for the QNMs derived in section \ref{sec:QNMs} from the analytic formula for the NLO polarization function.

Section \ref{sec:numpol} is devoted to the comparison of the exact (numerically computed) polarization function with the LO and NLO analytic results. The numerical correlator is obtained by solving the fluctuation equations with ingoing boundary conditions at the horizon, and extracting the source and response coefficients from the near-boundary asymptotics of the solution. The numerical procedure follows the shooting method described in detail in \cite{Apostolidis:2026efq}. As in \cite{Apostolidis:2026efq}, we focus on the imaginary part of the polarization function, i.e. the spectral function, which contains the same physical information as the full correlator (due to the Kramers-Kronig relation). This quantity is also particularly convenient for the comparison because it does not need to be regularized, as it is not affected by the logarithmically divergent contact terms discussed in section~\ref{sec:holomodel}.

\subsection{Quasinormal modes}
\label{qnm}

\begin{figure}[htb]
	\begin{subfigure}{\textwidth}
		\centering
		\begin{subfigure}{0.45\textwidth}
			\centering
			\includegraphics[width=\textwidth]{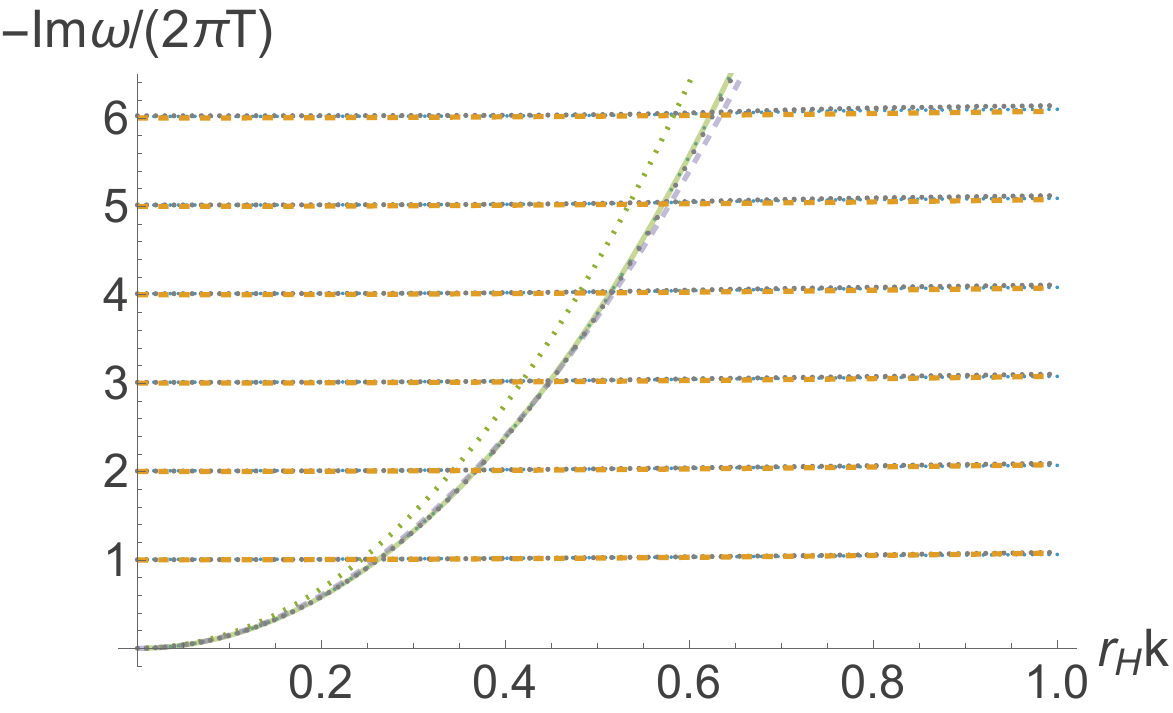}
		\end{subfigure}
		\hspace{1cm}
		\begin{subfigure}{0.45\textwidth}
			\centering
			\includegraphics[width=\textwidth]{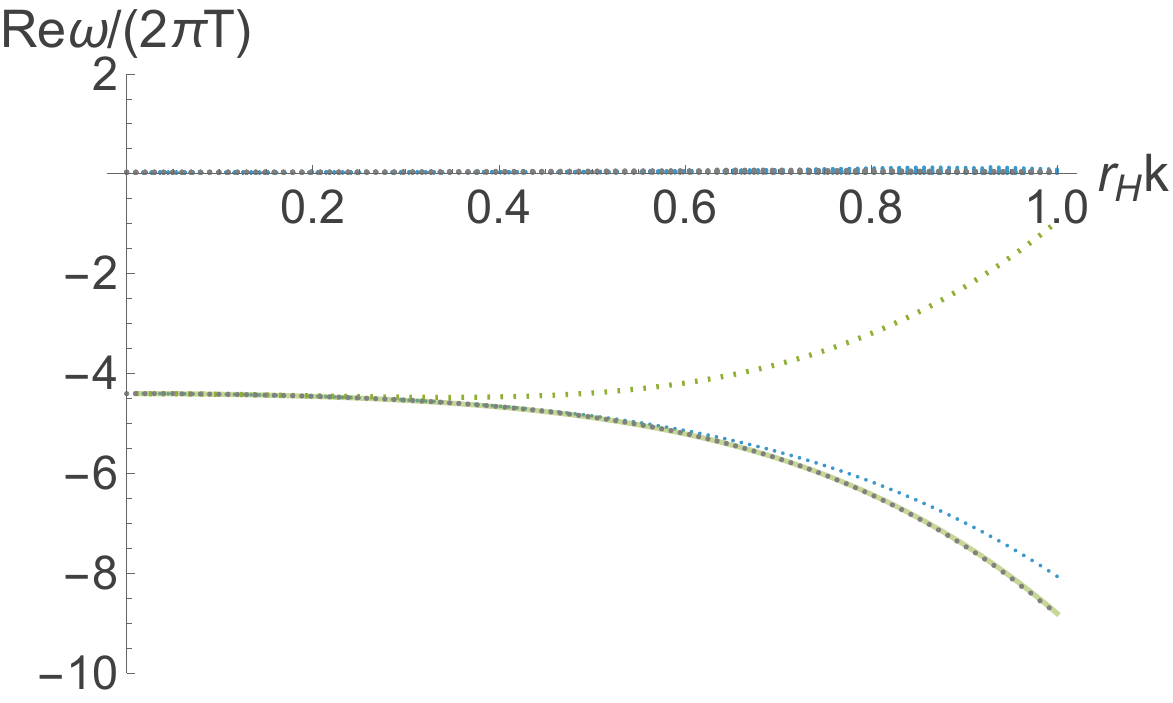}
		\end{subfigure}
	\end{subfigure}
	\caption{Imaginary parts (left) and real parts (right) of the first few longitudinal QNM's at $\mu_q/T = 300$ and $\mu_3/\mu_q=1.2/13$, as a function of momentum. We compare the numerical results for the full QNM's computed in \cite{Apostolidis:2026efq} (grey squares), with the zeros of \eqref{q1} (blue dots). Also shown are the AdS$_2$ modes \eqref{q2} (orange dashed), together with the LO (dashed) and NLO (green) zero-temperature dispersion relations of the hydro-like modes, given in \eqref{eq:long:QNMLO} and \eqref{eq:QNMphysicalT0}. The green curve with decreasing real part is the hydro-like pole, given by the principal determination of the logarithm in \eqref{eq:QNMphysicalT0}. The green curve with increasing real part is obtained with the other determination and describes the second hydro-like pole of section~\ref{sec:QNMT0}. It is drawn dotted where its real part is negative, i.e.\ where it lies on the second sheet of the logarithm and is not a pole of the retarded correlator; at $\mu_q/T=300$ this is the case in the whole range shown.}
	\label{fq1}
\end{figure}
\begin{figure}[htb]
	\begin{subfigure}{\textwidth}
		\centering
		\begin{subfigure}{0.45\textwidth}
			\centering
			\includegraphics[width=\textwidth]{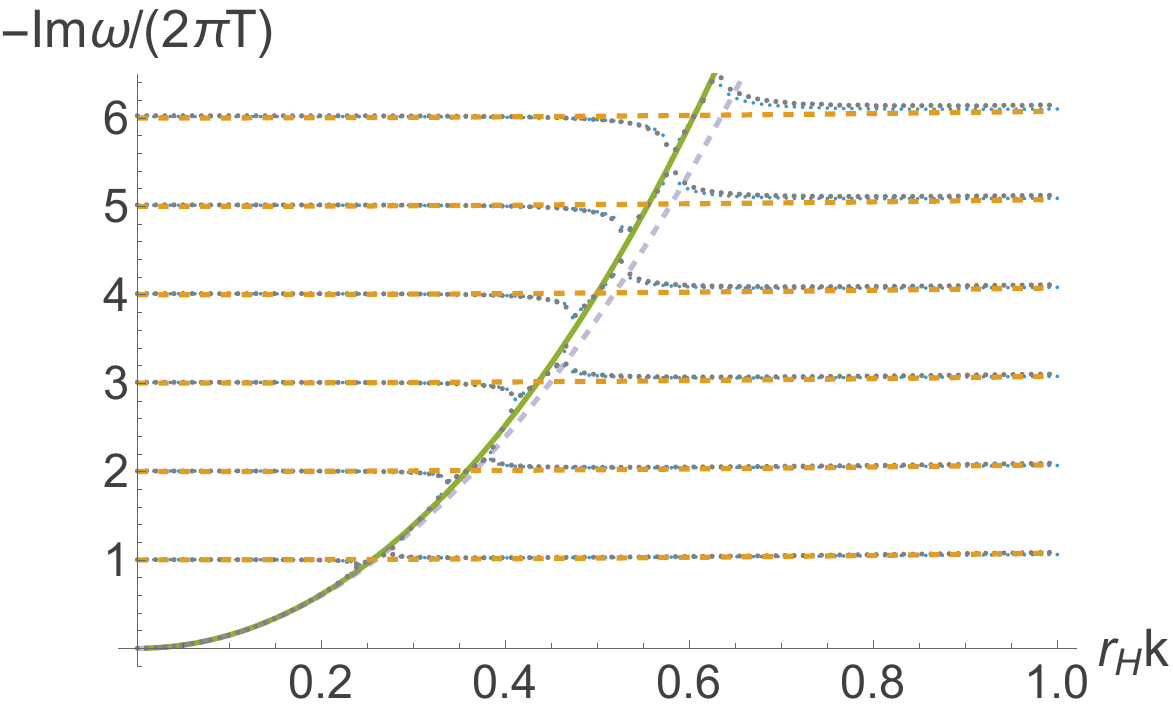}
		\end{subfigure}
		\hspace{1cm}
		\begin{subfigure}{0.45\textwidth}
			\centering
			\includegraphics[width=\textwidth]{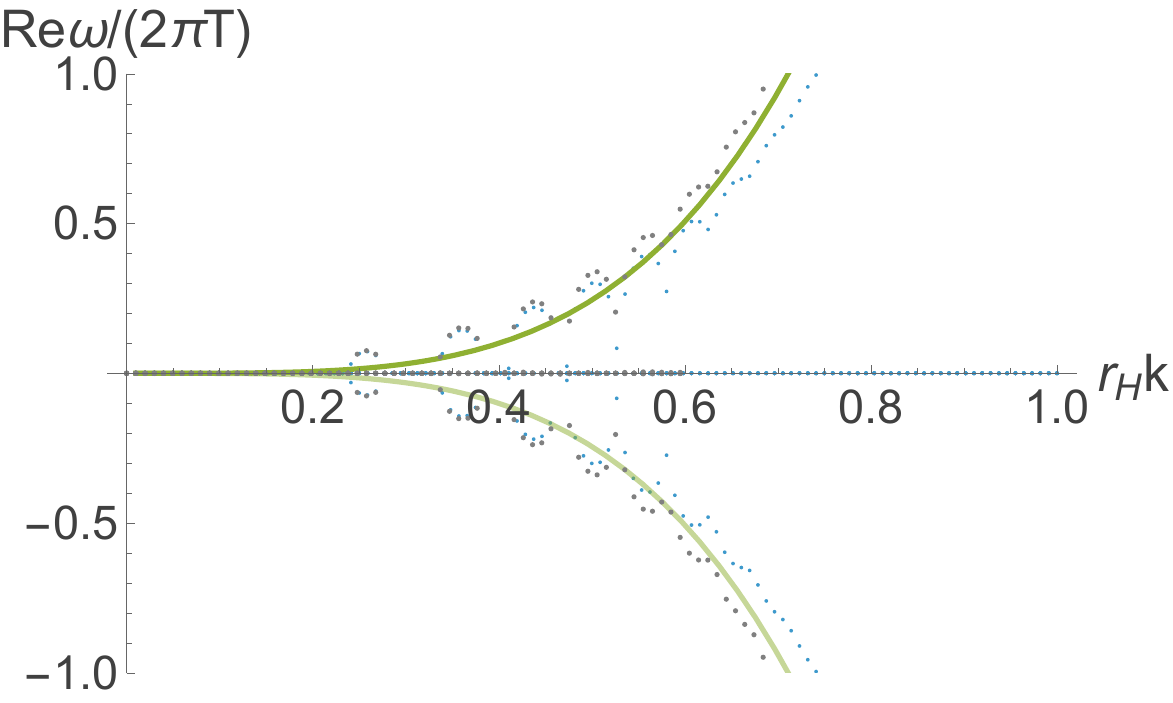}
		\end{subfigure}
	\end{subfigure}
	\caption{Same as figure \ref{fq1} but for $\mu_3 = 0$. The two green curves are the hydro-like poles \eqref{eq:twopoles}, obtained with the two determinations of the logarithm; at $\mu_3=0$ both are poles of the retarded correlator at all momenta, and their imaginary parts coincide.}
	\label{fq4}
\end{figure}

\begin{figure}[htb]
	\begin{subfigure}{\textwidth}
		\centering
		\begin{subfigure}{0.45\textwidth}
			\centering
			\includegraphics[width=\textwidth]{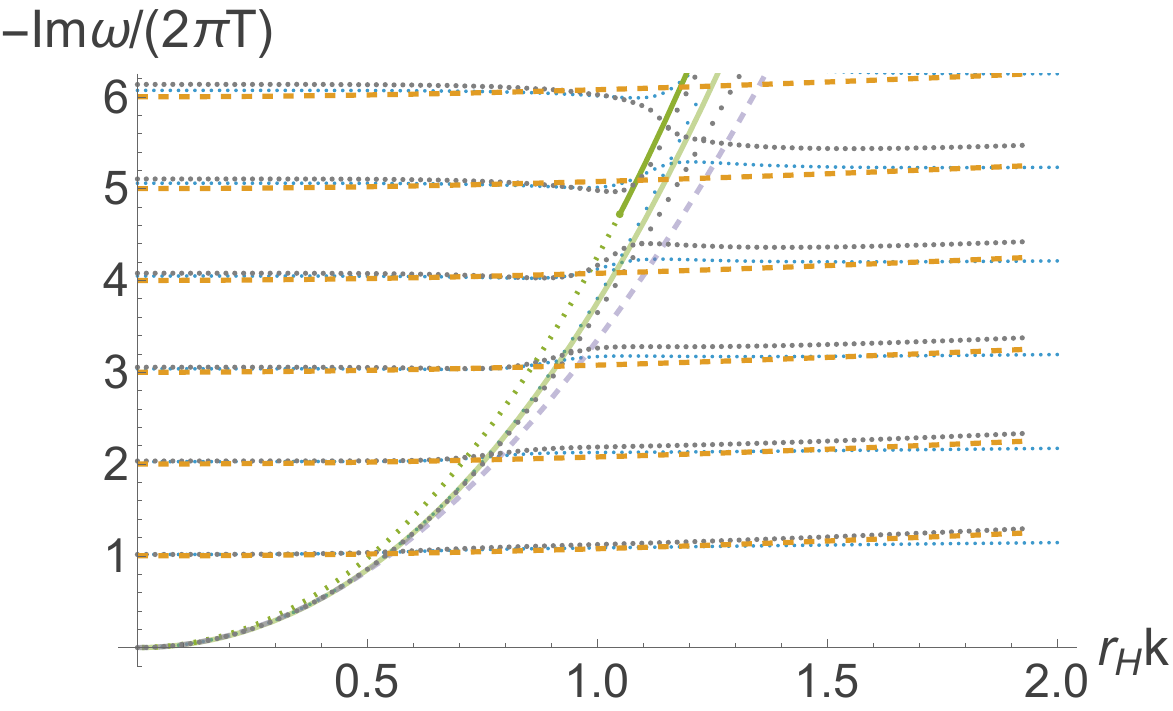}
		\end{subfigure}
		\hspace{1cm}
		\begin{subfigure}{0.45\textwidth}
			\centering
			\includegraphics[width=\textwidth]{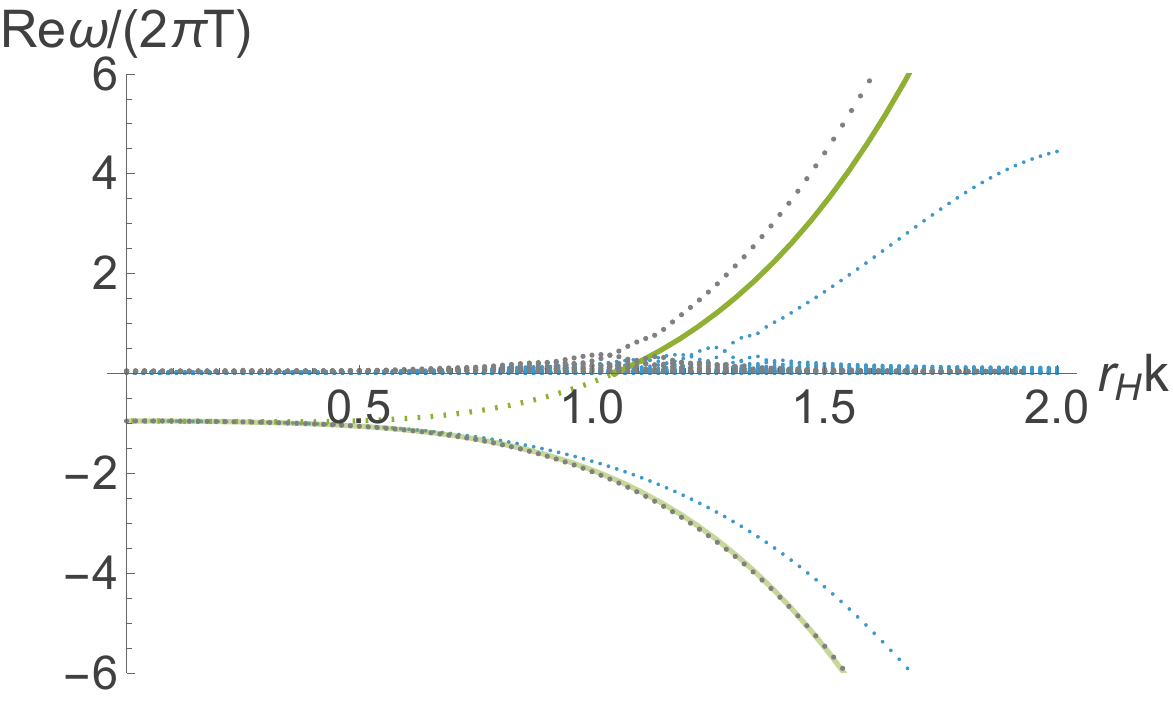}
		\end{subfigure}
	\end{subfigure}
	\caption{Same as figure \ref{fq1} ($\mu_3/\mu_q =1.2/13$) but for $\mu_q/T = 65$. Here the green curve with increasing real part becomes positive for $r_Hk\gtrsim1.05$, where it is drawn solid and describes the second hydro-like pole.}
	\label{fq2}
\end{figure}

\begin{figure}[htb]
	\begin{subfigure}{\textwidth}
		\centering
		\begin{subfigure}{0.45\textwidth}
			\centering
			\includegraphics[width=\textwidth]{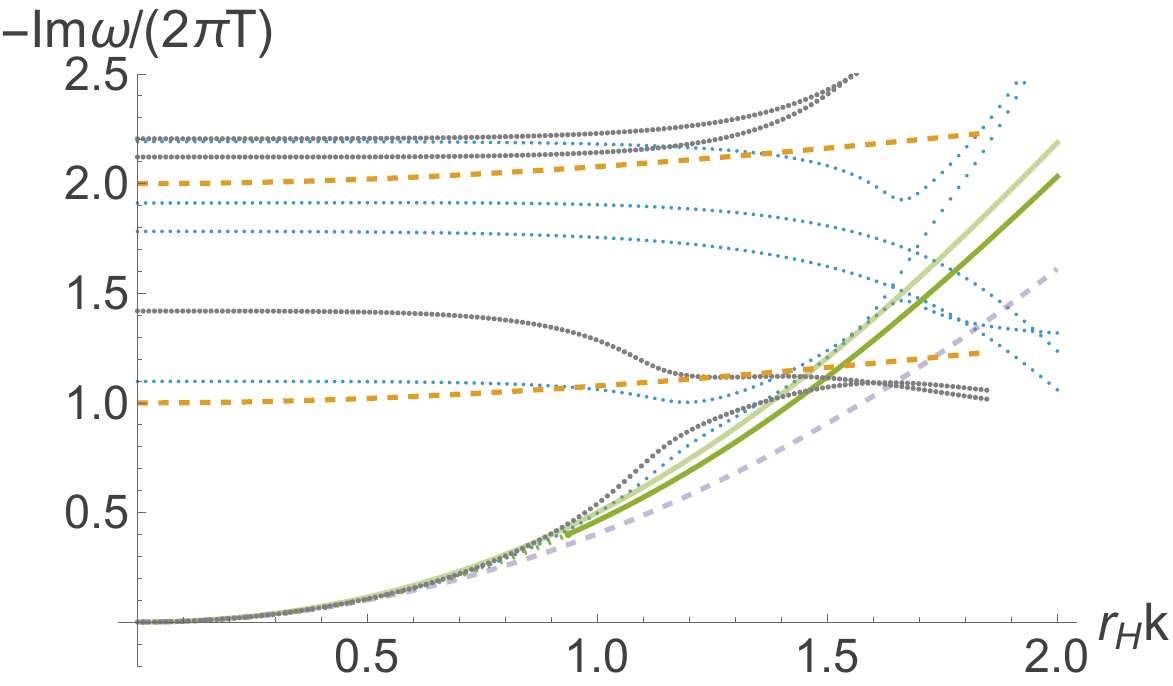}
		\end{subfigure}
		\hspace{1cm}
		\begin{subfigure}{0.45\textwidth}
			\centering
			\includegraphics[width=\textwidth]{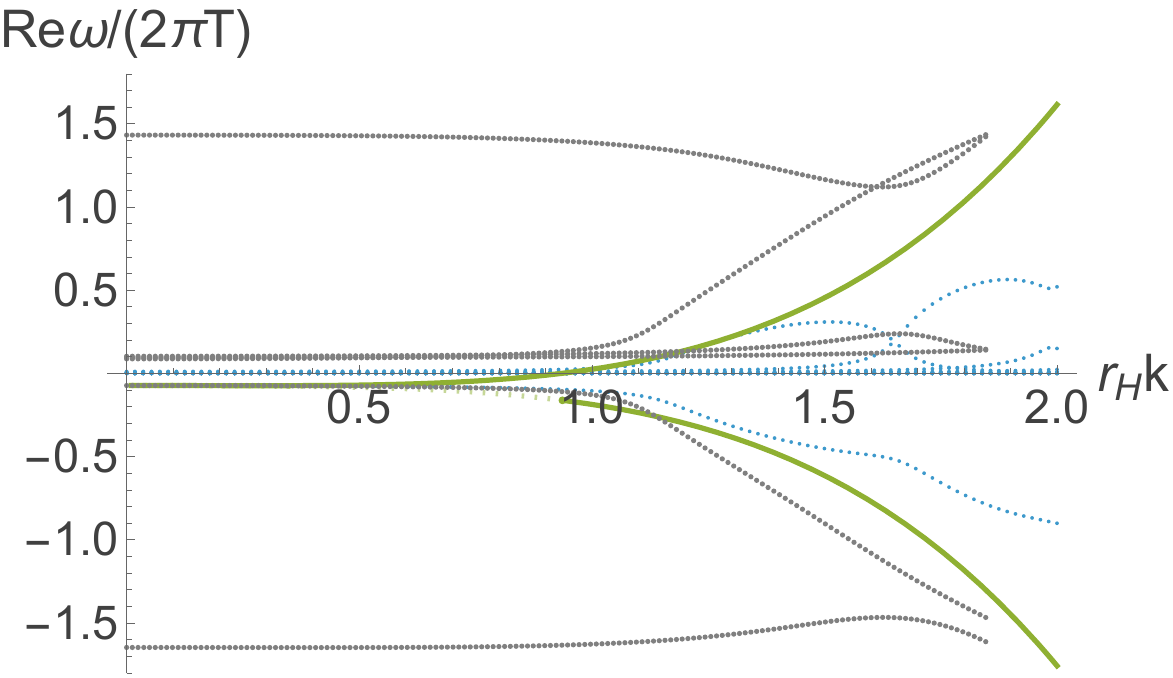}
		\end{subfigure}
	\end{subfigure}
	\caption{Same as figure \ref{fq1} but for $\mu_q/T = 5$.}
	\label{fq3}
\end{figure}

\begin{figure}[htb]
	\begin{subfigure}{\textwidth}
		\centering
		\begin{subfigure}{0.45\textwidth}
			\centering
			\includegraphics[width=\textwidth]{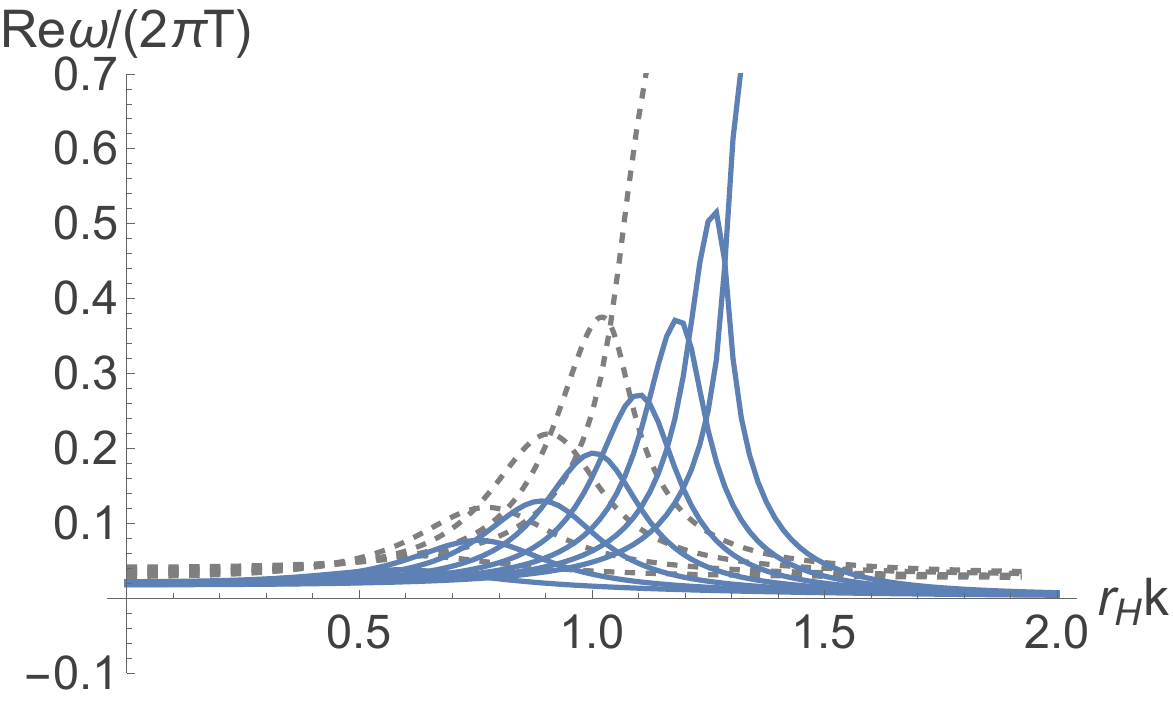}
		\end{subfigure}
		\hspace{1cm}
		\begin{subfigure}{0.45\textwidth}
			\centering
			\includegraphics[width=\textwidth]{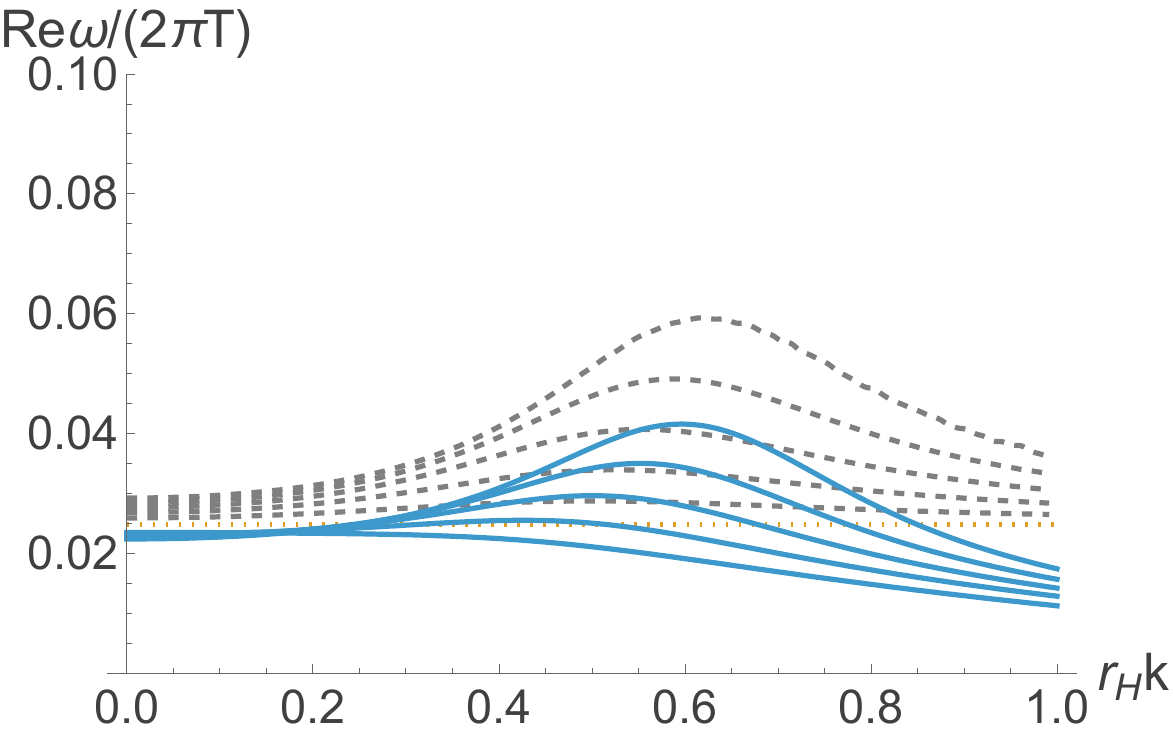}
		\end{subfigure}
	\end{subfigure}
	\caption{Zoom on the positive real parts in figures \ref{fq2} and \ref{fq1}, with the full QNM's shown in dashed grey, and the zeros of \eqref{q1} in blue. The orange dotted line shows the real part of the AdS$_2$ modes \eqref{q2}. The left figure ($\mu_q/T = 65$) shows eight modes and the right figure ($\mu_q/T = 300$) five modes. Note that the level at which a mode has an unbounded real part and which corresponds to the second hydro-like mode of section~\ref{sec:QNMT0} (called hydro-like in \cite{Apostolidis:2026efq}) is the fifth and the eighth for the exact and the NLO QNM spectrum respectively. Such a mode is also expected to exist at $\mu_q/T = 300$, but it is predicted to start around $n_c\simeq19$ (of the exact QNM spectrum) according to \eqref{eq:nc}.}
	\label{fq5}
\end{figure}

We first present the comparison with numerics at the level of the quasinormal spectrum. As mentioned before, the QNM's are poles of the retarded polarization function \eqref{eq:long:matchedcorrelator}, which correspond to values of the frequency for which the source $\AA^\pm$ in \eqref{eq:NBexp} is zero. The NLO approximation to the QNM's are thus given by the zeros of the NLO source \eqref{NLOA}. We use the form \eqref{eq:BAtilde} of the correlator, so that the condition reads
\begin{equation}
\label{q1} \left(\bm{\omega}_{\text{QNM}}^{\text{NLO}}\pm \mathfrak{m}\right) + \frac{i}{2}\bm{k}^2 + \left(\bm{\omega}_{\text{QNM}}^{\text{NLO}}\pm \mathfrak{m}\right)\widetilde{\mathfrak{D}}^\pm(\bm{\omega}_{\text{QNM}}^{\text{NLO}},\bm{k}^2,\mathfrak{m},\tau) = 0 \, ,
\end{equation}
with $\widetilde{\mathfrak{D}}^\pm$ defined in \eqref{eq:NDtilde}. As discussed in section~\ref{sec:QNMs}, the analytic formulas for the QNMs are the same whether they are derived from \eqref{BAepsilon} or from \eqref{eq:BAtilde}. The exact zeros of the two truncated denominators, instead, differ at $\mathcal O(\epsilon^2)$: a rescaling $\mathcal A^\pm\to(1+\epsilon\,G)\mathcal A^\pm$, with $G$ any function of the frequency, leaves the NLO matching unchanged but shifts the zeros at $\mathcal O(\epsilon^2)$. We use \eqref{eq:BAtilde} because its numerator and denominator have finite zero-temperature limits and its zeros do not contain the spurious branches of the denominator of \eqref{BAepsilon} in the range shown (see appendix \ref{app:otherdef}). The zeros of the denominator of \eqref{BAepsilon} are compared with those of \eqref{q1} in appendix~\ref{app:otherdef}. In this section, we consider the case of charge $+$, while results for the $-$ case are obtained via $\mu_3\to -\mu_3$. Equation \eqref{q1} does not admit closed form solutions, but it can be solved numerically to arbitrary precision. These (very precise) 
numerical results are indicated in figures \ref{fq1} to \ref{fq5} by the blue dots. Note that, unlike the LO result where the only visible pole was the LO diffusive-like mode $\bm{\omega}_D(\bm{k}) = -i\bm{k}^2/2 + \OO(\bm{k}^4)$, the NLO result not only accounts for NLO corrections to the dispersion relation $\bm{\omega}_D(\bm{k})$, but also includes a series of additional QNM's related to the IR modes.
For small $\tau$, these modes lie close to the AdS$_2$ modes, 
\begin{equation}
\label{q2} \omega_n^{\mathrm{AdS_2}} = -i2\pi T(\D(\bm{k},\mathfrak{m}) + n) \pm \frac{1}{3} \pi T\mathfrak{m} \, ,
\end{equation}
\begin{equation}
\label{q3} \Delta(\bm{k},\mathfrak{m}) = \frac{1}{2} + \frac{1}{2}\sqrt{1+\frac{1}{3}\bm{k}^2-\frac{1}{9}\mathfrak{m}^2} \, ,
\end{equation}
which are the poles of the pure AdS$_2$ correlator at small $\bm{\omega}$ and $\tau$ (see \cite{Apostolidis:2026efq} for more details). 
Figures \ref{fq1} to \ref{fq5} also show, indicated as grey squares, numerical results for the all order QNM's, that were computed in \cite{Apostolidis:2026efq} with the Mathematica package of \cite{Jansen:2017oag}.

The comparison with the blue dots in figures \ref{fq1} ($\mu_q/T=300\,;\,\mu_3/\mu_q=1.2/13$), \ref{fq4} ($\mu_q/T=300\,;\,\mu_3=0$) and \ref{fq2} ($\mu_q/T=65\,;\,\mu_3/\mu_q=1.2/13$), confirms that the NLO QNM's computed from \eqref{q1} give a good approximation to the first few modes at small $T/\mu_q$, until the momentum $\bm{k} = r_H k$ becomes of order one. The NLO zero-temperature dispersion relation \eqref{eq:QNMphysicalT0} (green line) also gives a good approximation to the hydro-like pole up to $\bm{k}\simeq 1$ for $\mu_q/T = 65$ and $300$, which is much better than the LO expression (grey dashed line). At $\mu_q/T=300$ the real part of the hydro-like zero of \eqref{q1} follows the numerical data up to $r_Hk\simeq0.7$ and lies slightly above them at larger momenta. This residual deviation is a consequence of the $\mathcal O(\epsilon^2)$ ambiguity of the exact zeros discussed above: the consistent NLO dispersion relation \eqref{QNMphysical}, whose $T\to0$ limit is the green line, follows the numerical data more closely. In the range shown, the zeros of \eqref{q1} do not display spurious branches. At $\mu_3=0$ (figure~\ref{fq4}), the zeros of \eqref{q1} reproduce the successive collisions of the hydro-like mode with the IR levels and the opposite real parts acquired by the colliding pairs. Beyond the collision region they follow the two hydro-like poles \eqref{eq:twopoles} (green curves).

As discussed in section~\ref{sec:QNMT0}, at non-zero isospin a second hydro-like pole exists in a given charge channel only above the threshold momentum $k_c$ of \eqref{eq:kc}, and at finite temperature it detaches from the IR tower near the level $n_c$ of \eqref{eq:nc}. This explains why it is hard to observe close to extremality. For $\mu_3/\mu_q=1.2/13$ and $\mu_q/T=300$, $r_Hk_c\simeq1.03$ and $n_c\simeq19$, both outside the range of figure~\ref{fq1}: indeed no such mode is visible there, and the green curve with increasing real part is dotted over the whole range. At $\mu_q/T=65$, instead, $n_c\simeq4.1$, and the numerical QNMs in figure~\ref{fq2} show a second hydro-like mode. Around $r_Hk\simeq1$ it leaves the IR tower between the fourth and fifth levels, consistent with the estimate of section~\ref{sec:QNMT0}, and its real part becomes positive and grows rapidly with momentum. The solid part of the green curve with increasing real part, i.e.\ \eqref{eq:cutexp} with $\eta=-1$, starts at $r_Hk\simeq1.05$ and follows the imaginary part of this mode closely, while it underestimates the growth of its real part. The zeros of \eqref{q1} show a corresponding branch in the right panel of figure~\ref{fq2}: a zero whose real part becomes positive at a somewhat larger momentum and grows more slowly than the numerical mode. On the negative side, the numerical hydro-like mode follows the green curve over the whole range, while the hydro-like zero of \eqref{q1} departs from it for $r_Hk\gtrsim1.3$. For these parameters $\mu_3/T\simeq6$, so the threshold is smoothed over $\Delta k/k_c\simeq0.17$ by the discreteness of the IR tower, and $r_Hk_c\simeq1$ lies at the edge of the regime of validity of the expansion: the agreement can therefore only be qualitative.

At higher temperature, figure \ref{fq3} ($\mu_q/T=5\,;\,\mu_3/\mu_q=1.2/13$) indicates that the NLO approximation is much farther from the actual modes, apart from the leading behavior of the hydro-like mode at small momentum. Nonetheless, it still reproduces some qualitative features of the QNM's, like the crossing of the hydro-like pole's imaginary part with the first IR mode visible in left figure \ref{fq3}, beyond which the imaginary parts of the two modes both decrease in absolute value with momentum.

The scale in right figures \ref{fq1} and \ref{fq2} is actually too large to see well the region of small positive real parts, close to the AdS$_2$ real part in \eqref{q2}. To extend our analysis to these modes, we show in figure \ref{fq5} a close-up on the small positive real part region of right figures \ref{fq1} and \ref{fq2}.

We observe that the corresponding NLO modes are qualitatively similar to the exact modes, and the quantitative agreement improves as $\mu_q/T$ increases. The approximation worsens for momenta $\bm{k}$ larger than one, which is also as expected. The real parts of the IR zeros of \eqref{q1} are somewhat smaller than the numerical ones. Since these real parts arise entirely at NLO, they are particularly sensitive to the $\mathcal O(\epsilon^2)$ ambiguity of the exact zeros: the zeros of the denominator of \eqref{BAepsilon} give larger real parts, closer to the numerical data (appendix~\ref{app:otherdef}). In the left figure~\ref{fq5}, the NLO IR modes do not follow the rapid growth of the real part of the fifth mode (at the same QNM level), which is the second hydro-like mode. In the spectrum of \eqref{q1}, the corresponding zero detaches from the IR tower at slightly larger momentum and from the eighth level. Its growing real part is visible in the right figure~\ref{fq2} and in its zoomed version, i.e. left plot of figure \ref{fq5}.

We conclude this section by testing our formula for the local interacting poles \eqref{int:roots} against the numerical results. In order to do so, we focus on a background with $\mu_q/T = 300$ for which the preceding comparisons show good agreement between the low-lying numerical QNMs and the NLO spectrum, and consider both $\mu_3=0$ and $\mu_3 =1.2/13$. We first compare the two roots of the local
quadratic equation with the isolated-pole expressions
\eqref{QNMphysical} and~\eqref{eq:NLOIR}, and then compare them with QNMs obtained from the full bulk fluctuation equation. These are distinct tests: the first checks consistency between the analytic approximations, while the second assesses their accuracy against the numerical spectrum. Also for the following plots we focus on the plus charged correlator.

\begin{figure}[htb]
	\begin{subfigure}{\textwidth}
		\centering
		\begin{subfigure}{0.45\textwidth}
			\centering
			\includegraphics[width=\textwidth]{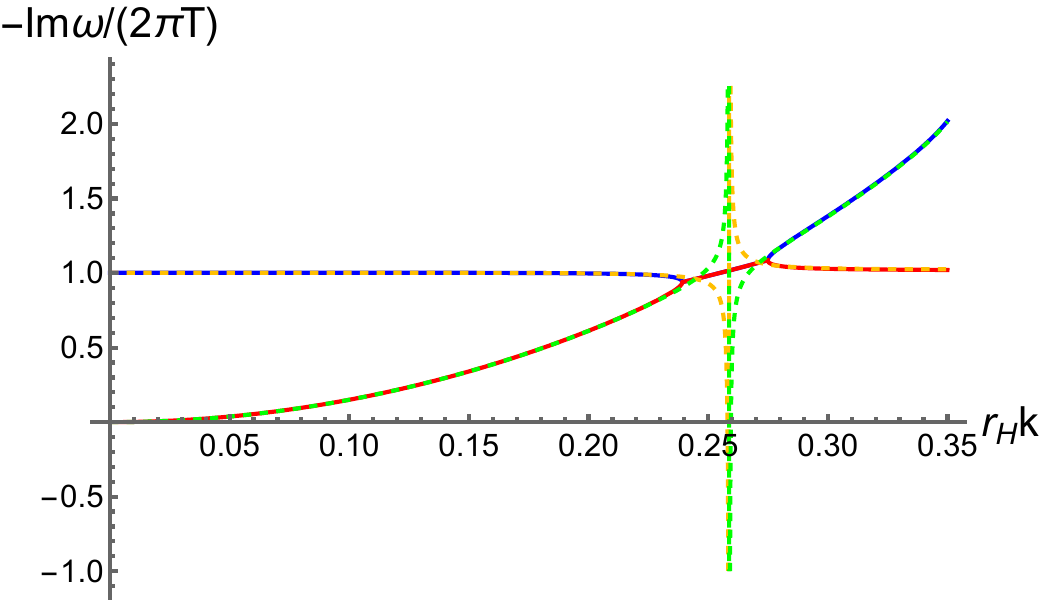}
		\end{subfigure}
		\hspace{1cm}
		\begin{subfigure}{0.45\textwidth}
			\centering
			\includegraphics[width=\textwidth]{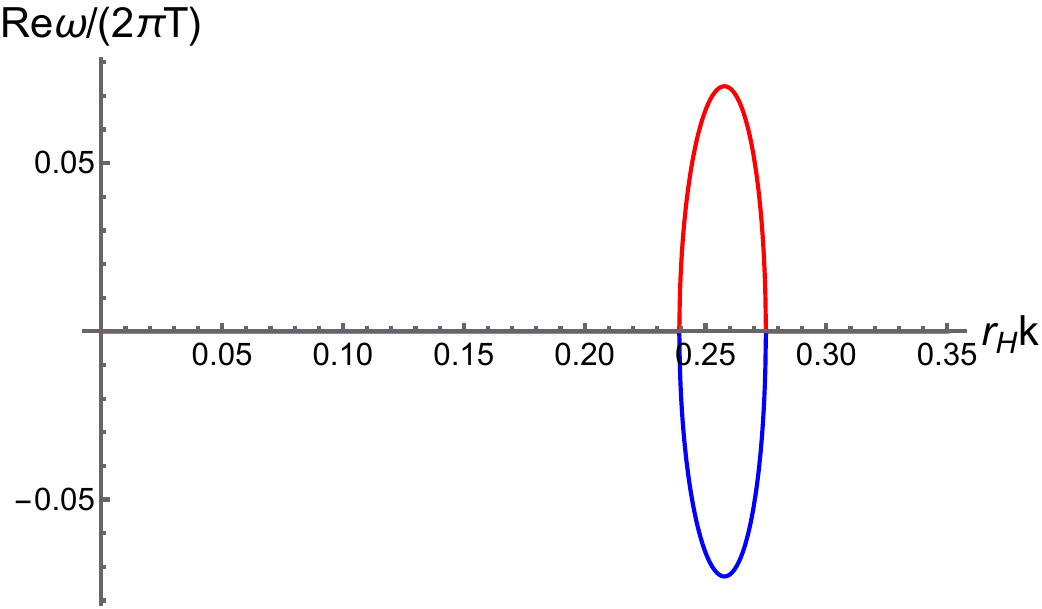}
		\end{subfigure}
	\end{subfigure}
	\caption{Imaginary (left panel) and real (right panel) part of the following dispersion relations: the solid blue and red solid lines corresponds to the two interacting roots \eqref{int:roots} for $n =1$; the dashed green line corresponds to the NLO hydro-like dispersion relation \eqref{QNMphysical}; the dashed yellow line shows the NLO IR dispersion relation \eqref{eq:NLOIR} for $n=1$. The background is characterized by $\mu_q/T=300$ and $\mu_3=0$.}\label{fig:A}
\end{figure}

We start with the case $\mu_3=0$. In figure \ref{fig:A}, we present the imaginary (left panel) and real (right panel) part of the following dispersion relations: the solid blue and red solid lines corresponds to the two interacting roots \eqref{int:roots} for $n=1$, the first one is the root connected to the IR pole while the second is connected to the hydro-like pole; the dashed green line corresponds to the NLO hydro-like dispersion relation \eqref{QNMphysical}; the dashed yellow line shows the NLO IR dispersion relation \eqref{eq:NLOIR} for $n = 1$. We stress that there are no exact numerical data in these plots.

Away from the interaction region, the interacting pole formula reproduce the behavior of the isolated poles. The isolated-pole corrections
become singular at $r_Hk^2=4\pi T$ (recall formula \eqref{eq:psidiv}), whereas
the interacting roots remain finite. This singularity marks the
failure of the non-degenerate expansion; it should not be identified with the collision points of the corrected spectrum. In the local two-pole approximation, the roots coalesce, acquire opposite real parts over a finite momentum interval, and then return to the imaginary axis. Within that interval they have the same imaginary part.

\begin{figure}[htb]
	\begin{subfigure}{\textwidth}
		\centering
		\begin{subfigure}{0.45\textwidth}
			\centering
			\includegraphics[width=\textwidth]{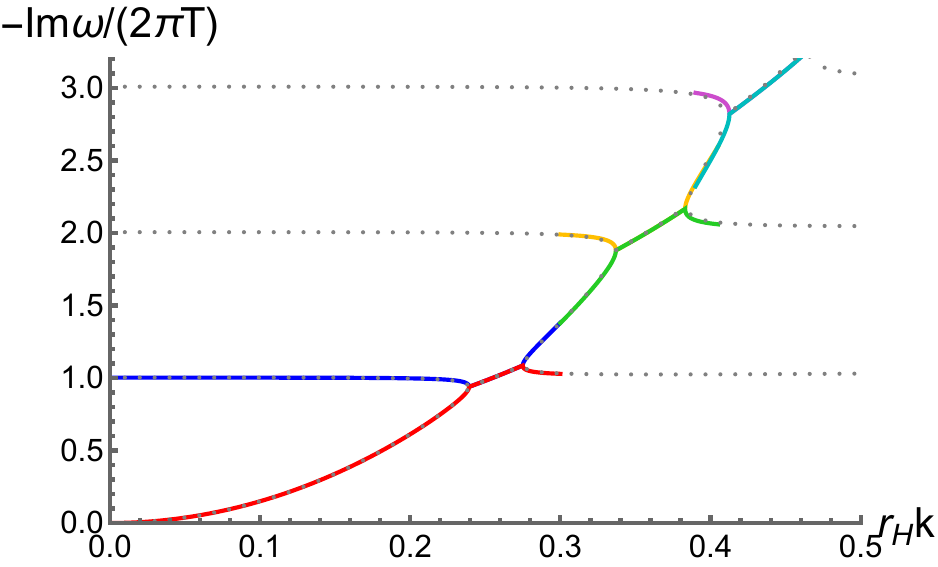}
		\end{subfigure}
		\hspace{1cm}
		\begin{subfigure}{0.45\textwidth}
			\centering
			\includegraphics[width=\textwidth]{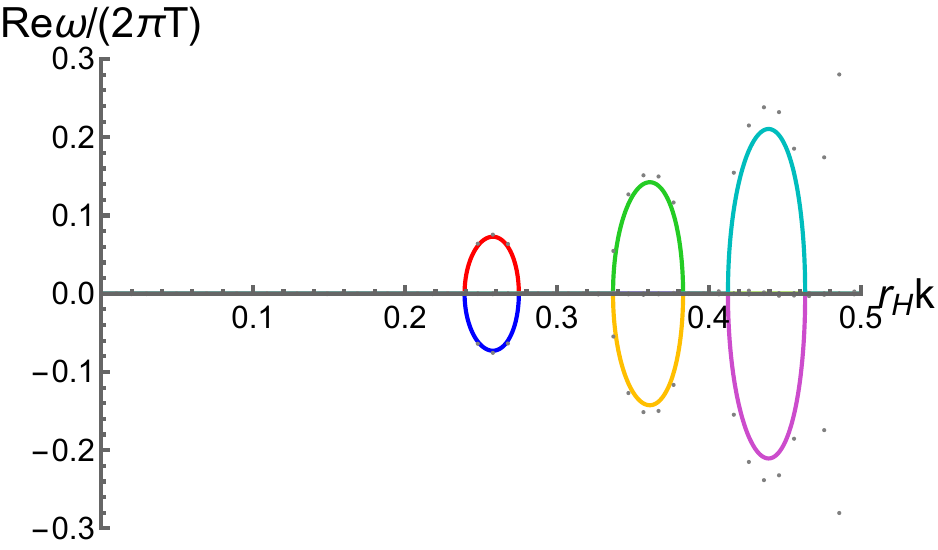}
		\end{subfigure}
	\end{subfigure}
	\caption{Imaginary (left panel) and real (right panel) part of the first longitudinal QNMs at $\mu_q/T=300$ and $\mu_3=0$. We compare the numerical results (grey squares) with the analytic formula for the interacting poles \eqref{int:roots}. The blue--red, yellow--green and purple--cyan pairs correspond to separate local approximations for $n=1,2,3$, respectively, shown around the corresponding interaction regions.}\label{fig:B}
\end{figure}

Then, we compare formula \eqref{int:roots} with the exact numerical results for the quasinormal mode frequencies at $\mu_3=0$. The results are presented in figure \ref{fig:B} where the left and right panel show the imaginary and real part respectively of the first QNMs. The numerical results are indicated with grey squares. The blue--red, yellow--green and
purple--cyan pairs correspond to $n=1,2,3$, respectively. Each pair is a separate local
approximation around the selected IR level; the curves do not arise from a simultaneous treatment of all three levels. The agreement is particularly good for the imaginary parts and for the first interaction. Differences are more visible in the real parts, already for $n=2$, and become more pronounced for $n=3$. Thus the formula captures the first collision and the qualitative pattern of the subsequent interactions, but does not reproduce all higher-level trajectories with the same accuracy. These discrepancies indicate the limits of the combined NLO and local two-pole approximation.

\begin{figure}[htb]
	\begin{subfigure}{\textwidth}
		\centering
		\begin{subfigure}{0.45\textwidth}
			\centering
			\includegraphics[width=\textwidth]{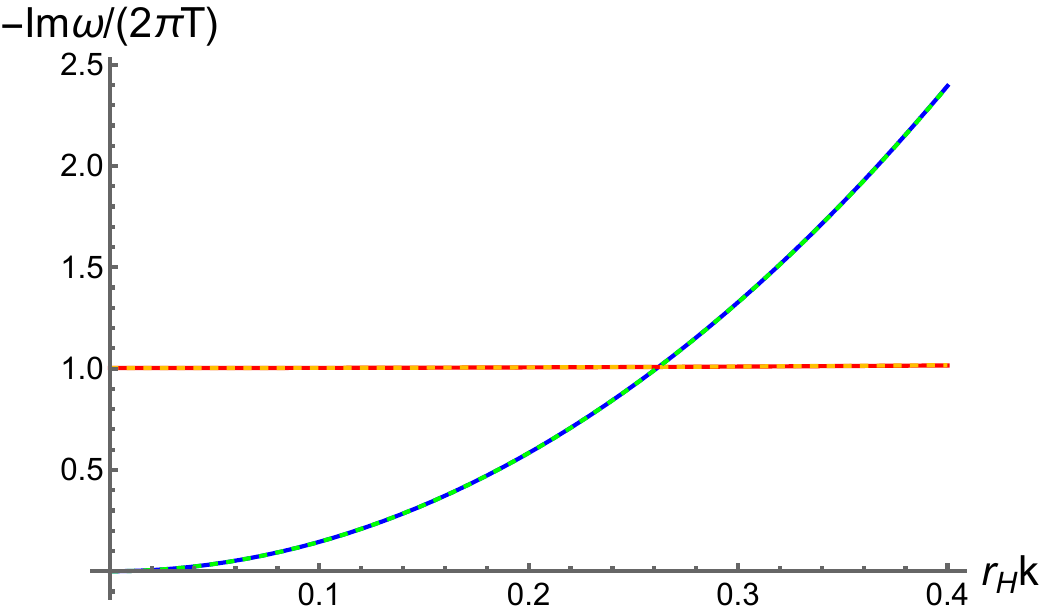}
		\end{subfigure}
		\hspace{1cm}
		\begin{subfigure}{0.45\textwidth}
			\centering
			\includegraphics[width=\textwidth]{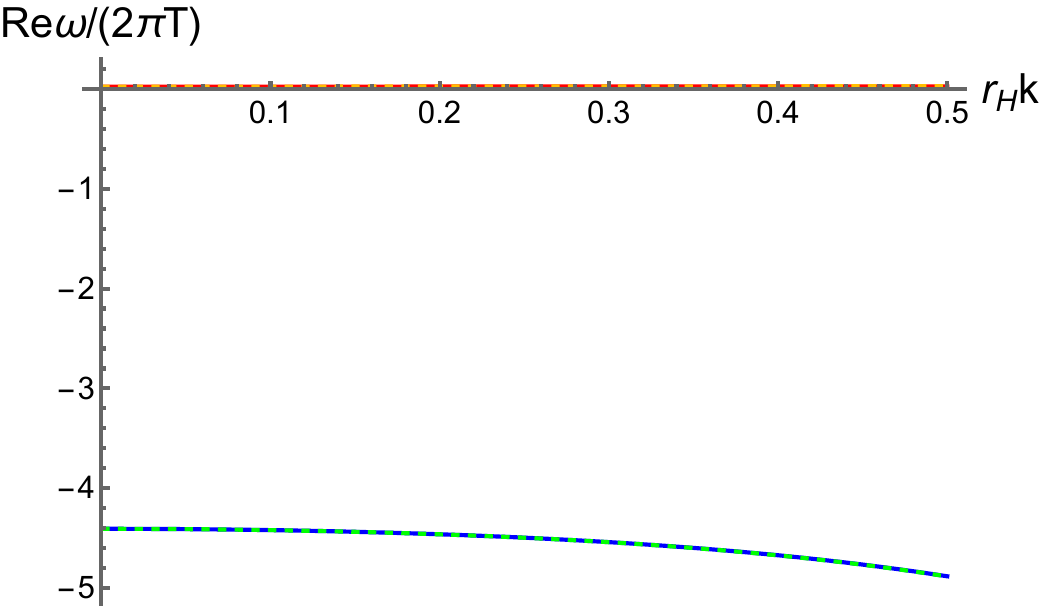}
		\end{subfigure}
	\end{subfigure}
	\caption{Same as figure \ref{fig:A} but for $\mu_3/\mu_q = 1.2/13$.}\label{fig:C}
\end{figure}

We now consider the case of non-zero isospin chemical potential $\mu_3/\mu_q =1.2/13$ while we keep $\mu_q/T=300$. Figure \ref{fig:C} again compares the interacting roots \eqref{int:roots} for $n=1$ with the isolated NLO expressions \eqref{QNMphysical}, \eqref{eq:NLOIR}, without displaying numerical data.  In
this figure, the solid blue curve follows the hydro-like branch (dashed green line) and the solid red curve
the IR-like branch (yellow dashed line). The solid and dashed curves are nearly indistinguishable at the
resolution of the plot. Although the imaginary parts cross, the real parts remain separated, so this crossing is not a collision in the complex frequency plane. For these values of the parameters, the isolated-pole expressions therefore remain useful through the crossing of the imaginary parts,
and the real-momentum collision seen at $\mu_3=0$ is lifted.

\begin{figure}[htb]
	\begin{subfigure}{\textwidth}
		\centering
		\begin{subfigure}{0.45\textwidth}
			\centering
			\includegraphics[width=\textwidth]{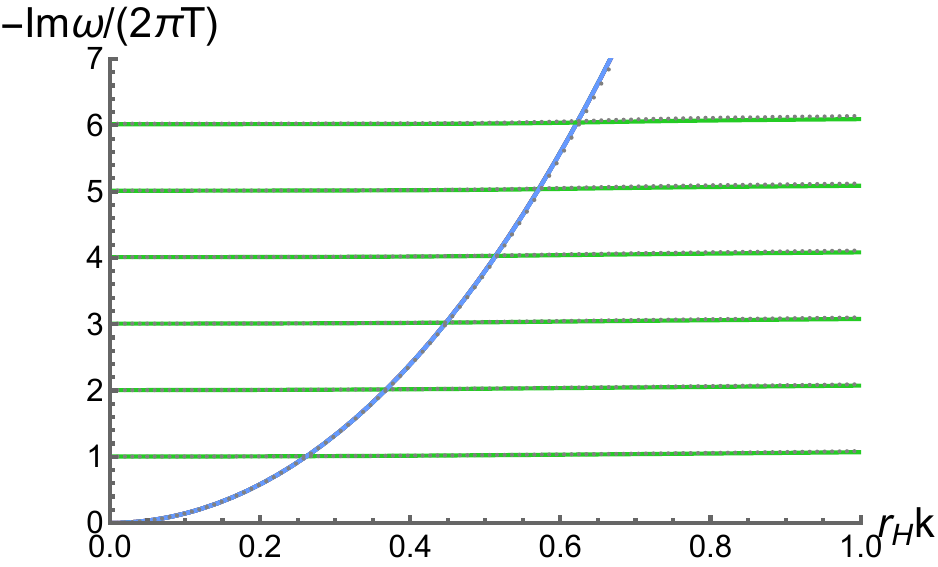}
		\end{subfigure}
		\hspace{1cm}
		\begin{subfigure}{0.45\textwidth}
			\centering
			\includegraphics[width=\textwidth]{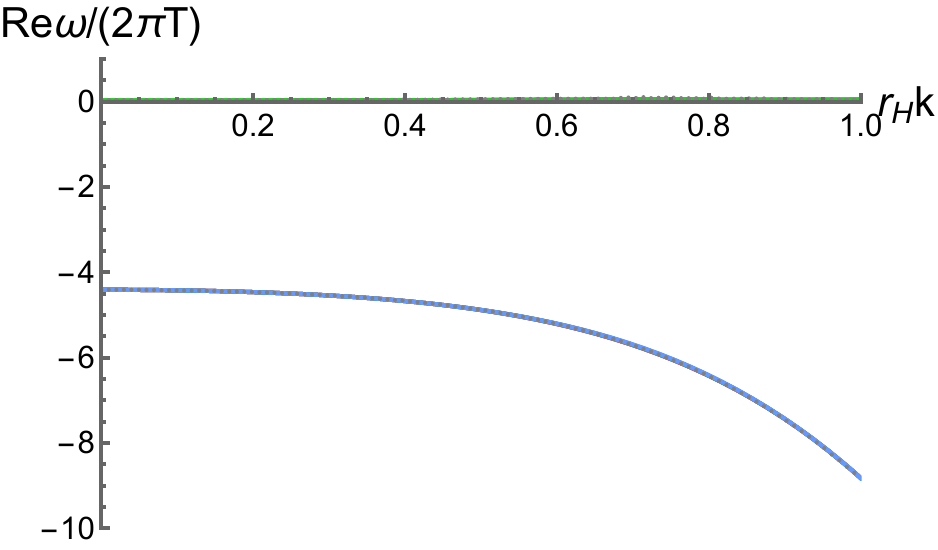}
		\end{subfigure}
	\end{subfigure}
	\caption{Imaginary (left panel) and real (right panel) part of the first longitudinal QNMs at $\mu_q/T=300$ and $\mu_3/\mu_q=1.2/13$. We compare the numerical results (grey squares) with the analytic formula for the interacting poles \eqref{int:roots}. The green solid lines correspond to one of the roots of \eqref{int:roots} (the one connected with the IR poles) for $n=1,\ldots,6$. The cyan solid lines correspond to the other root of \eqref{int:roots} for the same values of $n$ and they nearly coincide. }\label{fig:D}
\end{figure}

Figure \ref{fig:D} shows the comparison of the analytic formula \eqref{int:roots}  with the numerical results represented by the grey dots. Also for these plots the isospin chemical potential is set to $\mu_3/\mu_q =1.2/13$. In the plots, the green solid lines correspond to one of the roots of \eqref{int:roots} (the one connected with the IR poles) for $n=1,\ldots,6$. The cyan solid line corresponds to the other root of \eqref{int:roots} (the one connected with the hydro-like pole) for the same values of $n$. They correspond to six distinct values of the root for the first six values of $n$, and they nearly coincide on the scale shown. Also this plot confirms that for a background close to extremality as $\mu_q/T=300$ our analytic results continue to reproduce very well the numerical data 
at non-zero isospin chemical potential.

Taken together,
figures~\ref{fig:A}--\ref{fig:D} support the use of the local formula
to describe the first pole interaction and its lifting at finite
isospin, while also displaying its limitations for higher IR levels.

\subsection{Polarization function}\label{sec:numpol}
We now compare the exact numerical result for the imaginary part of the longitudinal polarization function \eqref{eq:long:matchedcorrelator} with the LO and NLO analytic expressions derived in section~\ref{sec:NLOlong}. For convenience, we report the two formulas here:
\begin{align}\label{num:LO}
	&\Pi_\text{LO}^{\parallel,\pm}(\omega,k)
	=
	\frac{(M\ell)^3}{8r_H^2}w_0^2N_c\,
	\dfrac{
		i\,\bm{\omega}\left(\bm{k}^2-(\bm{\omega}\pm\mathfrak{m})^2\right)
	}{
		(\bm{\omega}\pm\mathfrak{m})
		\left(\bm{\omega}\pm\mathfrak{m}+\frac{i}{2}\bm{k}^2\right)
	}\,,\\
	\label{num:BAdimensionful}
	&\Pi_\text{NLO}^{\parallel,\pm}(\omega,\bm{k})
	=
	\frac{(M\ell)^3}{8r_H^2}w_0^2N_c\,
	\left(1-\dfrac{(\bm{\omega}\pm\mathfrak{m})^2}{\bm{k}^2}\right)
	\frac{
		i\bm{k}^2\bm{\omega}-
		\left(
		\mathfrak N^\pm
		-i\,\bm{\omega}(\bm{\omega}\pm\mathfrak{m})^2
		\right)
	}{
		(\bm{\omega}\pm\mathfrak{m})^2
		\left[
		1+\dfrac{i\bm{k}^2}{2(\bm{\omega}\pm\mathfrak{m})}
		+\mathfrak D^\pm
		\right]
	}\,.
\end{align}
Here $\mathfrak N^\pm$ and $\mathfrak D^\pm$ are defined in \eqref{Npm} and \eqref{Dpm}, respectively.

In order to quantify the agreement with the numerical solution, we define the percentage relative difference
\begin{equation}\label{eq:percrelfdiff}
	\text{relative difference}
	=
	100\left|
	1-
	\dfrac{
		\operatorname{Im}\Pi^{\parallel,+}_{\text{LO,NLO}}
	}{
		\operatorname{Im}\Pi^{\parallel,+}_{\text{numerical}}
	}
	\right|\,.
\end{equation}
where $\operatorname{Im}\Pi^{\parallel,+}_{\text{numerical}}$ is the exact, numerically computed imaginary part of the longitudinal polarization function. Note that we have selected the \textit{plus charged} polarization functions for the plots.

In the following we present several density plots of the formula \eqref{eq:percrelfdiff} as function of $\omega/\mu$ and $k/\mu$ with both ranging from 0 to 1. This regime of parameters corresponds to the extended hydrodynamic regime \cite{Apostolidis:2026efq}. The yellow, red and dark green contours appearing in the following figures correspond, respectively, to relative differences of $1\%$, $5\%$ and $10\%$. They provide a convenient way of comparing the regions of quantitative validity of the LO and NLO approximations independently of the details of the color scale. We also show a black line corresponding to the locus $\omega=k+\mu_3$ for the charged sector considered in the plots. This curve is useful as a kinematic reference and reflects the shifted frequency appearing in the longitudinal correlator. It should not, however, be interpreted as a boundary of validity of the analytic expansion. As we shall see, the region of good agreement can extend on both sides of this line.

Following \cite{Apostolidis:2026efq}, we consider three different values of the ratio between the quark chemical potential and the temperature,
\begin{equation}
	\frac{\mu_q}{T}=\left\{10^4,\,65,\,5\right\}
\end{equation}
and study both vanishing isospin chemical potential and a representative non-zero value
\begin{equation}
	\frac{\mu_3}{\mu_q}=-0.1\,.
\end{equation}
The first two values of $\mu_q/T$ correspond to backgrounds close to extremality, whereas the case $\mu_q/T=5$ provides a useful test of the robustness of the analytic approximation when the hierarchy between the temperature and the chemical potential is substantially reduced.

Another point which is important when comparing the different values of $\mu_q/T$ is that the maximum value of the color scale is not the same in all figures. The LO and NLO panels belonging to a given figure (and hence the same value of $\mu_q/T$ and $\mu_3/\mu_q$) always use the same scale and can therefore be compared directly. Comparisons between different backgrounds, instead, should be based primarily on the numerical values of the contours and on the size of the low-error regions rather than on the color alone.

\subsubsection*{Zero isospin chemical potential}

We start from the isospin-symmetric case, $\mu_3=0$. This provides the cleanest setting in which to isolate the improvement generated by the NLO terms, since there is no additional frequency scale associated with isospin precession.

\begin{figure}[htb]
	\centering
	\begin{subfigure}{0.45\textwidth}
		\centering
		\caption*{LO}
		\includegraphics[width=\textwidth]{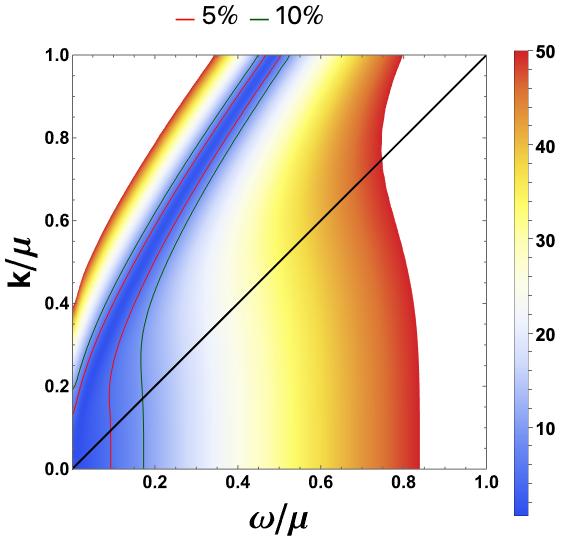}
	\end{subfigure}
	\hspace{1cm}
	\begin{subfigure}{0.45\textwidth}
		\centering
		\caption*{NLO}
		\includegraphics[width=\textwidth]{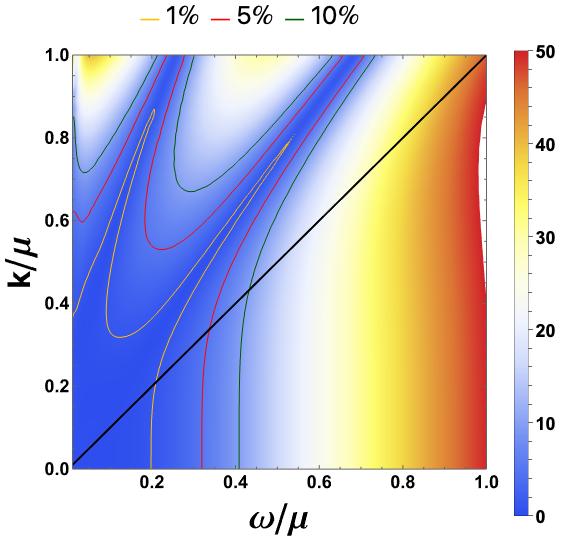}
	\end{subfigure}
	
	\caption{The percentage relative difference \eqref{eq:percrelfdiff} of the imaginary part of the longitudinal charged current polarization function with respect to the LO analytic result \eqref{num:LO} (left panel) and the NLO result \eqref{num:BAdimensionful} (right panel) for $\mu_q/T=10^4$ and $\mu_3=0$. The black line describes the locus $\omega=k$.}
	\label{fig:RelDiffNum_long_mu104mu30}
\end{figure}

Figure~\ref{fig:RelDiffNum_long_mu104mu30} shows our results for the most strongly near-extremal background considered here, $\mu_q/T=10^4$. The comparison between the two panels gives a particularly clear illustration of the effect of the NLO corrections. At leading order, the region with a small relative difference is concentrated in a comparatively narrow band extending from the origin towards larger values of $k/\mu$. Away from this band, the error grows rather rapidly, and a substantial portion of the $(\omega/\mu,k/\mu)$ plane lies outside the $10\%$ contour.

The NLO result modifies this pattern significantly. The low-error region becomes much broader, in particular at small and intermediate frequencies, and extends to substantially larger values of momentum. Correspondingly, both the $5\%$ and the $10\%$ contours are displaced away from the narrow region selected by the LO approximation. This is important because it shows that the improvement is not restricted to the immediate neighborhood of the diffusive regime at very small $k$: the NLO terms improve the approximation also at momenta which are no longer parametrically smaller than the frequency scales visible in the plot. Moreover, in the right panel of figure~\ref{fig:RelDiffNum_long_mu104mu30} we  have drawn also the $1\%$ contour which covers a considerable region of the extended hydrodynamic regime.

At the same time, figure~\ref{fig:RelDiffNum_long_mu104mu30} also makes clear the limitations of the truncated expansion. Towards the largest values of $\omega/\mu$, the relative difference increases strongly even in the NLO result. Thus, the main effect of the NLO correction should be characterized as a substantial enlargement of the region of quantitative agreement, rather than as a uniform improvement over the entire square $0\leq\omega/\mu,k/\mu\leq1$. In particular, the remaining discrepancy at large frequency is consistent with the fact that the analytic result is obtained as a low-frequency expansion in units of the chemical-potential scale.

\begin{figure}[htb]
	\begin{subfigure}{\textwidth}
		\centering
		\begin{subfigure}{0.45\textwidth}
			\centering
			\caption*{LO}
			\includegraphics[width=\textwidth]{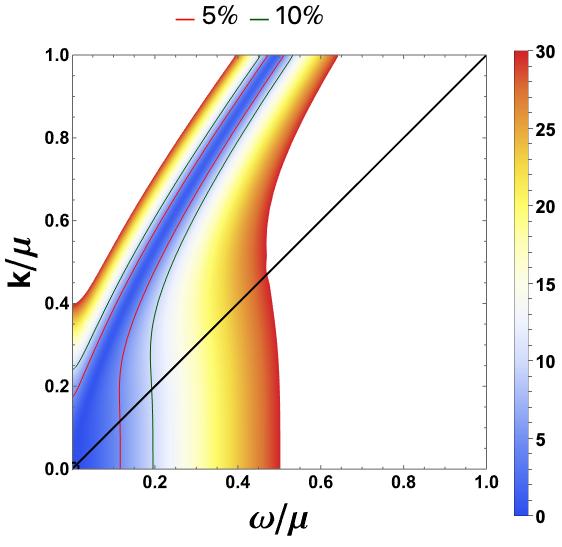}
		\end{subfigure}
		\hspace{1cm}
		\begin{subfigure}{0.45\textwidth}
			\centering
			\caption*{NLO}
			\includegraphics[width=\textwidth]{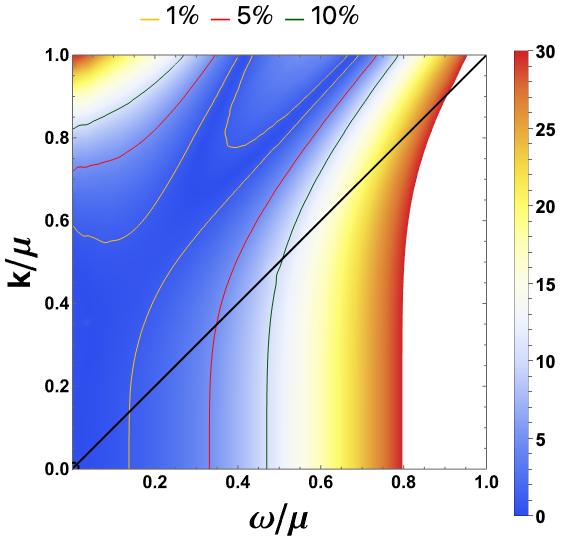}
		\end{subfigure}
	\end{subfigure}
	
	\caption{Same as figure \ref{fig:RelDiffNum_long_mu104mu30} but for $\mu_q/T=65$. The dashed square in bottom-left corner of the plots represents the standard hydrodynamic region $\omega/\mu,k/\mu\in[0,T/\mu]$.}
	\label{fig:RelDiffNum_long_mu65mu30}
\end{figure}

The same comparison is performed in figure~\ref{fig:RelDiffNum_long_mu65mu30} for $\mu_q/T=65$. The qualitative behavior is very similar to that found at $\mu_q/T=10^4$. At LO, the region of best agreement again follows a relatively narrow, momentum-dependent band. The NLO correction considerably enlarges this region and shifts the $5\%$ and $10\%$ contours towards larger frequencies over a broad range of momenta. Also for this value of $\mu_q/T$, the $1\%$ contour on the right panel encloses a large portion of the $\omega/\mu, k/\mu$ plane, especially for small frequencies.

One can also observe that, in the NLO panel, the error becomes less strongly dependent on momentum over a sizable portion of the plot. In particular, the low-error region has a more nearly vertical boundary than in the LO panel. This indicates that, once the leading momentum-dependent corrections have been incorporated, the residual discrepancy is increasingly controlled by the frequency expansion. We stress that this statement refers to the broad structure of the density plot and not to every individual contour feature.

The comparison between figures~\ref{fig:RelDiffNum_long_mu104mu30} and \ref{fig:RelDiffNum_long_mu65mu30} also shows that the improvement due to the NLO terms is not peculiar to an extremely small value of $T/\mu_q$. It remains clearly visible for the less extreme, but still strongly degenerate, background $\mu_q/T=65$.

\begin{figure}[htb]
	\begin{subfigure}{\textwidth}
		\centering
		\begin{subfigure}{0.45\textwidth}
			\centering
				\caption*{LO}
			\includegraphics[width=\textwidth]{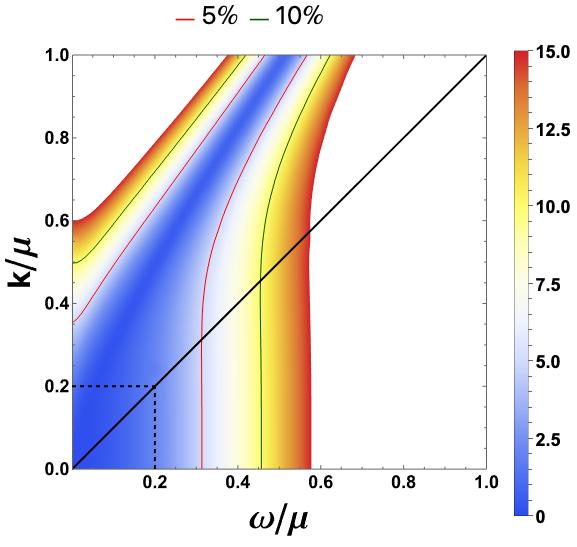}
		\end{subfigure}
		\hspace{1cm}
		\begin{subfigure}{0.45\textwidth}
			\centering
				\caption*{NLO}
			\includegraphics[width=\textwidth]{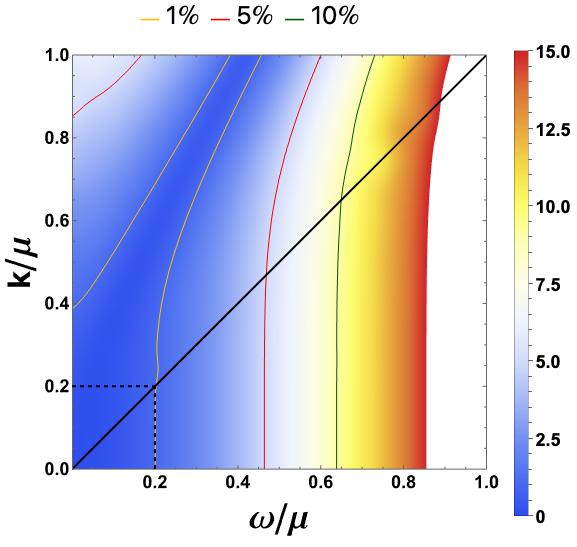}
		\end{subfigure}
	\end{subfigure}
	
	\caption{Same as figure \ref{fig:RelDiffNum_long_mu104mu30} but for $\mu_q/T=5$. The dashed square in bottom-left corner of the plots represents the standard hydrodynamic region $\omega/\mu,k/\mu\in[0,T/\mu]$}
	\label{fig:RelDiffNum_long_mu5mu30}
\end{figure}

Figure~\ref{fig:RelDiffNum_long_mu5mu30} corresponds to $\mu_q/T=5$. The overall relative differences are smaller than in the two more strongly near-extremal examples, as can also be seen from the smaller range of the color scale. Already at LO, a relatively large portion of the low- and intermediate-frequency region is described with an error of order a few percent. Nevertheless, the effect of the NLO terms remains clearly visible: the low-error region is enlarged further and the $5\%$ and $10\%$ contours are displaced towards larger $\omega/\mu$ for a broad interval of momenta. Also here, the comparison with the NLO analytic result is below $1\%$ error for large band of the extended hydrodynamic regime.

The right panel is also characterized by an approximately vertical organization of the relative difference over much of the plane. This again suggests that, after including the NLO momentum corrections, the dominant limitation of the approximation in this range is associated with increasing frequency rather than increasing momentum. The improvement is therefore particularly clear at moderate and large $k/\mu$ while $\omega/\mu$ remains sufficiently small.

It is interesting that the absolute relative differences in figure~\ref{fig:RelDiffNum_long_mu5mu30} are smaller than in the backgrounds with larger $\mu_q/T$. This observation should be interpreted with some care. It does not imply that the formal near-extremal expansion becomes parametrically better when $T/\mu_q$ is increased. Indeed, the analytic derivation assumes $T/\mu\ll1$. At fixed values of $\omega/\mu$ and $k/\mu$, however, increasing $T/\mu$ simultaneously decreases the corresponding ratios $\omega/T$ and $k/T$. The numerical trend therefore reflects the interplay between the several scales entering the problem and cannot be reduced to a monotonic statement about the near-extremal expansion parameter alone.

Taken together, figures~\ref{fig:RelDiffNum_long_mu104mu30}--\ref{fig:RelDiffNum_long_mu5mu30} establish two features which will remain important below. First, the NLO result provides a substantial improvement with respect to the LO approximation for all three backgrounds considered. Second, the improvement is especially pronounced in extending the small-error region towards larger momenta at small and intermediate frequencies.

\subsubsection*{Non-zero isospin chemical potential $\mu_3/\mu_q=-0.1$}

We now turn on a small isospin asymmetry and consider $\mu_3/\mu_q=-0.1$. The comparison between zero and nonzero $\mu_3$ is particularly useful because the isospin chemical potential enters the problem in two conceptually different ways. It contributes to the total chemical potential
\begin{equation}
	\mu=\sqrt{\mu_q^2+\mu_3^2}
\end{equation}
which determines the background geometry and the horizon radius, and it also enters explicitly in the charged fluctuation equations through the shifted frequency $\Omega_\pm$. The latter effect shifts the hydro-like pole away from the imaginary axis and introduces a genuine new dependence of the charged correlator on $\mu_3$. One of the questions relevant for the present work is whether these additional effects spoil the improvement obtained by including the NLO terms.

\begin{figure}[htb]
	\begin{subfigure}{\textwidth}
		\centering
		\begin{subfigure}{0.45\textwidth}
			\centering
			\caption*{LO}
			\includegraphics[width=\textwidth]{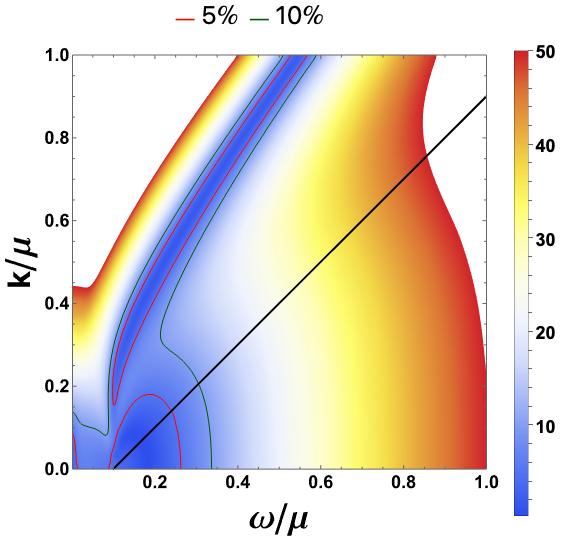}
		\end{subfigure}
		\hspace{1cm}
		\begin{subfigure}{0.45\textwidth}
			\centering
			\caption*{NLO}
			\includegraphics[width=\textwidth]{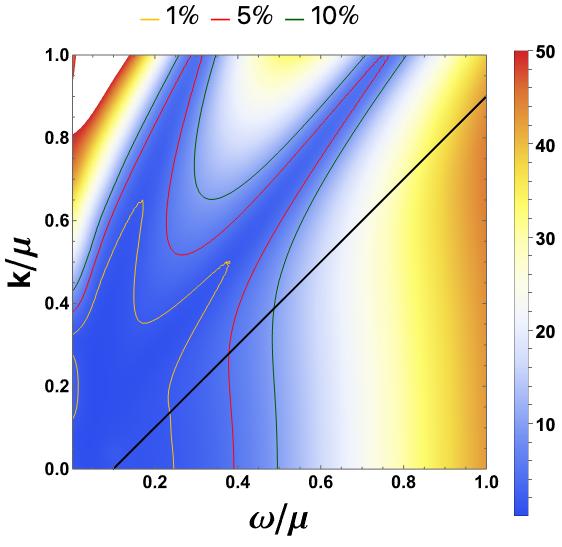}
		\end{subfigure}
	\end{subfigure}
	
	\caption{The percentage relative difference \eqref{eq:percrelfdiff} of the imaginary part of the longitudinal charged current polarization function with respect to the LO analytic result \eqref{num:LO} (left panel) and the NLO result \eqref{num:BAdimensionful} (right panel) for $\mu_q/T=10^4$ and $\mu_3/\mu_q=-0.1$. The black line describes the locus $\omega=k-\mu_3$.}
	\label{fig:RelDiffNum_long_mu104mu301}
\end{figure}

Figure~\ref{fig:RelDiffNum_long_mu104mu301} shows the result for the most strongly near-extremal background with $\mu_q/T=10^4$. Comparing first the LO and NLO panels at fixed $\mu_3$, we find the same main effect already observed at zero isospin chemical potential: the NLO approximation substantially enlarges the region in which the relative difference is small. The improvement is especially evident at low and intermediate frequencies, where the blue region extends over a significantly larger interval in momentum in the NLO panel.

The finite isospin chemical potential produces an additional deformation of the contours at low frequency, which is absent in figure~\ref{fig:RelDiffNum_long_mu104mu30}.
This is expected because the relevant frequency entering the charged equations is shifted by the background isospin gauge field. In particular, we recall that at finite $\mu_3$, the polarization function is strongly peaked at $\omega=\pm\mu_3$, $k=0$ due to the presence of the hydro-like pole. Importantly, the NLO approximation reproduces the numerical result well over a broad region despite this additional scale. The gain obtained by including the NLO corrections is therefore not tied to the isospin-symmetric limit.

The black line $\omega=k+\mu_3$ passes through regions with rather different relative errors and does not delimit the region in which the approximation is accurate. In particular, the NLO low-error domain extends broadly around and across this locus. This confirms that the improvement should be attributed to the additional terms retained in the near-extremal expansion rather than to an accidental agreement along a special kinematic curve.

As in the zero-isospin case, the agreement deteriorates towards the largest frequencies displayed. The NLO result thus considerably extends, but does not remove, the finite range of validity associated with the low-frequency expansion.

\begin{figure}[htb]
	\begin{subfigure}{\textwidth}
		\centering
		\begin{subfigure}{0.45\textwidth}
			\centering
				\caption*{LO}
			\includegraphics[width=\textwidth]{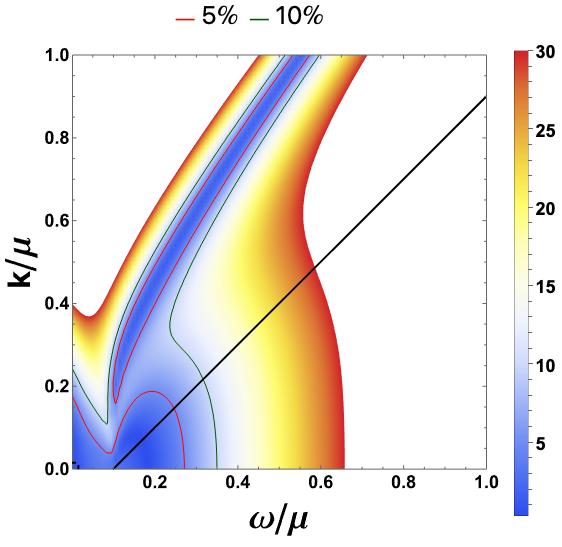}
		\end{subfigure}
		\hspace{1cm}
		\begin{subfigure}{0.45\textwidth}
			\centering
			\caption*{NLO}
			\includegraphics[width=\textwidth]{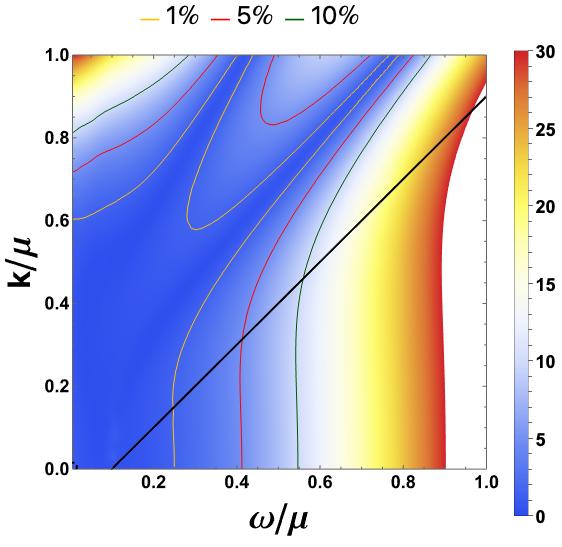}
		\end{subfigure}
	\end{subfigure}
	
	\caption{Same as figure \ref{fig:RelDiffNum_long_mu104mu301} but for $\mu_q/T=65$. The dashed square in bottom-left corner of the plots represents the standard hydrodynamic region $\omega/\mu,k/\mu\in[0,T/\mu]$}
	\label{fig:RelDiffNum_long_mu65mu301}
\end{figure}

The same pattern is visible in figure~\ref{fig:RelDiffNum_long_mu65mu301}, corresponding to $\mu_q/T=65$. At LO, the relative-difference contours show a non-trivial momentum dependence at small and intermediate $\omega/\mu$, including a curved low-error region associated with the finite isospin scale. After the NLO corrections are included, a broad region with small relative difference occupies most of the low- and intermediate-frequency part of the plot. The $1\%$ contour on the right panel enclose a large portion of the small frequency region.

The $5\%$ and $10\%$ contours are displaced towards larger frequencies for a substantial range of $k/\mu$. As at $\mu_3=0$, the residual NLO error is comparatively weakly dependent on momentum over much of this region and grows more clearly as one moves towards the right-hand side of the plot. This provides further evidence that the NLO momentum corrections are capturing an important part of the deviation of the LO result from the exact correlator.

Comparing figures~\ref{fig:RelDiffNum_long_mu65mu30} and \ref{fig:RelDiffNum_long_mu65mu301}, the introduction of $\mu_3/\mu_q=-0.1$ visibly reshapes the contours, especially at low frequency. Nevertheless, it does not spoil the NLO approximation on the scale resolved by these plots. The dominant observation remains the substantial enlargement of the region with a relative difference below $5\%$--$10\%$ when passing from LO to NLO.

\begin{figure}[htb]
	\begin{subfigure}{\textwidth}
		\centering
		\begin{subfigure}{0.45\textwidth}
			\centering
				\caption*{LO}
			\includegraphics[width=\textwidth]{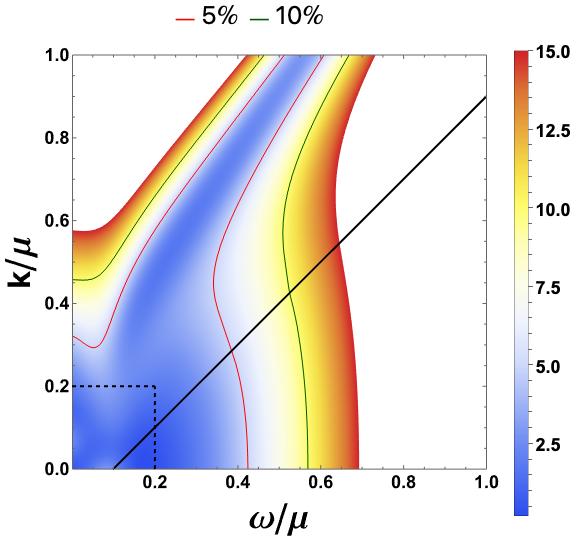}
		\end{subfigure}
		\hspace{1cm}
		\begin{subfigure}{0.45\textwidth}
			\centering
				\caption*{NLO}
			\includegraphics[width=\textwidth]{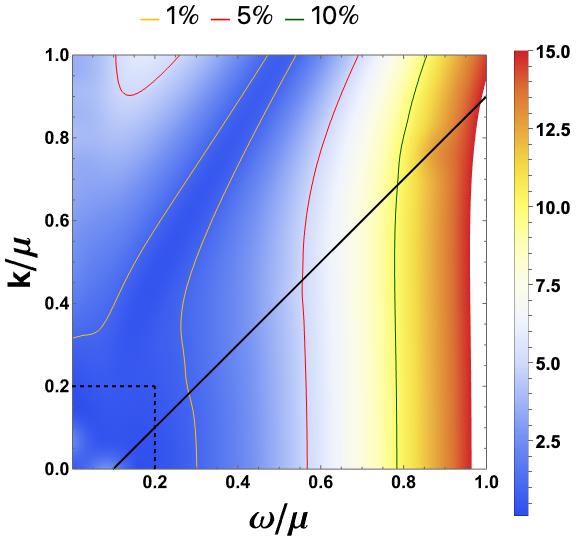}
		\end{subfigure}
	\end{subfigure}
	
	\caption{Same as figure \ref{fig:RelDiffNum_long_mu104mu301} but for $\mu_q/T=5$. The dashed square in bottom-left corner of the plots represents the standard hydrodynamic region $\omega/\mu,k/\mu\in[0,T/\mu]$}
	\label{fig:RelDiffNum_long_mu5mu301}
\end{figure}

Finally, figure~\ref{fig:RelDiffNum_long_mu5mu301} presents the finite-isospin result at $\mu_q/T=5$. As in the corresponding zero-isospin plot, figure~\ref{fig:RelDiffNum_long_mu5mu30}, the overall relative differences are smaller than in the more strongly near-extremal backgrounds. The LO result already describes a significant part of the low-frequency domain with an error below $10\%$, but the NLO correction further extends this region towards larger $\omega/\mu$.

The displacement of the $5\%$ and $10\%$ contours is particularly clear over intermediate and large momenta. At the same time, some deformation of the contours remains visible at large $k/\mu$ and small frequency, reflecting the presence of the additional isospin scale. Therefore, although the NLO result improves the agreement over most of the displayed low-frequency region, the improvement should not be regarded as strictly pointwise or uniform.

It is useful to summarize our findings and compare between the $\mu_3=0$ and
$\mu_3/\mu_q=-0.1$ results:
\begin{itemize}
	\item \textbf{Effect of the isospin chemical potential.}
	At fixed $\mu_q/T$, introducing the finite isospin chemical
	potential considered here modifies the relative-difference
	contours, most visibly at low frequencies. This is consistent
	with the isospin dependence of both the hydro-like pole and its
	NLO correction. Nevertheless,
	figures~\ref{fig:RelDiffNum_long_mu104mu301}--\ref{fig:RelDiffNum_long_mu5mu301}
	show that the improvement from LO to NLO persists at non-zero $\mu_3$: in all three
	backgrounds, the NLO result retains a substantially larger
	low-error region than the LO result.
	
	\item \textbf{Dependence on $\mu_q/T$.}
	At both zero and finite isospin chemical potential, the
	magnitudes of the relative differences decrease as $\mu_q/T$
	is reduced from $10^4$ to $5$ over the displayed parameter
	range. As discussed above, this should not be interpreted as
	evidence that the formal near-extremal expansion becomes more
	accurate as $T/\mu$ increases. Rather, the comparison shows
	that the NLO expression remains numerically useful in all
	three backgrounds, including $\mu_q/T=5$, where the hierarchy
	of scales is less pronounced.
	
	\item \textbf{Extent and limitations of the NLO improvement.}
	For all backgrounds considered, at both zero and finite
	isospin, the NLO correction considerably enlarges the region
	in the $(\omega/\mu,k/\mu)$ plane where the imaginary part of
	the longitudinal polarization function agrees with the
	numerical result within a few percent. The improvement is
	particularly pronounced at small and intermediate frequencies
	and extends to larger momenta. It is not uniform across the
	displayed domain: at sufficiently large $\omega/\mu$, the
	discrepancy increases again, consistently with the limitations
	of the truncated low-energy expansion.
\end{itemize}
These comparisons establish a quantitative improvement of the NLO result over the
leading approximation in the parameter ranges examined, without
implying uniform validity outside the near-extremal, low-energy
regime.

\section*{Acknowledgments}

We want to thank B. Gouteraux, C. Hoyos and C.Rosen for very helpful discussions.

This work was partially supported by  the H.F.R.I. call ``Basic research Financing" (Horizontal support of all Sciences)
under the National Recovery and Resilience Plan ``Greece 2.0" funded by the European Union -NextGenerationEU
(H.F.R.I. Project Number: 15384), by the In2p3 grant ``Extreme Dynamics", the ANR grant ``XtremeHolo"
(ANR project n.284452), by the H.F.R.I. Project Number: 23770  of the H.F.R.I call
``3rd Call for H.F.R.I.'s Research Projects to Support Faculty Members \& Researchers", the ERC starting grant 101078061 SINGinGR, under the European Union's Horizon Europe program for research and innovation"
and the UoC grant number 12030.

\appendix

\section{Details on the matching calculation}\label{app:details}
In this section, we provide a detailed analysis of the matching calculation. In particular we present the techniques used to solve perturbatively the fluctuation equations of motion in the inner region (section \ref{app:inner}) and in the outer region (section \ref{app:outer}). Then, in section \ref{app:expansions}, we present the series expansion of the solutions to the equations of motion in the overlapping region.

\subsection{Inner region}\label{app:inner}

In this  section, we provide a detailed derivation of the solution to the fluctuation equation in the inner region. We recall that the general solution of the equation of motion \eqref{eq:EOM} in the inner region can be written as a leading term and a next-to-leading contribution which is of order $\epsilon$ in the expansion
\begin{equation}
	\varphi^\pm_I(\zeta) = \varphi^{\pm,(0)}_{I}(\zeta) + \epsilon\,\boldsymbol{k}^2\varphi^{\pm,(1)}_{I,\boldsymbol{k}}(\zeta)\pm \epsilon\,\mathfrak{m}\,\varphi^{\pm,(1)}_{I,\,\mathfrak{m}}(\zeta)+\epsilon\,\tau\,\varphi^{\pm,(1)}_{I,\tau}(\zeta) +\mathcal{O}(\epsilon^2)\,.
\end{equation}

The leading order solution is the solution of
\begin{equation}\label{app:L0eq}
	\mathcal{L}_0\left[\varphi^{\pm,(0)}_{I}\right] =0\,,
\end{equation}
where we recall that
\begin{equation}\label{app:eq:L0}
	\mathcal{L}_0[\chi]\equiv  \partial_\zeta\left(h(\zeta)\partial_\zeta \chi(\zeta)\right) +\dfrac{\hat\omega^2}{4h(\zeta)}\chi(\zeta)\,.
\end{equation}

The ingoing leading order solution normalized to approach one at the boundary $\zeta\to \infty$ is
\begin{equation}\label{app:eq:INNleadinglon}
	\varphi^{\pm,(0)}_{I}= \left(\dfrac{3\zeta}{1+3\zeta}\right)^{-i\,\hat\omega/2}\,.
\end{equation}

The next-to-leading equation of motion can be written in a compact way as follows
\begin{equation}
	\mathcal{L}_0\left[\boldsymbol{k}^2\varphi_{I,\boldsymbol{k}}^{\pm,(1)}\pm \mathfrak{m}\,\varphi_{I,\,\mathfrak{m}}^{\pm,(1)}+\tau\,\varphi_{I,\tau}^{\pm,(1)}\right] = \boldsymbol{k}^2\,S_{\boldsymbol{k}}\left[\varphi_I^{\pm,(0)}\right]\pm\mathfrak{m}\,S_\mathfrak{m}\left[\varphi_I^{\pm,(0)}\right]+\tau\,S_\tau\left[\varphi_I^{\pm,(0)}\right].
\end{equation}
where
\begin{align}
	&S_{\boldsymbol{k}}\left[\varphi_I^{\pm,(0)}\right] = \dfrac{1}{4}\varphi_{I,(0)}(\zeta)-\dfrac{h(\zeta)(1+6\zeta)}{\hat\omega^2}\partial_\zeta\varphi_I^{\pm,(0)}(\zeta)\,,\\
	&S_\mathfrak{m}\left[\varphi_I^{\pm,(0)}\right] = -\dfrac{\hat\omega}{2(1+3\zeta)}\varphi_I^{\pm,(0)}(\zeta)\,,\\
	&S_\tau\left[\varphi_I^{\pm,(0)}\right] = \dfrac{g(\zeta)}{8h(\zeta)^2}\varphi_I^{\pm,(0)}(\zeta)+\dfrac{\zeta^2(7+8\zeta)(2+3\zeta)}{2h(\zeta)}\partial_\zeta\varphi_I^{\pm,(0)}(\zeta)\,,
\end{align}
are the source terms.

We use the linearity of the differential operator $\mathcal{L}_0$, and solve three independent equations of motion, since the parameters $\boldsymbol{k}^2,\mathfrak{m},\tau$ are independent:
\begin{align}
	&\mathcal{L}_0\left[\varphi_{I,\boldsymbol{k}}^{\pm,(1)}\right] = S_{\boldsymbol{k}}\left[\varphi_I^{\pm,(0)}\right]\,,& &\mathcal{L}_0\left[\varphi_{I,\,\mathfrak{m}}^{\pm,(1)}\right] = S_\mathfrak{m}\left[\varphi_I^{\pm,(0)}\right]\,,& &\mathcal{L}_0\left[\varphi_{I,\tau}^{\pm,(1)}\right] = S_\tau\left[\varphi_I^{\pm,(0)}\right]\,.&
\end{align}

In the following, we  report the procedure to solve the three differential equations separately.

\subsubsection*{NLO momentum correction}
Let us write the generic solution as
\begin{equation}
	\varphi_{I,\boldsymbol{k}}^{\pm,(1)}(\zeta) = \varphi_{I}^{\pm,(0)}(\zeta)F_{\boldsymbol{k}}^\pm(\zeta)\,.
\end{equation}
Using $\mathcal{L}_0\left[\varphi_{I}^{\pm,(0)}\right] =0$, and
\begin{equation}
	\partial_\zeta\varphi_{I}^{\pm,(0)}(\zeta) = -i\,\hat\omega\dfrac{\varphi_{I}^{\pm,(0)}(\zeta)}{2h(\zeta)}\,,
\end{equation}
we can rewrite the equation of motion as
\begin{equation}\label{eq:eomFK}
	\partial_\zeta(h(\zeta)\varphi_{I}^{\pm,(0)}(\zeta)^2\partial_\zeta F_{\boldsymbol{k}}^\pm(\zeta))-\left(\dfrac{1}{4}+\dfrac{i(1+6\zeta)}{2\hat\omega}\right)\varphi_{I}^{\pm,(0)}(\zeta)^2=0\,.
\end{equation}

Equation \eqref{eq:eomFK} admits the following integral solution, implementing  the ingoing condition and the fixed normalization
\begin{equation}
	F_{\boldsymbol{k}}^\pm(\zeta)= \int^\zeta_0\dfrac{d\eta}{h(\eta)\varphi_{I}^{\pm,(0)}(\eta)^2}\int^\eta_0d\rho\,\left(\dfrac{1}{4}+\dfrac{i(1+6\rho)}{2\hat\omega}\right)\varphi_{I}^{\pm,(0)}(\rho)^2\,,
\end{equation}
We have fixed the two integration constants such that the ingoing condition is not spoiled, and the normalization at the horizon to be the one set by the leading solution. In formulae,
\begin{align}
	&\underset{\zeta\to 0}{\lim}h(\zeta)\varphi_{I}^{\pm,(0)}(\zeta)^2\partial_\zeta F_{\boldsymbol{k}}^\pm=0\,,&  &\underset{\zeta\to 0}{\lim} F_{\boldsymbol{k}}^\pm=0\,.&
\end{align}
The first condition removes the outgoing homogeneous mode: the second prevents an additional multiple of $\varphi_{I}^{\pm,(0)}$, fixing the horizon normalization.

The complete solution reads
\begin{equation}
	\varphi_{I,\boldsymbol{k}}^{\pm,(1)}(\zeta)= \varphi_{I}^{\pm,(0)}(\zeta)\int^\zeta_0\dfrac{d\eta}{h(\eta)\varphi_{I}^{\pm,(0)}(\eta)^2}\int^\eta_0d\rho\,\left(\dfrac{1}{4}+\dfrac{i(1+6\rho)}{2\hat\omega}\right)\varphi_{I}^{\pm,(0)}(\rho)^2\,.
\end{equation}
We can perform the integral analytically and we obtain
\begin{align}
	\varphi_{I,\boldsymbol{k}}^{\pm,(1)}= \left(\frac{3\zeta}{1+3\zeta}\right)^{-i\hat\omega/2}\dfrac{i\,\zeta}{4\hat\omega}\left(2+\hat\omega^2\dfrac{\Gamma(-i\,\hat\omega)}{\Gamma(2-i\,\hat\omega)}{}_2F_1(1,1,2-i\,\hat\omega;-3\zeta)\right)\,.
\end{align}

\subsubsection*{NLO isospin correction}
We repeat the procedure carried out for the momentum correction. We set
\begin{equation}
	\varphi_{I,\,\mathfrak{m}}^{\pm,(1)}(\zeta) = \varphi_{I}^{\pm,(0)}(\zeta)F_\mathfrak{m}^\pm(\zeta)\,.
\end{equation}
Using $\mathcal{L}_0\left[\varphi_{I}^{\pm,(0)}\right] =0$, we can rewrite the equation of motion as
\begin{equation}
	\partial_\zeta(h(\zeta)\varphi_I^{\pm,(0)}(\zeta)^2\partial_\zeta F_\mathfrak{m}^\pm(\zeta))+\dfrac{\hat\omega}{2(1+3\zeta)}\varphi_I^{\pm,(0)}(\zeta)^2=0\,.
\end{equation}
The ingoing and fixed normalization conditions select the following integral solution,
\begin{equation}
	F_\mathfrak{m}^\pm(\zeta)= -\dfrac{\hat\omega}{2}\int^\zeta_0\dfrac{d\eta}{h(\eta)\varphi_I^{\pm,(0)}(\eta)^2}\int^\eta_0d\rho\,\dfrac{\varphi_I^{\pm,(0)}(\rho)^2}{1+3\rho}\,.
\end{equation}
The two integration constants of the general solution are automatically set to zero by writing the solution in the above form. The first one ensures that the ingoing condition is not spoiled, the second fixes the normalization at the horizon to be the one set by the leading solution. As before, this is equivalent to
\begin{align}
	&\underset{\zeta\to 0}{\lim}h(\zeta)\varphi_{I}^{\pm,(0)}(\zeta)^2\partial_\zeta F_\mathfrak{m}^\pm=0\,,& &\underset{\zeta\to 0}{\lim} F_\mathfrak{m}^\pm=0\,.&
\end{align}

Hence the total solution can be written as
\begin{equation}
	\varphi_{I,\,\mathfrak{m}}^{\pm,(1)}(\zeta)= -\dfrac{\hat\omega}{2}\varphi_I^{\pm,(0)}(\zeta)\int^\zeta_0\dfrac{d\eta}{h(\eta)\varphi_I^{\pm,(0)}(\eta)^2}\int^\eta_0d\rho\,\dfrac{\varphi_I^{\pm,(0)}(\rho)^2}{1+3\rho}\,.
	\label{a1}\end{equation}

After performing the first integral in (\ref{a1}), we obtain
\begin{equation}
	-\dfrac{\hat\omega}{2}\int^\zeta_0\dfrac{d\eta}{h(\eta)\varphi_I^{\pm,(0)}(\eta)^2}\int^\eta_0d\rho\,\dfrac{\varphi_I^{\pm,(0)}(\rho)^2}{1+3\rho} = -\frac{\hat\omega}{2(1-i \hat\omega  )}\int^\zeta_0d\eta\frac{\, _2F_1(1,1;2-i \hat\omega ;-3 \eta)}{1+3 \eta}\,.
\end{equation}
We then perform the following change of coordinates
\begin{equation}\label{eq:coordtransf}
	x = \dfrac{3\eta}{1+3\eta}
\end{equation}
and rewrite the integral as
\begin{align}\nonumber
	& F_\mathfrak{m}^\pm \equiv  -\dfrac{\hat\omega}{2}\int^\zeta_0\dfrac{d\eta}{h(\eta)\varphi_I^{\pm,(0)}(\eta)^2}\int^\eta_0d\rho\,\dfrac{\varphi_I^{\pm,(0)}(\rho)^2}{1+3\rho}  = \\ \nonumber
	&= -\frac{\hat\omega}{2(1-i \hat\omega  )}\int^\zeta_0d\eta\frac{\, _2F_1(1,1;2-i \hat\omega ;-3 \eta)}{1+3 \eta}=\\ \nonumber
	&=-\frac{\hat\omega}{6(1-i \hat\omega  )}\int^{x_\zeta}_0dx\frac{\, _2F_1(1,1;2-i \hat\omega ;-\frac{x}{1-x})}{1-x} =\\
	& -\frac{\hat\omega}{6(1-i \hat\omega  )}\int^{x_\zeta}_0dx\,_2F\, _1(1,1-i\hat\omega;2-i \hat\omega ;x)
\end{align}
with $x_\zeta = 3\zeta/(1+3\zeta)$ and where in the last equality we used the Pfaff's transformation
\begin{equation}\label{eq:Pfaff}
	\, _2F_1\left(1,1;2-i \hat\omega ;-\frac{x}{1-x}\right) = (1-x)\,_2F_1(1,1-i\hat\omega;2-i \hat\omega ;x)
\end{equation}

Then, the integral can be performed and we finally obtain
\begin{align}\nonumber
	\varphi_{I,\,\mathfrak{m}}^{\pm,(1)}(\zeta)
	=
	&\frac{i}{6} \left(\frac{3 \zeta }{3 \zeta +1}\right)^{-i\hat\omega/2 } \bigg(\frac{3\zeta}{(1+3\zeta)^{i\hat\omega}}\frac{1}{1-i\hat{\omega}}\,
	{}_2F_1\!\left(
	1-i\hat{\omega},\,1-i\hat{\omega};
	\,2-i\hat{\omega};
	\,-3\zeta
	\right)\\
	&-\log \left(1+3\zeta\right)\bigg)\,.
\end{align}

\subsubsection*{NLO temperature correction}
As for the momentum and isospin corrections, we set
\begin{equation}
	\varphi_{I,\tau}^{\pm,(1)}(\zeta) =\varphi_I^{\pm,(0)}(\zeta)F_\tau^\pm(\zeta)\,.
\end{equation}
Using $\mathcal{L}_0\left[\varphi_I^{\pm,(0)}\right] =0$, and
\begin{equation}
	\partial_\zeta\varphi_I^{\pm,(0)}(\zeta) = -i\,\hat\omega\dfrac{\varphi_I^{\pm,(0)}(\zeta)}{2h(\zeta)}\,,
\end{equation}
we can rewrite the equation of motion as
\begin{equation}
	\partial_\zeta(h(\zeta)\varphi_I^{\pm,(0)}(\zeta)^2\partial_\zeta F_\tau^\pm(\zeta))+\dfrac{\hat\omega^2(9+14\zeta)+i\,\hat\omega(7+16\zeta+24\zeta^2)}{4(1+3\zeta)^2}\varphi_I^{\pm,(0)}(\zeta)^2=0\,.
\end{equation}

The ingoing and fixed normalization select the following integral solution
\begin{equation}
	F_\tau^\pm(\zeta)= -\int^\zeta_0\dfrac{d\eta}{h(\eta)\varphi_I^{\pm,(0)}(\eta)^2}\int^\eta_0d\rho\,\dfrac{\hat\omega^2(9+14\rho)+i\,\hat\omega(7+16\rho+24\rho^2)}{4(1+3\rho)^2}\varphi_I^{\pm,(0)}(\rho)^2\,.
\end{equation}
Note that we set to zero the two integration constants. The first one ensures that the ingoing condition is not spoiled, the second fixes the normalization at the horizon to be the one set by the leading solution. Hence
\begin{align}\nonumber
	\varphi_{I,\tau}^{\pm,(1)}(\zeta)=&- \varphi_I^{\pm,(0)}(\zeta)\int^\zeta_0\dfrac{d\eta}{h(\eta)\varphi_I^{\pm,(0)}(\eta)^2}\times\\
	&\times\int^\eta_0d\rho\,\dfrac{\hat\omega^2(9+14\rho)+i\,\hat\omega(7+16\rho+24\rho^2)}{4(1+3\rho)^2}\varphi_I^{\pm,(0)}(\rho)^2\,.
\end{align}

We can perform a similar procedure as the one we carried out for the NLO isospin correction. We change coordinate as in \eqref{eq:coordtransf} and use the Pfaff's transformation \eqref{eq:Pfaff}. We finally obtain
\begin{align}\nonumber
	\varphi_{I,\tau}^{\pm,(1)}(\zeta)= &\dfrac{i\,\hat\omega}{6}\left(\frac{3\zeta}{1+3\zeta}\right)^{-i\hat\omega/2}\bigg[\frac{3\zeta}{(1+3\zeta)^{i\hat\omega}}\frac{1}{1-i\hat{\omega}}\,
	{}_2F_1\!\left(
	1-i\hat{\omega},\,1-i\hat{\omega};
	\,2-i\hat{\omega};
	\,-3\zeta
	\right)+\\
	&-\dfrac{13}{2}\frac{\zeta}{1+3\zeta}-\dfrac{7}{3}\log(1+3\zeta)\bigg]\,,
\end{align}
which satisfies the required boundary conditions.

\subsection{Outer region}\label{app:outer}
In the outer region, we write the general solution as a series in $\epsilon$
\begin{equation}
	\varphi_O^\pm
	=
	\varphi_O^{\pm,(0)}
	+\epsilon\,\varphi_O^{\pm,(1)}+\epsilon^2\,\varphi_O^{\pm,(2)}
	+\mathcal{O}(\epsilon^3)\,,
\end{equation}

We recall that the outer LO and NLO solutions are written in the main text in \eqref{eq:long:LOout} and \eqref{NLOouterexplicit} respectively. It is useful to discuss the integral representation of the NLO solution. The NLO equation has the generic form
\begin{equation}
	\partial_u
	\left[
	\dfrac{\boldsymbol{\Omega}_\pm(u)^2}{u}
	\partial_u\varphi_O^{\pm,(1)}(u)
	\right]
	=
	\mathcal{S}^{\pm,(1)}(u)\,,
\end{equation}
where $	\mathcal{S}^{\pm,(1)}$ is defined in \eqref{eq:S1pm}. Choosing an arbitrary reference point $u_\star$ in a regular part of the outer region, the general solution can be written as
\begin{align}
	\varphi_O^{\pm,(1)}(u)
	={}&
	A_\star^\pm
	+
	B_\star^\pm
	\int_{u_\star}^{u}
	dx\,
	\dfrac{x}{\boldsymbol{\Omega}_\pm(x)^2}+
	\int_{u_\star}^{u}
	dx\,
	\dfrac{x}{\boldsymbol{\Omega}_\pm(x)^2}
	\int_{u_\star}^{x}
	dy\,
	\mathcal{S}^{\pm,(1)}(y)\,.
	\label{eq1}
\end{align}
Here $A_\star^\pm$ and $B_\star^\pm$ are the two independent integration constants. The choice of $u_\star$ has no physical significance, since changing the reference point amounts to a redefinition of these constants.
It is tempting to set $u_\star=0$ directly. However, the source $\mathcal{S}^{\pm,(1)}$ is singular at the asymptotic boundary:
\begin{equation}
	\mathcal{S}^{\pm,(1)}(u)
	=
	\dfrac{\alpha_\pm}{u}
	+
	\mathcal{O}(u)\,,
	\qquad
	\alpha_\pm^{(1)}
	\equiv
	\boldsymbol{k}^2a_O^{\pm,(0)}
	(\boldsymbol{\omega}\pm\mathfrak{m})^2\,.
	\label{eq2}
\end{equation}
Therefore, a naive attempt to integrate,
\begin{equation}
	\int_0^xdy\,\mathcal{S}^{\pm,(1)}(y) \,,
\end{equation}
leads to a logarithmic divergence. This divergence is the origin of the usual logarithmic term in the near-boundary expansion of a gauge-field fluctuation in AdS$_5$.
In order to use the asymptotic boundary as the integration point, we first subtract the singular part of the source and define
\begin{equation}
	\widehat{\mathcal{S}}^{\pm,(1)}(u)
	\equiv
	\mathcal{S}^{\pm,(1)}(u)
	-
	\dfrac{\alpha_\pm^{(1)}}{u}\,.
\end{equation}
Then close to the boundary, by construction, $\widehat{\mathcal{S}}^{\pm,(1)}(u)
\sim
\mathcal{O}(u)$ and its integral from the boundary is therefore finite. After the first integration, the solution can consequently be written as
\begin{align}\nonumber
	\varphi_O^{\pm,(1)}(u)
	={}&
	a_O^{\pm,(1)}
	+
	b_O^{\pm,(1)}
	\dfrac{u^2}
	{2(\boldsymbol{\omega}\pm\mathfrak{m})\boldsymbol{\Omega}_\pm(u)}+\\ \nonumber
	&	\int_0^u
	dx\,
	\dfrac{x}{\boldsymbol{\Omega}_\pm(x)^2}
	\Bigg[
	\alpha_\pm^{(1)}\,\log x
	+
	\beta_\pm^{(1)}+\\
	&+
	\int_0^x
	dy\,
	\left(
	\mathcal{S}^{\pm,(1)}(y)
	-
	\dfrac{
		\boldsymbol{k}^2(\boldsymbol{\omega}\pm\mathfrak{m})^2a_O^{\pm,(0)}
	}{y}
	\right)
	\Bigg]\,,
	\label{eq3}
\end{align}
where $\beta_\pm^{(1)}$ is an arbitrary finite integration constant of the integration that we performed.
All the integrals appearing in \eqref{eq3} are now convergent at $u=0$. Moreover, since the outermost integrand behaves at most as $x\log x$, its contribution vanishes as $u\to0$. Therefore,
\begin{equation}
	\lim_{u\to0}\varphi_O^{\pm,(1)}(u)
	=
	a_O^{\pm,(1)}\,,
\end{equation}
as required by our convention for the source.

It is important to stress that the boundary condition above, does not by itself fix the finite constant $\beta_\pm^{(1)}$. The value of $\beta_\pm^{(1)}$ only specifies the convention used to separate the homogeneous and particular pieces of the NLO solution. It is convenient to choose
\begin{equation}\label{eq4}
	\beta_\pm^{(1)}
	=
	\dfrac{
		a_O^{\pm,(0)}\boldsymbol{k}^2\boldsymbol{\omega}^2
	}{6}
	+
	\dfrac{
		b_O^{\pm,(0)}\boldsymbol{k}^2\boldsymbol{\omega}
	}{
		12(\boldsymbol{\omega}\pm\mathfrak{m})
	}
	-
	\dfrac{
		b_O^{\pm,(0)}
		(\boldsymbol{\omega}\pm\mathfrak{m})^2
	}{\boldsymbol{k}^2}\,.
\end{equation}
With this choice, \eqref{eq3} reproduces exactly the explicit solution \eqref{NLOouterexplicit}, with the same definitions of $a_O^{\pm,(1)}$ and $b_O^{\pm,(1)}$.
Finally, the term proportional to $\alpha_\pm^{(1)}\log x$
immediately reproduces the expected UV logarithm. Indeed, expanding the solution close to the boundary gives
\begin{equation}
	\varphi_O^{\pm,(1)}(u)
	=
	a_O^{\pm,(1)}
	+
	\dfrac{\boldsymbol{k}^2a_O^{\pm,(0)}}{2}
	u^2\log u
	+
	\mathcal{O}(u^2)\,,
\end{equation}
in agreement with the general near-boundary expansion of the longitudinal fluctuation.

Then, we provide the full expression for the NNLO solution in the outer region. In the main text, we wrote the solution as
\begin{align}
	\varphi^{\pm,(2)}_O(u)
	={}&
	a^{\pm,(2)}_O
	+
	b^{\pm,(2)}_O
	\frac{u^2}{2(\boldsymbol{\omega}\pm\mathfrak{m})\boldsymbol{\Omega}_\pm(u)}+
	\nonumber\\
	&+
	a^{\pm,(1)}_O\,\mathcal F_{a_1}^\pm(u)
	+
	b^{\pm,(1)}_O\,\mathcal F_{b_1}^\pm(u)+
	b^{\pm,(0)}_O\,\mathcal F_{b_0}^\pm(u)
	+
	a^{\pm,(0)}_O\,\mathcal F_{a_0}^\pm(u)\,.
	\label{app:eq:NNLOoutersol}
\end{align}
Here, we provide the explicit expressions for $\mathcal F_{a_1}^\pm\,,\mathcal F_{b_1}^\pm\,,\mathcal F_{b_0}^\pm\,,\mathcal F_{a_0}^\pm$ which read
\begin{align}
	\mathcal F_{a_1}^\pm(u) = &- \frac{\boldsymbol{k}^{2}}{36 (\boldsymbol{\omega}\pm\mathfrak{m}) \boldsymbol{\Omega}_{\pm}(u)}\Bigg\{6 \mathfrak{m}^{2} u^{2} \log{\left(2 u^{2} + 1 \right)} - 9 (\boldsymbol{\omega}\pm\mathfrak{m})^{2} u^{2} \log{\left(u^{2} \right)}+
	\nonumber\\ &- 3 \mathfrak{m}^{2} \log{\left(u^{2} - 1 \right)}+ 3 \mathfrak{m}^{2} \log{\left(2 u^{2} + 1 \right)}+ 3 i \pi \mathfrak{m}^{2}\pm 2 \mathfrak{m} u^{2} \boldsymbol{\omega} \log{\left(u^{2} - 1 \right)}
	\nonumber\\ &+ 3 \left(\pm\mathfrak{m}\right) u^{2} \boldsymbol{\omega}- 2 i \pi \left(\pm\mathfrak{m}\right) u^{2} \boldsymbol{\omega} - 5 \left(\pm\mathfrak{m}\right) \boldsymbol{\omega} \log{\left(u^{2} - 1 \right)}
	\nonumber\\&+ 4 u^{2} \boldsymbol{\omega}^{2} \log{\left(2 u^{2} + 1 \right)}+ u^{2} \boldsymbol{\omega} \left(\pm6\mathfrak{m} + 5 \boldsymbol{\omega}\right) \log{\left(1 - u^{2} \right)}- 2 \boldsymbol{\omega}^{2} \log{\left(u^{2} - 1 \right)}\nonumber\\
	&\pm 5 i \pi \,\mathfrak{m} \,\boldsymbol{\omega}+ 2 \boldsymbol{\omega}^{2} \log{\left(2 u^{2} + 1 \right)}\pm 10\, \mathfrak{m}\,u^{2} \boldsymbol{\omega} \log{\left(2 u^{2} + 1 \right)}+\nonumber\\
	&\pm 5 \mathfrak{m}\,\boldsymbol{\omega} \log{\left(2 u^{2} + 1 \right)}+ 3 \mathfrak{m}^{2} u^{2} \log{\left(u^{2} - 1 \right)}- 3 i \pi \mathfrak{m}^{2} u^{2}+ 2 i \pi \boldsymbol{\omega}^{2}\Bigg\} \label{eq:Fa1NNLOouter}\, .
\end{align}

\begin{align}
	\mathcal F_{b_1}^\pm(u) ={}& - \frac{1}{72 (\boldsymbol{\omega}\pm\mathfrak{m})^{2} \boldsymbol{\Omega}_{\pm}(u) \boldsymbol{k}^{2} \left(u^{2} - 1\right)}\times\nonumber\\
	&\times\Bigg\{ (\boldsymbol{\omega}\pm\mathfrak{m}) \left(u^{2} - 1\right) \left(- 8 (\boldsymbol{\omega}\pm\mathfrak{m})\boldsymbol{\Omega}_{\pm}(u) + \boldsymbol{k}^{4} \left(2 u^{2} + 1\right)\right) \log{\left(u^{2} - 1 \right)}
	\nonumber\\ & - (\boldsymbol{\omega}\pm\mathfrak{m})  \left(u^{2} - 1\right) \left(- 8 (\boldsymbol{\omega}\pm\mathfrak{m})  \boldsymbol{\Omega}_{\pm}(u) + \boldsymbol{k}^{4} \left(2 u^{2} + 1\right)\right) \log{\left(2 u^{2} + 1 \right)}
	\nonumber\\ & + i (4 (\boldsymbol{\omega}\pm\mathfrak{m})^{2} \boldsymbol{\Omega}_{\pm}(u) \left(3 i u^{2} + 2 \pi \left(u^{2} - 1\right)\right) \nonumber\\
	&+ \boldsymbol{k}^{4} \left(u^{2} - 1\right) \left(- 3 i \left(\pm\mathfrak{m}\right) u^{2} + \pi \left(\pm\mathfrak{m}\right) \left(u^{2} - 1\right) - \pi \boldsymbol{\omega}\right))\Bigg\}\label{eq:Fb1NNLOouter}\, .
\end{align}



Let us conclude this section by showing how to write the NNLO solution in integral form, similarly to what we did for the  NLO solution. In particular, the source behaves close to the AdS boundary as
\begin{equation}
	\mathcal S^{\pm,(2)}(u)
	=
	\frac{\alpha_\pm^{(2)}}{u}
	+\mathcal O(u\log u)\,,
\end{equation}
where
\begin{equation}
	\alpha_\pm^{(2)}
	=
	(\boldsymbol{\omega}\pm\mathfrak{m})^2
	\left[
	\boldsymbol{k}^2a_O^{\pm,(1)}
	-
	(\boldsymbol{\omega}\pm\mathfrak{m})^2a_O^{\pm,(0)}
	\right]\,.
\end{equation}
Therefore, subtracting the singular part of the source as in the NLO case, the general NNLO solution can be written as
\begin{align}
	\varphi_O^{\pm,(2)}(u)
	={}&
	a_O^{\pm,(2)}
	+
	b_O^{\pm,(2)}
	\dfrac{u^2}
	{2(\boldsymbol{\omega}\pm\mathfrak{m})\boldsymbol{\Omega}_\pm(u)}
	\nonumber\\
	&+
	\int_0^u
	dx\,
	\dfrac{x}{\boldsymbol{\Omega}_\pm(x)^2}
	\left[
	\alpha_\pm^{(2)}\log x
	+
	\beta_\pm^{(2)}
	\right.
	\nonumber\\
	&\hspace{3.7cm}\left.
	+
	\int_0^x
	dy\,
	\left(
	\mathcal S^{\pm,(2)}(y)
	-
	\dfrac{\alpha_\pm^{(2)}}{y}
	\right)
	\right]\,.
	\label{eq:NNLOouterintegral}
\end{align}
Here $\beta_\pm^{(2)}$ is a finite integration constant whose value depends on the convention chosen for $b_O^{\pm,(2)}$. As at NLO, a finite shift of $\beta_\pm^{(2)}$ can be absorbed into a redefinition of $b_O^{\pm,(2)}$. The subtraction in the inner integral makes the integrand regular at the AdS boundary and implements the convention
\begin{equation}
	\varphi_O^{\pm,(2)}(0)=a_O^{\pm,(2)}\,.
\end{equation}
Substituting the explicit LO and NLO solutions into $\mathcal S^{\pm,(2)}$ and evaluating the two integrals gives the explicit NNLO solution reported in this section.

\subsection{Series expansion in the overlapping region}\label{app:expansions}

\subsubsection*{Inner solution}
In this section, we report the expansion of the inner solution in the overlapping region. The leading order solution  has the following expansion
\begin{equation}\label{app:eq:leadinINNERexp}
	\varphi_I^{\pm,(0)}(\zeta) = 1+\frac{i \,\hat\omega }{6 \zeta }-\frac{\hat\omega  (\hat\omega +2 i)}{72 \zeta ^2}+\mathcal{O}\left(\left(\frac{1}{\zeta }\right)^3\right)\,,
\end{equation}
where we normalized the solution to approach one at $\zeta\to\infty$. Then, we go to the NLO solutions. We start from the momentum correction, which can be expanded as
\begin{align}\nonumber
	\varphi_{I,\boldsymbol{k}}^{\pm,(1)}(\zeta)=&\frac{i \zeta }{2 \hat\omega }+\frac{1}{12} \left(\left(\psi\left(1-i\,\hat\omega\right)+\gamma_E\right)-1-\log (3\zeta)\right)+\\ \label{eq:INNERKexp}
	&-\frac{i \hat\omega  \left(2(\psi(-i\,\hat\omega)+\gamma_E)-3-2\log (3\zeta)\right)}{144 \zeta }+O\left(\frac{\log\zeta}{\zeta ^2}\right)
\end{align}
where $\psi(z)= \Gamma'(z)/\Gamma(z)$ is the digamma function and $\gamma_E$ is the Euler-Mascheroni constant.

Then, we expand the isospin NLO correction:
\begin{align}\label{eq:INNERnuexp}
	\varphi_{I,\mathfrak{m}}^{\pm,(1)}(\zeta)=-\frac{i(\psi(1-i\,\hat\omega)+\gamma_E)}{6} -\frac{\hat\omega  \left((\psi(1-i\,\hat\omega)+\gamma_E)-2-2\log (3\zeta)\right)}{36 \zeta }+\mathcal{O}\left(\frac{\log\zeta}{\zeta ^2}\right),
\end{align}
and the temperature NLO correction
\begin{align}\nonumber
	\varphi_{I,\tau}^{\pm,(1)}(\zeta)=&-\frac{ i \hat\omega }{36} \left(6 (\psi(1-i\,\hat\omega)+\gamma_E)+13+8 \log (3\zeta)\right)+\\ \label{eq:INNERtauexp}
	&-\frac{\hat\omega  \left(-20 \hat\omega  \log (3\zeta )+6 \hat\omega  (\psi(1-i\,\hat\omega)+\gamma_E)-25 \hat\omega -10 i\right)}{216 \zeta }+\mathcal{O}\left(\frac{\log\zeta}{\zeta ^2}\right)\,.
\end{align}

In the previous expansions, we can recognize eight terms with different dependence on $\zeta$ and $\epsilon$ that have to be matched independently
\begin{align} &1\,,& &\dfrac{1}{\zeta}\,,& &\dfrac{1}{\zeta^2}\,,& &\epsilon\,\zeta\,,& &\epsilon\,\log\zeta\,,& &\epsilon\,,& &\epsilon\,\dfrac{\log\zeta}{\zeta}\,,& &\dfrac{\epsilon}{\zeta}\,.&
\end{align}

We can rewrite the expansion of the inner solution collecting the coefficients of the terms above
\begin{align}\nonumber
	\varphi_{I}^\pm(\zeta) = &1+\frac{i \,\hat\omega }{6}\dfrac{1}{\zeta} -\frac{\hat\omega  (\hat\omega +2 i)}{72}\dfrac{1}{\zeta^2}+\dfrac{i\boldsymbol{k}^2}{2 \hat\omega}\epsilon\,\zeta+ \left(-\frac{\boldsymbol{k}^2}{12}-\dfrac{2i\hat\omega\,\tau}{9}\right)\epsilon\,\log\zeta+\\ \nonumber
	&+\bigg[\frac{\boldsymbol{k}^2}{12}\left(\psi(1-i\,\hat\omega)+\gamma_E-1-\log (3)\right)\mp\mathfrak{m}\frac{i}{6}(\psi(1-i\,\hat\omega)+\gamma_E)\\ \nonumber
	& -\frac{i \hat\omega\,\tau}{36}   \left(6 (\psi(1-i\,\hat\omega)+\gamma_E)+13+8 \log (3)\right)\bigg]\epsilon+\\ \nonumber
	&+\dfrac{\epsilon\,\log\zeta}{\zeta}\left[\dfrac{i\hat\omega\,\boldsymbol{k}^2}{72}\pm\dfrac{\mathfrak{m}\,\hat\omega}{18}+\dfrac{5\hat\omega^2\,\tau}{54}\right]+\\ \nonumber
	&+\dfrac{\epsilon}{\zeta}\Bigg[\boldsymbol{k}^2\left(-\frac{i \hat\omega  \left(
		2 ((\psi(-i\,\hat\omega)+\gamma_E))-3-2\log (3)\right)}{144}\right) +\\ \nonumber
		&\pm\mathfrak{m}\left( -\frac{\hat\omega  \left(\psi(1-i\,\hat\omega)+\gamma_E-2-2\log (3)\right)}{36}\right)\\ \nonumber
	&+\tau\left(-\frac{\hat\omega  \left(-20 \hat\omega  \log (3)+6 \hat\omega  (\psi(1-i\,\hat\omega)+\gamma_E)-25 \hat\omega -10 i\right)}{216 }\right)\Bigg] +\\
	&+\mathcal{O}(\epsilon^2,\epsilon\,\zeta^{-2}\log\zeta,\zeta^{-3})\,.
\end{align}

\subsubsection*{Outer solution}
In this section we perform the near-horizon expansion of the outer solution, that is in the overlapping region where we want to match it with the inner solution.

The outer leading-order solution expanded around the horizon reads
\begin{equation}
	\varphi_O^{\pm,(0)}(u)
	=
	Y^{\pm,(0)}_O
	-
	\dfrac{b_O^{\pm,(0)}}{\boldsymbol{\omega}^2}(1-u)+b_O^{\pm,(0)}\dfrac{\boldsymbol{\omega}\pm 4\mathfrak{m}}{2\boldsymbol{\omega}^3}(1-u)^2
	+
	\mathcal{O}\left((1-u)^3\right)\,,
\end{equation}
where we have defined the combination
\begin{equation}
	Y^{\pm,(0)}_O
	\equiv
	a_O^{\pm,(0)}
	+
	\dfrac{b_O^{\pm,(0)}}
	{2\boldsymbol{\omega}(\boldsymbol{\omega}\pm\mathfrak{m})}\,,
\end{equation}

The near-horizon expansion of the next-to-leading order solution reads
\begin{align}
	\varphi_O^{\pm,(1)}(u)
	={}&
	-\dfrac{b_O^{\pm,(0)}}{12\boldsymbol{k}^2}\dfrac{1}{1-u}
	+
	\left(
	\dfrac{b_O^{\pm,(0)}}{9\boldsymbol{k}^2}
	-
	\dfrac{\boldsymbol{k}^2}{12}Y^{\pm,(0)}_O
	\right)\log(1-u)
	+
	\widetilde Y^{\pm,(1)}_O
	\nonumber\\
	&+
	\left[
	m^{\pm,(0)}_O\log(1-u)
	+
	n^{\pm,(1)}_O
	\right](1-u)
	+
	\mathcal{O}\left((1-u)^2\log(1-u)\right)\,,
	\label{eq:NLOouterhorizonexp}
\end{align}
where $\widetilde Y_{O,(1)}^\pm$ denotes the finite constant term of the full NLO solution near the horizon and is given by
\begin{align}
	\widetilde Y^{\pm,(1)}_O
	={}&
	Y^{\pm,(1)}_O
	-
	\dfrac{\boldsymbol{k}^2Y^{\pm,(0)}_O}{12}
	\left[
	\log 2
	+
	\left(
	2\pm\dfrac{3\mathfrak{m}}{\boldsymbol{\omega}}
	\right)\log 3
	\pm
	\dfrac{\mathfrak{m}}{\boldsymbol{\omega}\pm\mathfrak{m}}
	\right]
	\nonumber\\
	&+
	b_O^{\pm,(0)}
	\left[
	\dfrac{\boldsymbol{k}^2}{8\boldsymbol{\omega}^2}\log 3
	+
	\dfrac{1}{8\boldsymbol{k}^2}
	+
	\dfrac{1}{9\boldsymbol{k}^2}\log\left(\dfrac{2}{3}\right)
	\right]\,,
	\label{eq:ctildeNLOouter}
\end{align}
where
\begin{equation}
	Y^{\pm,(1)}_O
	\equiv
	a_O^{\pm,(1)}
	+
	\dfrac{b_O^{\pm,(1)}}
	{2\boldsymbol{\omega}(\boldsymbol{\omega}\pm\mathfrak{m})}\,.
\end{equation}
The coefficient of the $(1-u)\log(1-u)$ term is
\begin{equation}
	m^{\pm,(0)}_O
	=
	\boldsymbol{k}^2
	\left[
	\left(
	\dfrac{5}{18}
	\pm
	\dfrac{\mathfrak{m}}{3\boldsymbol{\omega}}
	\right)Y^{\pm,(0)}_O
	-
	\dfrac{b_O^{\pm,(0)}}{12\boldsymbol{\omega}^2}
	\right]\,,
	\label{eq:mNLOouter}
\end{equation}
while the regular coefficient multiplying $(1-u)$ reads
\begin{align}
	n^{\pm,(1)}_O
	={}&
	-\dfrac{b_O^{\pm,(1)}}{\boldsymbol{\omega}^2}
	\nonumber\\
	&+
	\boldsymbol{k}^2Y^{\pm,(0)}_O
	\left[
	-\dfrac{17}{72}
	+
	\left(
	\dfrac{5}{18}
	\pm
	\dfrac{\mathfrak{m}}{3\boldsymbol{\omega}}
	\right)\log 2
	+
	\dfrac{(2\boldsymbol{\omega}\pm3\mathfrak{m})^2}
	{18\boldsymbol{\omega}^2}\log 3
	\right]
	\nonumber\\
	&+
	b_O^{\pm,(0)}
	\left[
	\dfrac{31}{432\boldsymbol{k}^2}
	+
	\dfrac{\boldsymbol{k}^2}{12\boldsymbol{\omega}^2}
	\left(
	1-\log 2
	-
	\left(
	2\pm\dfrac{3\mathfrak{m}}{\boldsymbol{\omega}}
	\right)\log 3
	\right)
	\right]\,.
	\label{eq:nNLOouter}
\end{align}

Notice that $Y_O^{\pm,(1)}$ is the horizon value of the homogeneous part of the NLO solution, while $\widetilde{Y}_O^{\pm,(1)}$ is the finite constant appearing in the near-horizon expansion of the complete NLO solution, including the particular solution.

The near-horizon expansion of the NNLO solution reads
\begin{align}
	\varphi^{\pm,(2)}_O(u)
	={}&
	\frac{E_{-2}^{\pm,(0)}}{(1-u)^2}
	+
	\frac{
		E_{-1,\log}^{\pm,(0)}\log(1-u)
		+
		E_{-1}^{\pm,(1)}
	}{(1-u)}
	+
	\mathcal{O}\!\left(\log(1-u)\right),
	\label{NNLOouterHorizonExpansion}
\end{align}
with
\begin{align}\label{Eminus2}
	E_{-2}^{\pm,(0)}
	={}&
	\frac{\tau\,b^{\pm,(0)}_O }{72\boldsymbol{k}^2}
	-
	\frac{\boldsymbol{\omega}^2Y^{\pm,(0)}_O }{288},
	\\ \label{Eminus1log}
	E_{-1,\log}^{\pm,(0)}
	={}&
	\frac{1}{432}
	\left[
	-3b^{\pm,(0)}_O+2\boldsymbol{\omega} \,Y^{\pm,(0)}_O (6\pm\mathfrak{m}
	+
	5\boldsymbol{\omega})
	\right],
	\\ \nonumber
	E_{-1}^{\pm,(1)}
	={}&-\frac{b^{\pm,(1)}_O}{12 k^2}+
	\frac{b^{\pm,(0)}_O}{864}\left(-\frac{20 \tau }{\boldsymbol{k}^2}\mp\frac{18 \,\mathfrak{m}  \log (3)}{\boldsymbol{\omega} }-3 (3+\log (324))\right)+\\ \nonumber
	&+\frac{Y^{\pm,(0)}_O}{864} \bigg(-12 \boldsymbol{k}^2 \tau +36 \,\mathfrak{m} ^2 \log (3)\pm24 \,\mathfrak{m} \, \boldsymbol{\omega}  (1+\log (18))+\\ \label{Eminus1}
	&+\boldsymbol{\omega}^2 (9+20 \log (2)+16 \log (3))\bigg)\,.
\end{align}

We have to convert these expansions in terms of the holographic coordinate $\zeta$ describing the AdS$_2$ region of the background spacetime. In order to do so, we recall relation \eqref{eq:zetadef}
\begin{equation}
	u = 1-\epsilon\,\tau\,\zeta\,.
\end{equation}
The expansion of the outer solution in the overlapping region, that is the near-horizon region $u\to1$, is converted into an expansion around $\zeta =\infty$ in the $\zeta$ coordinate. Thus, we can group the LO, the NLO, and the NNLO outer solutions expansion in the overlapping region as
\begin{align}\nonumber
	\varphi_{O}^\pm(\zeta)&=	\varphi^{\pm,(0)}_O(\zeta)+	\epsilon\,\varphi^{\pm,(1)}_O(\zeta)+	\epsilon^2\varphi^{\pm,(2)}_O(\zeta) +\mathcal{O}(\epsilon^3)=\\ \nonumber
	&= Y^{\pm,(0)}_O
	-
	\dfrac{b_O^{\pm,(0)}}{\boldsymbol{\omega}^2}\epsilon\,\tau\,\zeta  +b_O^{\pm,(0)}\dfrac{\boldsymbol{\omega}\pm 4\mathfrak{m}}{2\boldsymbol{\omega}^3}(\epsilon\,\tau\,\zeta)^2+\\ \nonumber
	&-\dfrac{b_O^{\pm,(0)}}{12\boldsymbol{k}^2}\dfrac{1}{\tau\,\zeta}
	+
	\left(
	\dfrac{b_O^{\pm,(0)}}{9\boldsymbol{k}^2}
	-
	\dfrac{\boldsymbol{k}^2}{12}Y^{\pm,(0)}_O
	\right)\,\epsilon\,\log(\epsilon\,\tau\,\zeta)
	+
	\widetilde Y^{\pm,(1)}_O\,\epsilon\\ \nonumber
	&+
	\left[
	m^{\pm,(0)}_O\log(\epsilon\,\tau\,\zeta)
	+
	n^{\pm,(1)}_O
	\right]\,\epsilon^2\,\tau\,\zeta +\\
	&+\frac{E_{-2}^{\pm,(0)}}{\tau^2\,\zeta^2}
	+
	\frac{
		E_{-1,\log}^{\pm,(0)}\log(\epsilon\,\tau\,\zeta)
		+
		E_{-1}^{\pm,(1)}
	}{\tau\,\zeta}\,\epsilon
	+\mathcal{O}(\epsilon^3,\zeta^{-2}\log\zeta,\zeta^{-3})\,.
\end{align}

The expansions in the overlapping region displayed in this section are the ones that we use in the main text to perform the matching at LO and NLO to extract the connection coefficients.

\section{Residues of the NLO correlator}
\label{sec:NLO-residues}

In this section, we compute analytically the residues associated with the
quasinormal modes derived in section~\ref{sec:QNMs}.  Let
$\omega_n^\pm(k)$ be a simple pole of the longitudinal polarization function.
In a neighbourhood of the pole, its Laurent expansion takes the form
\begin{equation}
	\Pi^{\parallel,\pm}(\omega,k)
	=
	\frac{Z_n^\pm(k)}{\omega-\omega_n^\pm(k)}
	+\Pi_{\mathrm{reg}}^{\parallel,\pm}(\omega,k),
	\label{eq:Laurent_polarization}
\end{equation}
where $\Pi_{\mathrm{reg}}^{\parallel,\pm}$ is regular at
$\omega=\omega_n^\pm(k)$.  The residue is therefore
\begin{equation}
	Z_n^\pm(k)
	\equiv
	\underset{\omega=\omega_n^\pm(k)}{\operatorname{Res}}\,
	\Pi^{\parallel,\pm}(\omega,k)
	=
	\lim_{\omega\to\omega_n^\pm(k)}
	\left(\omega-\omega_n^\pm(k)\right)
	\Pi^{\parallel,\pm}(\omega,k).
	\label{eq:residue_definition}
\end{equation}

Before evaluating~\eqref{eq:residue_definition}, it is useful to distinguish
the residue of a quasinormal pole from a spectral weight.  In a unitary theory,
positivity follows most directly from the Wightman function of an operator
$A$ with its Hermitian conjugate: its spectral decomposition is a sum of
absolute squares with positive thermal weights.  Through the KMS relation,
this constrains the spectral density on the real-frequency axis
\cite{Kovtun:2012rj}.  For the convention
\begin{equation}
	\rho_A(\omega,k)=-2\,\operatorname{Im}G^R_{AA^\dagger}(\omega,k),
	\label{eq:spectral-density-convention}
\end{equation}
one has $\rho_A(\omega,k)\geq0$ at positive thermal frequency.  At non-zero
chemical potential, the latter is the frequency conjugate to evolution with
the grand-canonical Hamiltonian; in a convention based on the microscopic
Hamiltonian, it is shifted by the charge of $A$ times the chemical potential.

For the charged currents, $(J^+)^\dagger=J^-$.  The corresponding
spectral function is a matrix in Lorentz indices.  Positivity does not
require each component of this matrix to be positive separately.  Rather,
its restriction to the space of physical current perturbations must be
positive semidefinite.  Namely, for any physical linear combination
\begin{equation}
	A=v_\mu J^{\mu,+},
	\qquad
	A^\dagger=v_\mu^*J^{\mu,-},
\end{equation}
the associated Wightman function is non-negative.  Through the KMS
relation, this implies the corresponding sign condition
\begin{equation}
	v_\mu\,\rho_{+-}^{\mu\nu}(\omega,k)\,v_\nu^*
	\geq 0
\end{equation}
at positive thermal frequency, where
\begin{equation}
	\rho_{+-}
	=
	i\left(G_{+-}^R-G_{+-}^{R\dagger}\right).
\end{equation}
For a scalar channel in which the correlator is diagonal, this reduces,
up to conventions, to the usual relation
$\rho=-2\,\operatorname{Im}G^R$. Consequently, no sign condition should be imposed directly on an individual
residue of the scalar coefficient $\Pi^{\parallel,\pm}$, independently of the
projector and of the other contributions to the physical correlator.

Indeed, consider a simple pole at
$\omega_\star=\Omega-i\Gamma$, with $\Gamma>0$ and residue
$Z=Z_R+iZ_I$.  Its contribution to the spectral density is
\begin{equation}
	-2\,\operatorname{Im}
	\left(\frac{Z}{\omega-\Omega+i\Gamma}\right)
	=
	\frac{2\left[\Gamma Z_R-(\omega-\Omega)Z_I\right]}
	{(\omega-\Omega)^2+\Gamma^2}.
	\label{eq:single-pole-spectral-contribution}
\end{equation}
There is no general reason for this expression to have a definite sign.
This does not conflict with unitarity because a quasinormal mode is a damped
response mode, rather than an asymptotic state whose residue is a squared
matrix element.  The latter interpretation applies to stable real-axis poles
in the Lehmann representation, but not to generic poles of the analytically
continued retarded correlator \cite{Son:2002sd}.

A simple illustration is provided by the retarded response of a damped
oscillator,
\begin{equation}
	G^R_{\mathrm{osc}}(\omega)
	=
	\frac{1}{(\omega+i\Gamma)^2-\Omega^2}
	=
	\frac{1}{2\Omega}
	\left[
	\frac{1}{\omega-\Omega+i\Gamma}
	-
	\frac{1}{\omega+\Omega+i\Gamma}
	\right].
	\label{eq:damped-oscillator-poles}
\end{equation}
The two pole terms contribute with opposite signs, whereas their sum gives
\begin{equation}
	\rho_{\mathrm{osc}}(\omega)
	=
	\frac{\Gamma}{\Omega}
	\left[
	\frac{1}{(\omega-\Omega)^2+\Gamma^2}
	-
	\frac{1}{(\omega+\Omega)^2+\Gamma^2}
	\right]
	>0,
	\qquad \omega>0.
	\label{eq:damped-oscillator-spectral-density}
\end{equation}
Thus, even for a convergent pole decomposition, positivity need not hold term
by term.  It is a property of the complete physical spectral density.

This observation does not make positivity irrelevant for a low-energy
description.  If all poles retained in an effective correlator are soft, its
spectral density should have the correct sign within the regime of validity. In the present problem, this provides a useful
check on the complete NLO correlator, which contains the hydro-like pole, the
infrared tower, their interactions, and a regular part. We did check numerically that within our NLO approximation, the spectral function associated to the polarization function $\Pi^{\parallel,\pm}$ is indeed positive for positive thermal frequencies.

Finally, Hermitian conjugation relates the two charged channels.  For real
parameters and, for simplicity, at fixed $k$,
\begin{equation}
	\left[G_{+-}^R(\omega,k)\right]^*
	=
	G_{-+}^R(-\omega^*,k),
	\label{eq:charged-correlator-reality}
\end{equation}
so that related simple poles obey
\begin{equation}
	\omega_n^-(k)=-\left[\omega_n^+(k)\right]^*,
	\qquad
	Z_n^-(k)=-\left[Z_n^+(k)\right]^*.
	\label{eq:charged-pole-residue-pairing}
\end{equation}
These relations constrain the charged pair, but do not require either residue
to be separately real or positive. We did check a posteriori that our results satisfy these relations.

\subsection{Residue of the hydro-like pole}

For the kinematically completed LO correlator~\eqref{eq:Pi_long_LO}, the
residue associated with the hydro-like pole~\eqref{eq:long:QNMLO} is
\begin{equation}
	Z_{\mathrm{hydro,LO}}^\pm(k)
	=
	\frac{(M\ell)^3w_0^2N_c}{4r_H^2}
	\left(1+\frac{r_H^2k^2}{4}\right)
	\left(\pm\mu_3+i\frac{r_Hk^2}{2}\right).
	\label{eq:LO-hydro-residue}
\end{equation}
This result satisfies~\eqref{eq:charged-pole-residue-pairing}.  At
$\mu_3=0$, its spectral weight vanishes as $k\to0$.  The diffusion pole itself
is still present for non-zero momentum and approaches the conserved zero mode
as $k\to0$; what vanishes in this limit is its residue in the polarization
function.

We now consider the NLO correlator~\eqref{BAdimensionful}.  The hydro-like
frequency through NLO is
\begin{equation}
	\omega_{\mathrm{hydro}}^\pm
	=
	\omega_{\mathrm{hydro,LO}}^\pm
	+\delta\omega_{\mathrm{hydro}}^\pm
	+\mathcal O(\mathrm{NNLO}),
	\label{eq:hydro-frequency-expansion-residue}
\end{equation}
where
\begin{align}
	\omega_{\mathrm{hydro,LO}}^\pm
	={}&
	\mp\mu_3-\frac{i r_Hk^2}{2},
	\label{eq:hydro-LO-frequency-residue}
	\\
	\delta\omega_{\mathrm{hydro}}^\pm
	={}&
	\frac{r_H^2k^2}{24}
	\left(\pm2\mu_3+i r_Hk^2\right)
	\Bigg[
	\gamma_E
	+\psi\!\left(
	1+\frac{\pm2i\mu_3-r_Hk^2}{4\pi T}
	\right)
	+\log\!\left(\frac{2\pi T r_H}{9}\right)
	\Bigg].
	\label{eq:hydro-frequency-shift-residue}
\end{align}
The approximate frequency in
\eqref{eq:hydro-frequency-expansion-residue} sets the NLO-truncated
denominator to zero up to NNLO corrections.  Its residue is therefore
understood order by order through the perturbative Laurent expansion
\begin{equation}
	\Pi_{\mathrm{NLO}}^{\parallel,\pm}(\omega,k)
	=
	\frac{
		Z_{\mathrm{hydro,LO}}^\pm(k)
		+\delta Z_{\mathrm{hydro}}^\pm(k)
	}{
		\omega-\omega_{\mathrm{hydro,LO}}^\pm
		-\delta\omega_{\mathrm{hydro}}^\pm
	}
	+\Pi_{\mathrm{reg}}^{\parallel,\pm}(\omega,k)
	,
	\label{eq:perturbative-Laurent-residue}
\end{equation}
where the pole position and its residue are both understood through NLO,
while the regular part is finite at the hydro-like pole to the order under
consideration.

Expanding the numerator and the derivative of the QNM denominator
consistently through NLO gives
\begin{align}
	Z_{\mathrm{hydro,NLO}}^\pm(k)
	\equiv{}&
	Z_{\mathrm{hydro,LO}}^\pm(k)
	+\delta Z_{\mathrm{hydro}}^\pm(k)
	\nonumber\\
	={}&
	\frac{(M\ell)^3w_0^2N_c}{8r_H^4}
	\Bigg\{
	-2r_H^2\omega_{\mathrm{hydro,LO}}^\pm
	-i r_H^3k^2
	\mathfrak D^\pm
	\!\left(\omega_{\mathrm{hydro,LO}}^\pm,k\right)
	\nonumber\\
	&
	-\frac{
		\mathfrak N^\pm
		\!\left(\omega_{\mathrm{hydro,LO}}^\pm,k\right)
	}{
		\omega_{\mathrm{hydro,LO}}^\pm\pm\mu_3
	}
	-i r_H^3k^2\omega_{\mathrm{hydro,LO}}^\pm
	\left.
	\partial_\omega\mathfrak D^\pm(\omega,k)
	\right|_{\omega=\omega_{\mathrm{hydro,LO}}^\pm}
	\Bigg\}+\nonumber\\
	&+\mathcal O(\mathrm{NNLO}).
	\label{eq:nlo-hydro-residue-before-substitution}
\end{align}
All frequency-dependent NLO functions in
\eqref{eq:nlo-hydro-residue-before-substitution} are evaluated at the LO
frequency.  Evaluating them instead at the NLO-corrected frequency would only
modify terms beyond the controlled order.

Substituting~\eqref{Npm} and~\eqref{Dpm} and simplifying yields
\begin{align}
	Z_{\mathrm{hydro,NLO}}^\pm(k)
	={}&
	-\frac{(M\ell)^3w_0^2N_c}{4r_H^2}
	\left(\mp\mu_3-\frac{i r_Hk^2}{2}\right)
	\Bigg[
	1+\frac{r_H^2k^2}{4}
	\nonumber\\
	&
	+\frac{
		r_H^2k^2\left(r_Hk^2\mp2i\mu_3\right)
	}{48\pi T}
	\psi^{(1)}\!\left(
	1+\frac{\pm2i\mu_3-r_Hk^2}{4\pi T}
	\right)
	\Bigg]+\mathcal O(\mathrm{NNLO})
	,
	\label{eq:nlo-hydro-residue-final}
\end{align}
where $\psi^{(1)}(z)=\mathrm d\psi(z)/\mathrm dz$ is the trigamma
function.  The term $r_H^2k^2/4$ is the contribution of the kinematic
completion already present in~\eqref{eq:LO-hydro-residue}.  All terms
proportional to the undifferentiated digamma function and to
$\log(2\pi T r_H/9)$ cancel between the NLO numerator, the shift of the pole,
and the derivative of the denominator.

Equation~\eqref{eq:nlo-hydro-residue-final} is reliable only while the
hydro-like pole remains parametrically isolated from the infrared tower.  The
trigamma function diverges when
\begin{equation}
	r_H^2k^2
	=
	4\pi n\,r_HT\pm2i r_H\mu_3,
	\qquad n=1,2,\ldots\, .
	\label{eq:hydro-residue-interaction-locus}
\end{equation}
This is the same locus as~\eqref{eq:psidiv}, at which the non-degenerate NLO
correction to the hydro-like frequency also becomes singular.  Neither
divergence is a singularity of the exact correlator: both signal the failure
of perturbation theory around an isolated hydro-like pole.  The local
interacting-pole description must be used in this region.

\subsection{Residues of the infrared poles}

We next compute the residues associated with the infrared poles captured by
the NLO longitudinal polarization function.  We work in physical variables
and expand around the $n$th singularity of the digamma function,
\begin{equation}
	p^{(n)}=-2\pi iTn,
	\qquad n=1,2,\ldots, .
	\label{eq:IR-leading-frequency-residue}
\end{equation}
The frequencies $p^{(n)}$ are the leading locations around
which the infrared QNMs are generated; they are not by themselves exact poles
of the full NLO correlator.  Since the residue of the digamma function at each
non-positive integer is $-1$, its local expansion is
\begin{equation}
	\psi\!\left(1-\frac{i\omega}{2\pi T}\right)
	=
	-\frac{2\pi iT}{\omega-p^{(n)} }
	+\mathcal O\!\left(
	(\omega-p^{(n)} )^0
	\right).
	\label{eq:digamma-expansion-IR-residue}
\end{equation}
It follows that the two NLO functions entering the correlator have the local
expansions
\begin{align}
	\mathfrak D^\pm(\omega,k)
	&=
	\frac{d_{-1}^{\pm,(n)}}
	{\omega-p^{(n)} }
	+\mathfrak D_{\mathrm{reg}}^{\pm,(n)}(\omega,k),
	\label{eq:D-Laurent-IR-residue}
	\\
	\mathfrak N^\pm(\omega,k)
	&=
	\frac{a_{-1}^{\pm,(n)}}
	{\omega-p^{(n)} }
	+\mathfrak N_{\mathrm{reg}}^{\pm,(n)}(\omega,k),
	\label{eq:N-Laurent-IR-residue}
\end{align}
where the regular parts are finite at
$\omega=p^{(n)} $.  From~\eqref{Npm} and
\eqref{Dpm}, the singular coefficients are
\begin{align}
	d_{-1}^{\pm,(n)}
	={}&
	-2\pi iT
	\left[
	-\frac{i r_H(p^{(n)} \pm2\mu_3)}{12}
	+\frac{r_H^2k^2
		(p^{(n)} -i r_Hk^2)}
	{24(p^{(n)} \pm\mu_3)}
	\right],
	\label{eq:dminusone-physical-IR-residue}
	\\
	a_{-1}^{\pm,(n)}
	={}&
	-\frac{i\pi T r_H^4k^2p^{(n)} }{6}
	\left(p^{(n)} \pm2\mu_3+i r_Hk^2\right).
	\label{eq:aminusone-physical-IR-residue}
\end{align}

As derived in~\eqref{eq:NLOIR}, the corresponding infrared QNM is
\begin{equation}
	\omega_{\mathrm{IR}}^{\pm,(n)}
	=
	p^{(n)}
	+\delta\omega_{\mathrm{IR}}^{\pm,(n)}+\mathcal O(\mathrm{NNLO}),
	\label{eq:IR-frequency-split-residue}
\end{equation}
with
\begin{align}
	\delta\omega_{\mathrm{IR}}^{\pm,(n)}
	={}&
	-\frac{
		(p^{(n)} \pm\mu_3)
		d_{-1}^{\pm,(n)}
	}{
		p^{(n)} \pm\mu_3+i r_Hk^2/2
	}
	\nonumber\\
	={}&
	\pi T r_H
	\frac{
		r_H^2k^4+i r_Hk^2p^{(n)}
		+2(p^{(n)} \pm\mu_3)
		(p^{(n)} \pm2\mu_3)
	}{
		6\left[
		i r_Hk^2+2(p^{(n)} \pm\mu_3)
		\right]
	}.
	\label{eq:IR-frequency-correction-residue}
\end{align}
As for the hydro-like mode, this frequency sets the NLO-truncated
denominator to zero only to the controlled perturbative order.
Accordingly, the following expression is understood as an equality between
the Laurent data through the order retained:
\begin{equation}
	\Pi_{\mathrm{NLO}}^{\parallel,\pm}(\omega,k)
	=
	\frac{Z_{\mathrm{IR,NLO}}^{\pm,(n)}(k)}
	{\omega-p^{(n)}
		-\delta\omega_{\mathrm{IR}}^{\pm,(n)}}
	+\Pi_{\mathrm{reg}}^{\parallel,\pm,(n)}(\omega,k).
	\label{eq:IR-perturbative-Laurent}
\end{equation}

At the leading non-vanishing order of the infrared residue, the numerator of
the kinematically completed polarization function reduces consistently to
\begin{equation}
	i r_H^3k^2\omega
	-\mathfrak N^\pm(\omega,k).
	\label{eq:IR-effective-numerator}
\end{equation}
The remaining terms generated by the kinematic completion start one order
higher in the infrared residue.  Using~\eqref{eq:N-Laurent-IR-residue}, the
effective numerator at the corrected pole is
\begin{align}
	&i r_H^3k^2p^{(n)}
	-\frac{a_{-1,\mathrm{phys}}^{\pm,(n)}}
	{\delta\omega_{\mathrm{IR}}^{\pm,(n)}}=
	\frac{
		4i r_H^3k^2p^{(n)}
		(p^{(n)} \pm\mu_3)
	}{
		2(p^{(n)} \pm\mu_3)-i r_Hk^2
	}+\mathcal O(\mathrm{NNLO}).
	\label{eq:IR-numerator-at-pole}
\end{align}
Writing the QNM denominator as
\begin{equation}
	\mathcal H^\pm(\omega,k)
	=
	\omega\pm\mu_3+\frac{i r_Hk^2}{2}
	+(\omega\pm\mu_3)
	\mathfrak D^\pm(\omega,k),
	\label{eq:Hphys-IR-residue}
\end{equation}
its derivative is
\begin{equation}
	\partial_\omega\mathcal H^\pm
	=
	1+\mathfrak D^\pm
	+(\omega\pm\mu_3)
	\partial_\omega\mathfrak D^\pm.
	\label{eq:Hprime-IR-residue}
\end{equation}
The double pole in the last term of~\eqref{eq:Hprime-IR-residue} controls the
derivative at the shifted infrared root.  Combining it with
\eqref{eq:IR-numerator-at-pole} and with the additional factor
$1/(\omega\pm\mu_3)$ in the polarization function gives
\begin{align}
	Z_{\mathrm{IR}}^{\pm,(n)}(k)
	={}&
	-\frac{(M\ell)^3w_0^2N_c}{8r_H^4}
	\frac{
		\left(\delta\omega_{\mathrm{IR}}^{\pm,(n)}\right)^2
	}{
		(p^{(n)} \pm\mu_3)^2
		d_{-1}^{\pm,(n)}
	}
	\left[
	i r_H^3k^2p^{(n)}
	-\frac{a_{-1}^{\pm,(n)}}
	{\delta\omega_{\mathrm{IR}}^{\pm,(n)}}
	\right]+\mathcal O(\mathrm{NNLO}).
	\label{eq:IR-residue-intermediate}
\end{align}
Substitution of the explicit coefficients yields
\begin{equation}
	Z_{\mathrm{IR}}^{\pm,(n)}(k)
	=
	\frac{
		i\pi T(M\ell)^3w_0^2N_c\,k^2p^{(n)}
		\left[p^{(n)} \pm2\mu_3+i r_Hk^2\right]
	}{
		6\left[
		i r_Hk^2+2(p^{(n)} \pm\mu_3)
		\right]^2
	}+\mathcal O(\mathrm{NNLO}).
	\label{eq:IR-residue-final}
\end{equation}

The isolated infrared residue has a double divergence at
\begin{equation}
	r_Hk^2=4\pi Tn\pm2i\mu_3.
	\label{eq:IR-residue-interaction-locus}
\end{equation}
This is the same interaction locus at which the separate hydro-like and
infrared dispersion relations cease to be uniform approximations.  The
divergence is therefore not a divergence of the exact correlator.  It signals
that the two nearby modes cannot be assigned independent residues and must be
treated using the local interacting-pole description.

\subsection{Residues of the interacting poles}
\label{sec:interacting-pole-residues}

We now compute the residues of the two poles described by the local
interaction equation  \eqref{int:roots}.  The isolated hydro-like
and infrared residues derived above become singular when the corresponding
non-degenerate expansions cease to be valid.  In the interaction region, the
two roots must instead be kept together before their Laurent coefficients are
extracted.

We first identify the local form of the correlator.  It is useful to denote
the numerator of the kinematically completed NLO polarization function by
\begin{align}
	\mathcal U^\pm(\omega,k)
	\equiv{}&
	\left(1-\frac{(\omega\pm\mu_3)^2}{k^2}\right)
	\Big[
	i r_H^3k^2\omega
	-\mathfrak N^\pm(\omega,k)
	\nonumber\\
	&\hspace{4.2cm}
	+i r_H^3\omega(\omega\pm\mu_3)^2
	\Big].
	\label{eq:int-res-Uphysical}
\end{align}
The completed correlator~\eqref{BAdimensionful} can then be written as
\begin{equation}
	\Pi_{\mathrm{NLO}}^{\parallel,\pm}(\omega,k)
	=
	\frac{(M\ell)^3w_0^2N_c}{8r_H^4}
	\frac{\mathcal U^\pm(\omega,k)}
	{(\omega\pm\mu_3)\mathcal H^\pm(\omega,k)},
	\label{eq:int-res-completed-correlator}
\end{equation}
where
\begin{equation}
	\mathcal H^\pm(\omega,k)
	=
	\omega\pm\mu_3+\frac{i r_Hk^2}{2}
	+(\omega\pm\mu_3)
	\mathfrak D^\pm(\omega,k)
	\label{eq:int-res-Hphysical}
\end{equation}
is the QNM denominator used above.

For the selected infrared level, the same local approximation that leads to the quadratic QNM relation and its roots \eqref{int:roots}
 gives
\begin{align}
	&\left(\omega-p^{(n)}\right)
	\mathcal H^\pm(\omega,k)
	\nonumber\\
	&\hspace{1cm}=
	\left(\omega-\bar\omega_\text{hydro}^{\pm,(n)}\right)
	\left(\omega-p^{(n)}\right)
	+(\omega\pm\mu_3)
	d_{-1}^{\pm,(n)}
	+\mathcal{O}(\text{NNLO}),
	\label{eq:int-res-denominator-relation}
\end{align}
where $\bar\omega_\text{hydro}^{\pm,(n)}$ is defined in \eqref{int:regularhydro}. The remainder in \eqref{eq:int-res-denominator-relation} refers to the
local expansion of the QNM denominator before its zeros are determined.
Its effect on the roots is not uniform when the two roots become
degenerate. Although $\mathfrak N^\pm$ and therefore
$\mathcal U^\pm$ have a simple pole at
$\omega=p^{(n)}$, the product
$(\omega-p^{(n)})\mathcal U^\pm$ is regular
there.  Multiplying both the numerator and denominator of
\eqref{eq:int-res-completed-correlator} by
$\omega-p^{(n)}$, we obtain the local correlator
\begin{align}
	\Pi_{\mathrm{NLO}}^{\parallel,\pm}(\omega,k)\bigg|_\text{int.}
	\simeq{}&
	\frac{(M\ell)^3w_0^2N_c}{8r_H^4}\times	\nonumber\\
	&
\times	\frac{
		(\omega-p^{(n)})
		\mathcal U^\pm(\omega,k)
	}{
		(\omega\pm\mu_3)
		\Big[
		(\omega-\bar\omega_\text{hydro}^{\pm,(n)})
		(\omega-p^{(n)})
		+(\omega\pm\mu_3)
		d_{-1}^{\pm,(n)}
		\Big]
	}.
	\label{eq:int-res-local-correlator}
\end{align}
Here and below, $\simeq$ denotes equality within the local NLO
approximation.  The omitted terms are higher order in the local QNM
denominator before its roots are determined; they cannot in general be
represented by a uniform additive remainder for the correlator
arbitrarily close to a pole. The factor $\omega-p^{(n)}$ in the numerator is essential.  It
arises from the same multiplication that converts the meromorphic QNM
condition into the quadratic equation \eqref{int:localequation}.

We first assume that the two roots
$\omega_{\mathrm{int},\eta}^{\pm,(n)}$ in
\eqref{int:roots} are distinct.  Differentiating the
quadratic denominator gives
\begin{align}
	&\partial_\omega
	\Big[
	(\omega-\bar\omega_\text{hydro}^{\pm,(n)})
	(\omega-p^{(n)})
	+(\omega\pm\mu_3)
	d_{-1}^{\pm,(n)}
	\Big]=
	2\omega-p^{(n)}
	-\bar\omega_\text{hydro}^{\pm,(n)}
	+d_{-1}^{\pm,(n)}.
	\label{eq:int-res-quadratic-derivative}
\end{align}
At either interacting root, this derivative is
\begin{align}
	&2\omega_{\mathrm{int},\eta}^{\pm,(n)}
	-p^{(n)}
	-\bar\omega_\text{hydro}^{\pm,(n)}
	+d_{-1}^{\pm,(n)}
	\nonumber\\
	&\quad=
	\eta
	\left[
	\left(
	p^{(n)}-\bar\omega_\text{hydro}^{\pm,(n)}
	+d_{-1}^{\pm,(n)}
	\right)^2
	-4(p^{(n)}\pm\mu_3)
	d_{-1}^{\pm,(n)}
	\right]^{1/2}.
	\label{eq:int-res-derivative-at-root}
\end{align}
The two simple-pole residues are consequently
\begin{align}
	Z_{\mathrm{int},\eta}^{\pm,(n)}(k)
	={}&
	\frac{(M\ell)^3w_0^2N_c}{8r_H^4}
	\frac{
		\left(\omega_{\mathrm{int},\eta}^{\pm,(n)}
		-p^{(n)}\right)
		\mathcal U^\pm
		\!\left(\omega_{\mathrm{int},\eta}^{\pm,(n)},k\right)
	}{
		\eta\left(\omega_{\mathrm{int},\eta}^{\pm,(n)}\pm\mu_3\right)
	}
	\nonumber\\
	&\times
	\left[
	\left(
	p^{(n)}-\bar\omega_\text{hydro}^{\pm,(n)}
	+d_{-1}^{\pm,(n)}
	\right)^2
	-4(p^{(n)}\pm\mu_3)
	d_{-1}^{\pm,(n)}
	\right]^{-1/2},
	\qquad \eta=\pm1.
	\label{eq:int-res-simple-poles}
\end{align}
 Equation
\eqref{eq:int-res-simple-poles} is valid only when the square root is
non-zero. We can verify that~\eqref{eq:int-res-simple-poles} reproduces the NLO
hydro-like and IR residues away from the interaction region,  up to
corrections beyond the controlled order
\begin{align}
	Z_{\mathrm{int},+}^{\pm,(n)}(k)
	\to
	Z_{\mathrm{hydro,NLO}}^\pm(k)
	\,,
	\label{eq:int-res-hydro-recovery}\\
	Z_{\mathrm{int},-}^{\pm,(n)}(k)
	\to
	Z_{\mathrm{IR,NLO}}^{\pm,(n)}(k),
	\label{eq:int-res-IR-recovery}
\end{align}
where $Z_{\mathrm{hydro,NLO}}^{\pm,(n)}$, and  $Z_{\mathrm{IR,NLO}}^{\pm,(n)}$ are given explicitly in
\eqref{eq:nlo-hydro-residue-final}  and \eqref{eq:IR-residue-final} respectively.

The approach of the regularly corrected hydro-like level to
$p^{(n)}$ does not, by itself, produce a double pole.  Within the local quadratic approximation, the two interacting roots
coincide when
\begin{equation}
	\left(
	p^{(n)}-\bar\omega_\text{hydro}^{\pm,(n)}
	+d_{-1}^{\pm,(n)}
	\right)^2
	-4(p^{(n)}\pm\mu_3)
	d_{-1}^{\pm,(n)}
	=0.
	\label{eq:int-res-coalescence-condition}
\end{equation}
At this exceptional locus, the common root is
\begin{equation}
	\omega_\star^{\pm,(n)}
	=
	\frac{
		p^{(n)}+\bar\omega_\text{hydro}^{\pm,(n)}
		-d_{-1}^{\pm,(n)}
	}{2},
	\label{eq:int-res-coalesced-frequency}
\end{equation}
and the quadratic denominator becomes
$(\omega-\omega_\star^{\pm,(n)})^2$.  The local correlator then has the
Laurent expansion
\begin{equation}
	\Pi_{\mathrm{NLO}}^{\parallel,\pm}(\omega,k)\bigg|_\text{int.}
	=
	\frac{C_{-2}^{\pm,(n)}(k)}
	{(\omega-\omega_\star^{\pm,(n)})^2}
	+
	\frac{C_{-1}^{\pm,(n)}(k)}
	{\omega-\omega_\star^{\pm,(n)}}
	+\mathcal O\!\left((\omega-\omega_\star^{\pm,(n)})^0\right).
	\label{eq:int-res-double-pole-Laurent}
\end{equation}
Provided that
$\omega_\star^{\pm,(n)}\pm\mu_3\neq0$ and
\begin{equation}
	\left(\omega_\star^{\pm,(n)}-p^{(n)}\right)
	\mathcal U^\pm
	\!\left(\omega_\star^{\pm,(n)},k\right)
	\neq0,
	\label{eq:int-res-no-cancellation}
\end{equation}
the repeated zero of the denominator produces a genuine double pole.  Its
leading Laurent coefficient is
\begin{equation}
	C_{-2}^{\pm,(n)}(k)
	=
	\frac{(M\ell)^3w_0^2N_c}{8r_H^4}
	\frac{
		\left(\omega_\star^{\pm,(n)}-p^{(n)}\right)
		\mathcal U^\pm
		\!\left(\omega_\star^{\pm,(n)},k\right)
	}{
		\omega_\star^{\pm,(n)}\pm\mu_3
	}.
	\label{eq:int-res-double-pole-leading}
\end{equation}
The coefficient of the accompanying simple-pole term, namely the residue at
the second-order pole, is
\begin{align}
	C_{-1}^{\pm,(n)}(k)
	={}&
	\frac{(M\ell)^3w_0^2N_c}{8r_H^4}
	\left.
	\partial_\omega
	\left[
	\frac{
		(\omega-p^{(n)})
		\mathcal U^\pm(\omega,k)
	}{
		\omega\pm\mu_3
	}
	\right]
	\right|_{\omega=\omega_\star^{\pm,(n)}}.
	\label{eq:int-res-double-pole-residue}
\end{align}
For generic values of the parameters, the numerator in
\eqref{eq:int-res-no-cancellation} does not vanish.  The repeated zero
of the local denominator then produces a genuine double pole within
the local NLO approximation.  A reduction to a simple pole would
require an additional zero of the numerator at the same frequency;
such a pole-zero coincidence is an independent, non-generic
condition.

For non-zero discriminant, the two residues in
\eqref{eq:int-res-simple-poles} diverge with opposite signs as the roots
coalesce.  Their divergence is an artefact of resolving a double pole into
two simple poles.  The full local correlator has the finite limiting form
\eqref{eq:int-res-double-pole-Laurent}; at the coalescence point it must be
described by $C_{-2}^{\pm,(n)}$ and $C_{-1}^{\pm,(n)}$, rather than by two
separate simple-pole residues.

Equations~\eqref{eq:int-res-simple-poles},
\eqref{eq:int-res-double-pole-leading}, and
\eqref{eq:int-res-double-pole-residue} give the two complementary
descriptions required for distinct and coalescing roots.  The local
correlator~\eqref{eq:int-res-local-correlator} is the uniform object across
both regimes.  The construction remains local in the selected infrared
level and assumes that the other infrared poles are parametrically farther
away.  Moreover, the coalescence condition
\eqref{eq:int-res-coalescence-condition} identifies a double zero within the
local NLO approximation.

\section{Quasinormal modes from two definitions of the NLO correlator}\label{app:otherdef}

\begin{figure}[htb]
	\begin{subfigure}{\textwidth}
		\centering
		\begin{subfigure}{0.45\textwidth}
			\centering
			\includegraphics[width=\textwidth]{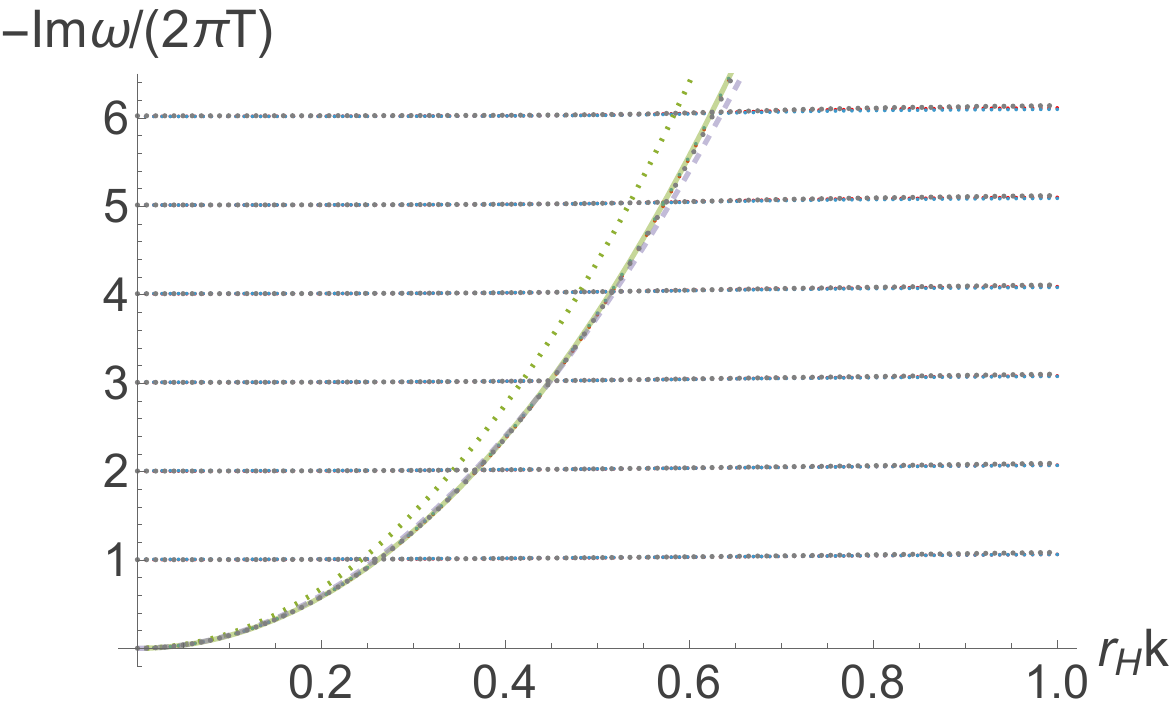}
		\end{subfigure}
		\hspace{1cm}
		\begin{subfigure}{0.45\textwidth}
			\centering
			\includegraphics[width=\textwidth]{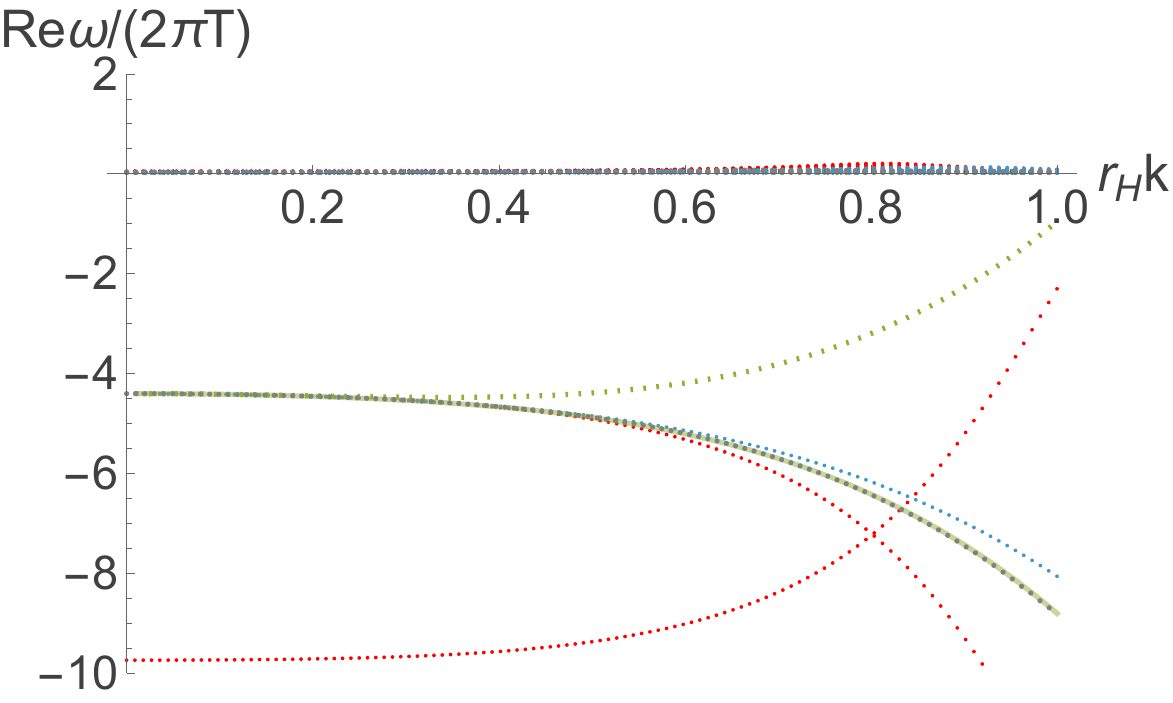}
		\end{subfigure}
	\end{subfigure}
	\caption{Real and imaginary parts of the first longitudinal QNMs at $\mu_q/T=300$ and $\mu_3/\mu_q=1.2/13$. We compare the zeros of the denominator of \eqref{BAepsilon}, eq.~\eqref{app:q1} (red dots), with the zeros of \eqref{q1} (blue dots) and with the numerical QNMs (grey squares). The other curves are as in figure~\ref{fq1}.}
	\label{app:fq1}
\end{figure}

\begin{figure}[htb]
	\begin{subfigure}{\textwidth}
		\centering
		\begin{subfigure}{0.45\textwidth}
			\centering
			\includegraphics[width=\textwidth]{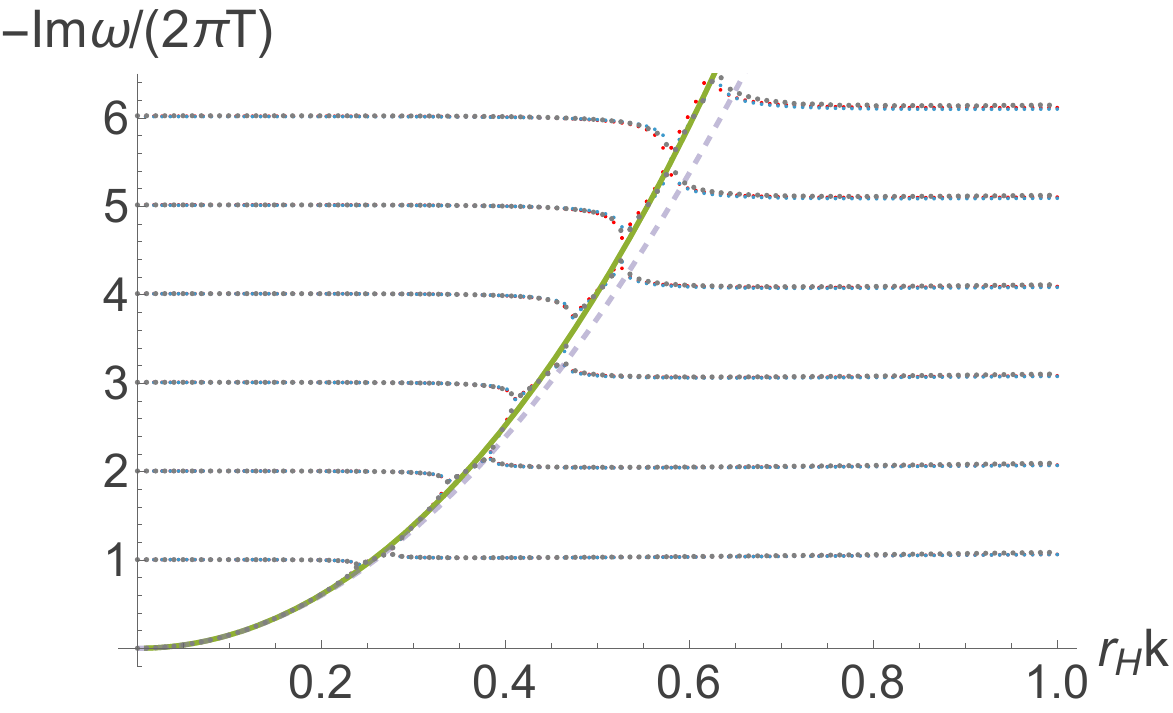}
		\end{subfigure}
		\hspace{1cm}
		\begin{subfigure}{0.45\textwidth}
			\centering
			\includegraphics[width=\textwidth]{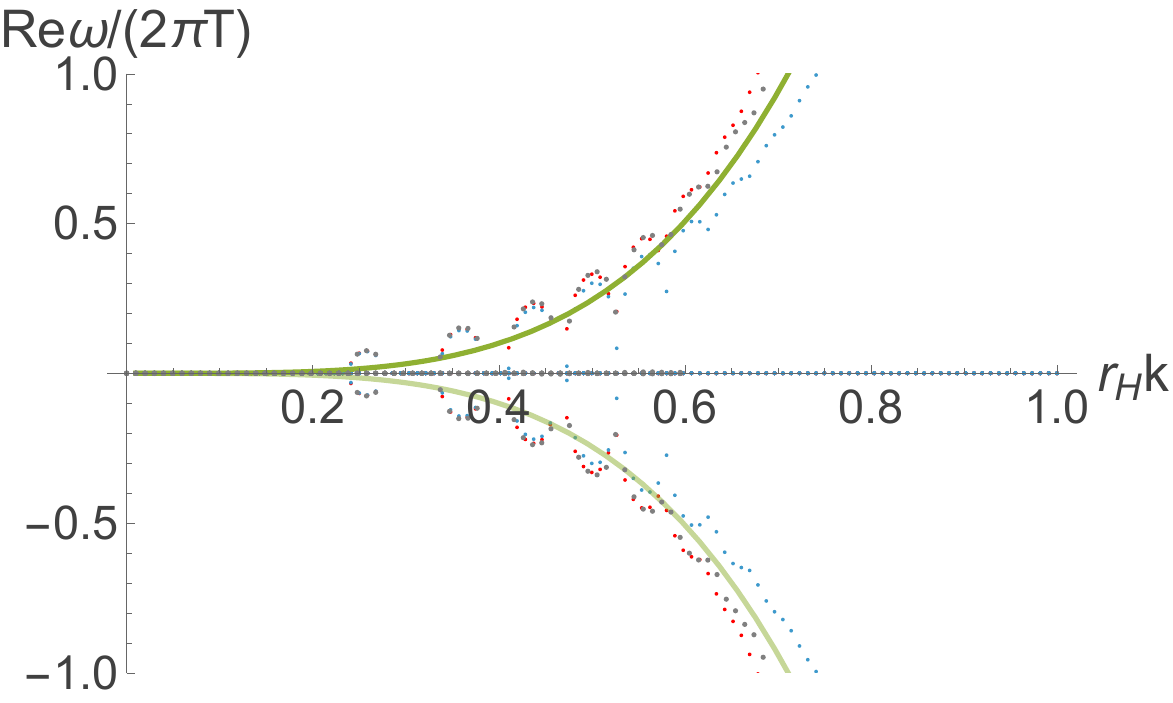}
		\end{subfigure}
	\end{subfigure}
	\caption{Same as figure~\ref{app:fq1} but for $\mu_3=0$.}
	\label{app:fq4}
\end{figure}

\begin{figure}[htb]
	\begin{subfigure}{\textwidth}
		\centering
		\begin{subfigure}{0.45\textwidth}
			\centering
			\includegraphics[width=\textwidth]{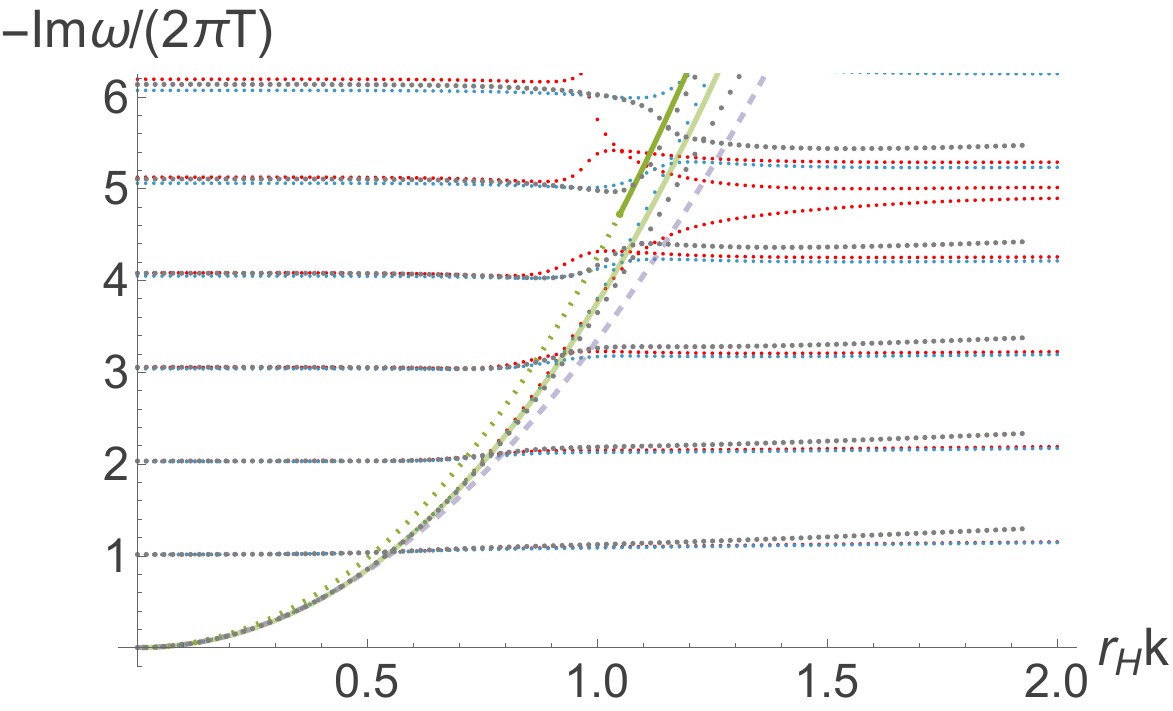}
		\end{subfigure}
		\hspace{1cm}
		\begin{subfigure}{0.45\textwidth}
			\centering
			\includegraphics[width=\textwidth]{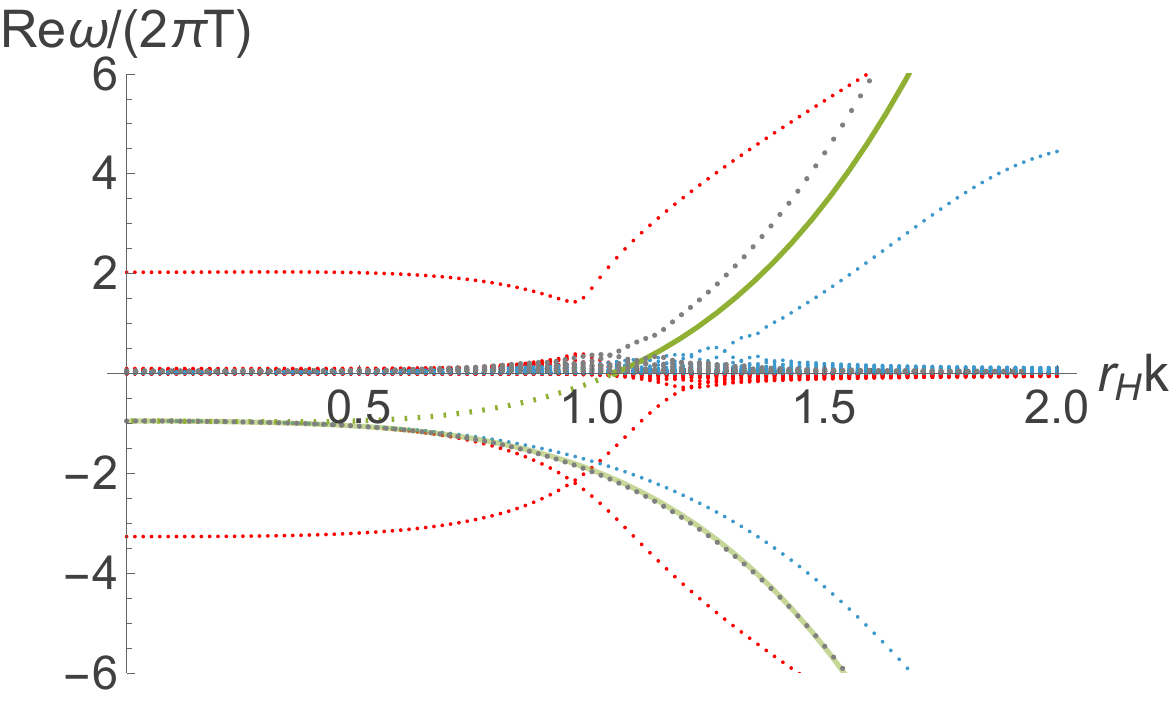}
		\end{subfigure}
	\end{subfigure}
	\caption{Same as figure~\ref{app:fq1} but for $\mu_q/T=65$.}
	\label{app:fq2}
\end{figure}

\begin{figure}[htb]
	\begin{subfigure}{\textwidth}
		\centering
		\begin{subfigure}{0.45\textwidth}
			\centering
			\includegraphics[width=\textwidth]{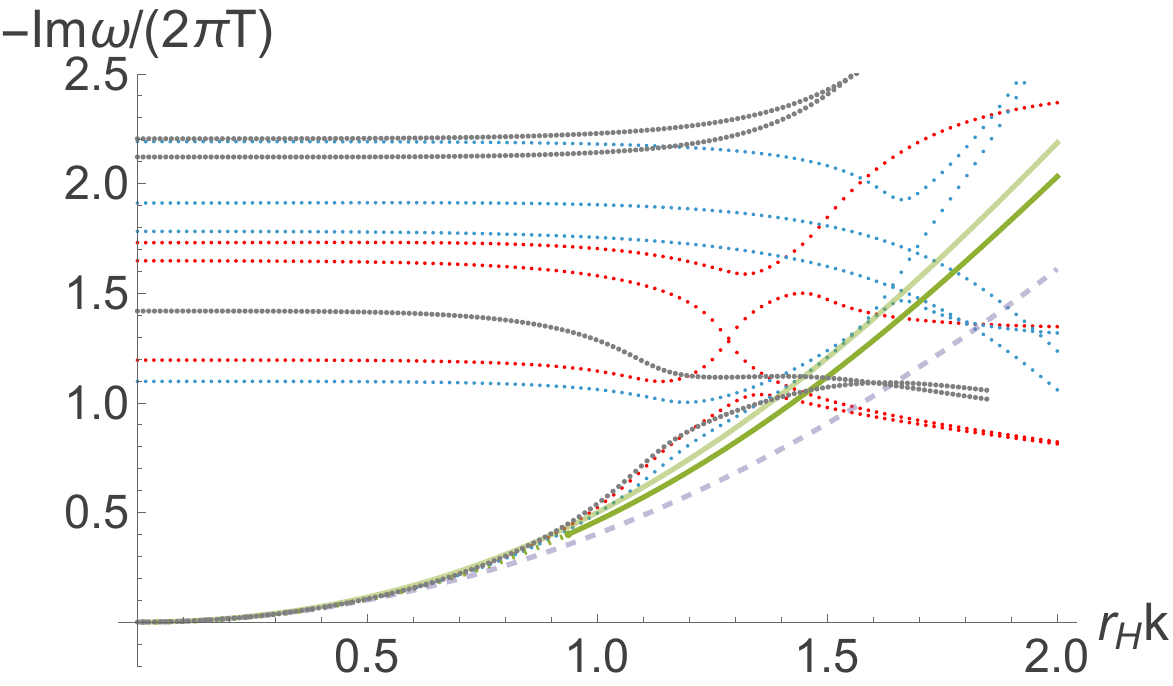}
		\end{subfigure}
		\hspace{1cm}
		\begin{subfigure}{0.45\textwidth}
			\centering
			\includegraphics[width=\textwidth]{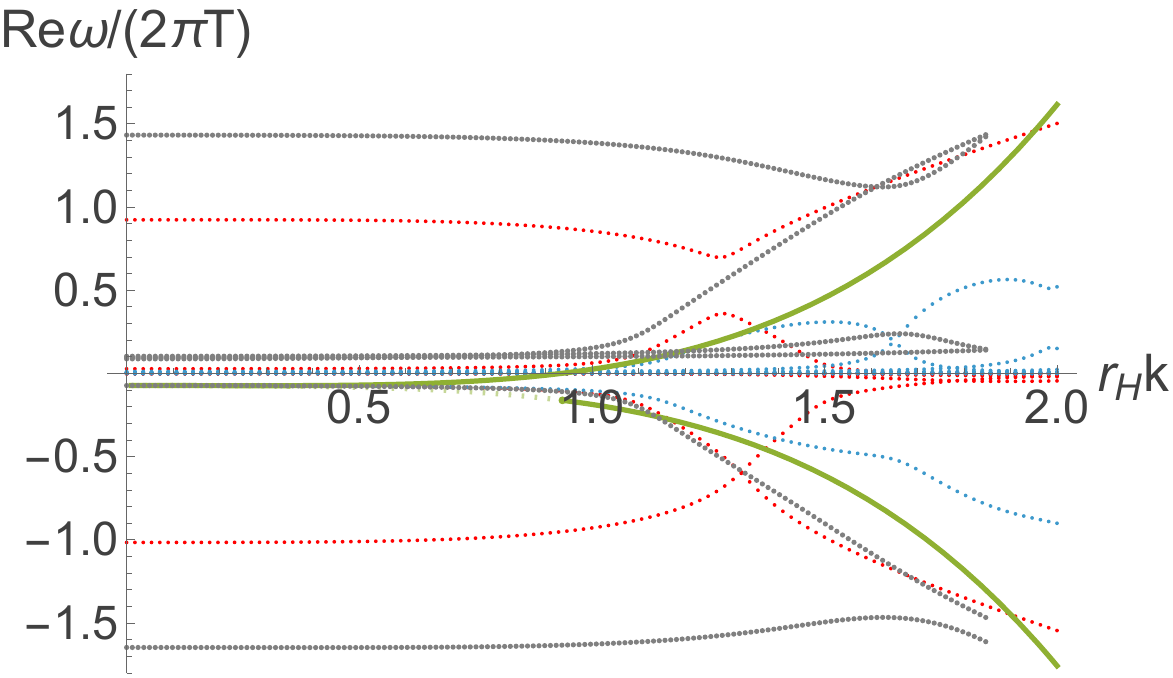}
		\end{subfigure}
	\end{subfigure}
	\caption{Same as figure~\ref{app:fq1} but for $\mu_q/T=5$.}
	\label{app:fq3}
\end{figure}

\begin{figure}[htb]
	\begin{subfigure}{\textwidth}
		\centering
		\begin{subfigure}{0.45\textwidth}
			\centering
			\includegraphics[width=\textwidth]{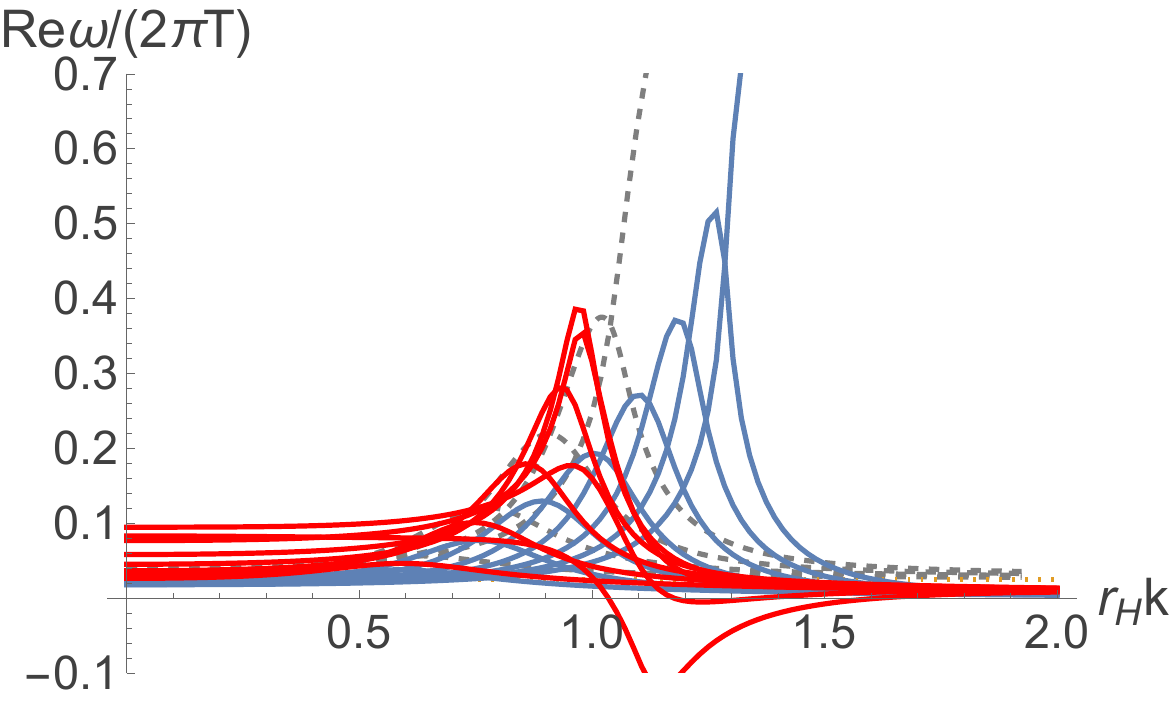}
		\end{subfigure}
		\hspace{1cm}
		\begin{subfigure}{0.45\textwidth}
			\centering
			\includegraphics[width=\textwidth]{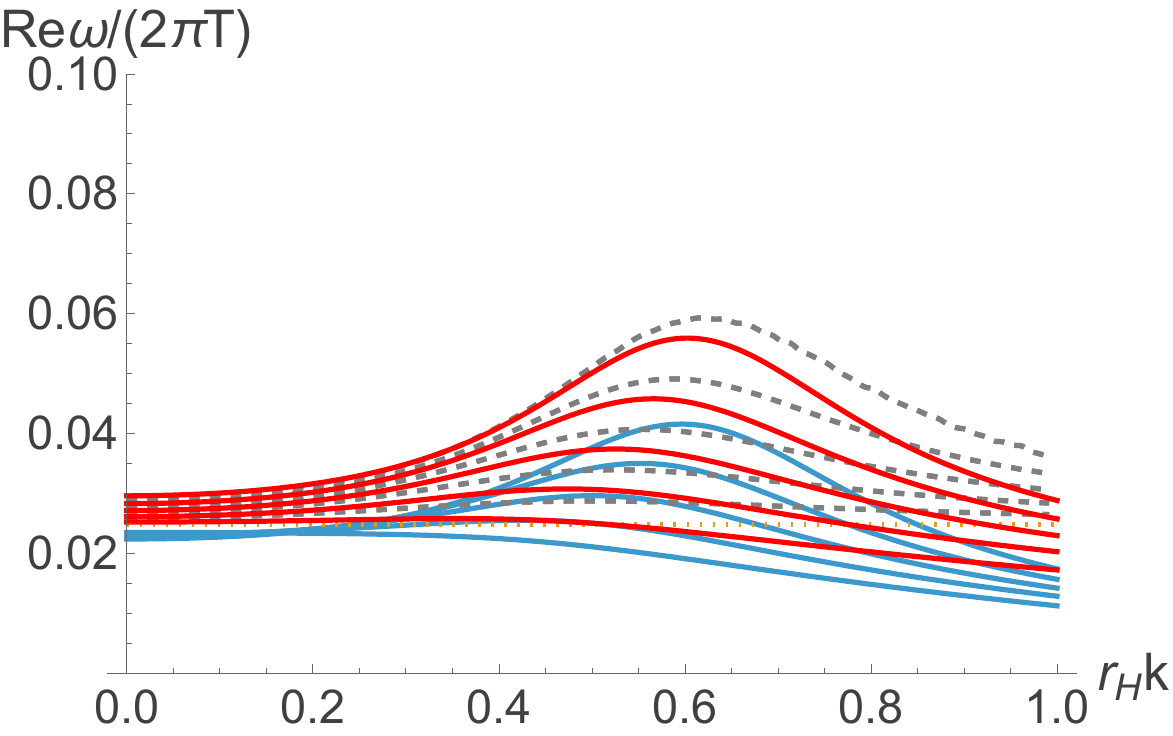}
		\end{subfigure}
	\end{subfigure}
	\caption{Zoom on the positive real parts at $\mu_q/T=65$ (left) and $\mu_q/T=300$ (right), as in figure~\ref{fq5}: numerical QNMs (dashed grey), zeros of \eqref{q1} (blue) and zeros of \eqref{app:q1} (red).}
	\label{app:fq5}
\end{figure}

In section~\ref{qnm} the NLO QNMs were computed as the zeros of \eqref{q1}, i.e.\ from the form \eqref{eq:BAtilde} of the correlator. In this appendix we compare them with the zeros of the denominator of the original form \eqref{BAepsilon},
\begin{equation}
	\label{app:q1} \left(\bm{\omega}_{\text{QNM}}^{\text{NLO}}\pm \mathfrak{m}\right) + \frac{i}{2}\bm{k}^2 + \left(\bm{\omega}_{\text{QNM}}^{\text{NLO}}\pm \mathfrak{m}\right)\mathfrak{D}^\pm(\bm{\omega}_{\text{QNM}}^{\text{NLO}},\bm{k}^2,\mathfrak{m},\tau) = 0 \, ,
\end{equation}
with $\mathfrak{D}^\pm$ defined in \eqref{Dpm}. As in the main text, we consider the channel $+$. The two conditions agree at NLO, and the analytic QNM formulas of section~\ref{sec:QNMs} follow from either of them, but their exact zeros differ at $\mathcal O(\epsilon^2)$. The comparison therefore illustrates the size of the corrections beyond the accuracy of the NLO matching. The zeros of \eqref{app:q1} are shown as red dots in figures~\ref{app:fq1}--\ref{app:fq5}, together with the zeros of \eqref{q1} (blue dots) and the numerical QNMs (grey squares).

At $\mu_q/T=300$ and $\mu_3/\mu_q = 1.2/13$ (figure~\ref{app:fq1}), the two sets of zeros are indistinguishable for the imaginary parts of the IR modes and of the hydro-like mode. The real part of the hydro-like zero of \eqref{app:q1}, however, departs from the numerical data already for $r_Hk\gtrsim0.5$, considerably more than the zero of \eqref{q1}. The zeros of \eqref{app:q1} also contain an additional branch, starting from large negative real part at $k=0$, which has no counterpart among the numerical QNMs nor among the zeros of \eqref{q1}. Along such branches the NLO term in \eqref{app:q1} is as large as the leading-order one, so they lie outside the domain of validity of the expansion and should not be regarded as approximations to physical modes. Branches of this kind appear for all the backgrounds considered.

At $\mu_q/T=300$ and $\mu_3=0$ (figure~\ref{app:fq4}) the two sets of zeros are very close. Both reproduce the collisions of the hydro-like mode with the IR levels and the opposite real parts acquired by the colliding pairs. They differ only in the real parts of the pairs emerging from the collisions, for $r_Hk\gtrsim0.6$: the zeros of \eqref{app:q1} grow slightly faster than the zero-temperature curves \eqref{eq:twopoles}, and those of \eqref{q1} slightly slower.

At $\mu_q/T=65$ and $\mu_3/\mu_q = 1.2/13$ (figure~\ref{app:fq2}) the differences are larger. Beyond $r_Hk\simeq1$, the imaginary part of the hydro-like zero of \eqref{app:q1} saturates, while the numerical hydro-like mode moves towards larger damping, as does the corresponding zero of \eqref{q1}. In the same region the zeros of \eqref{app:q1} contain a branch whose real part grows with momentum (red dots starting at Re$\omega/(2\pi T)\approx -3.25$ at $k=0$) and whose imaginary part also saturates. It belongs to the additional branches discussed above and cannot be identified with the numerical second hydro-like mode. The zeros of \eqref{q1}, instead, contain a second hydro-like branch that detaches from the IR tower, as discussed in section~\ref{qnm}. The IR zeros of the two conditions also differ visibly for the higher levels shown (see figure \ref{app:fq5}).

At $\mu_q/T=5$ and $\mu_3/\mu_q = 1.2/13$ (figure~\ref{app:fq3}) both sets of zeros are far from the numerical QNMs, apart from the hydro-like mode at small momentum, as expected since $T/\mu_q$ is not small. The IR zeros of the two conditions lie at visibly different positions, consistently with the fact that the corrections beyond NLO are not small for this background, and the zeros of \eqref{app:q1} again contain additional branches.

Figure~\ref{app:fq5} shows the real parts of the IR modes for $\mu_q/T=65$ (left panel) and $\mu_q/T=300$ (right panel) both at $\mu_3/\mu_q = 1.2/13$. Here the comparison is reversed: the zeros of \eqref{app:q1} have larger real parts than those of \eqref{q1}, closer to the numerical data, in particular at $\mu_q/T=300$ (right panel). However, the second hydro-like is captured only by \eqref{q1} and not from \eqref{app:q1}, even if it starts from a large level (the eighth instead of the fifth). Since the real parts of the IR modes arise entirely at NLO, they are especially sensitive to the $\mathcal O(\epsilon^2)$ terms by which the two conditions differ.

Taken together, these comparisons show that neither set of exact zeros is uniformly closer to the numerical QNMs: the zeros of \eqref{q1} describe better the hydro-like modes and are free of spurious branches, while those of \eqref{app:q1} give more accurate real parts for the IR modes. The difference between the two sets provides an estimate of the corrections beyond NLO, which are not controlled by the present matching.

\bibliographystyle{JHEP}
\addcontentsline{toc}{section}{References}
\bibliography{Ref.bib}

\end{document}